\documentclass[a4paper,fleqn,usenatbib,useAMS]{mnras}
\usepackage{float}
\usepackage{graphicx}
\usepackage{times}
\usepackage{amstext}
\usepackage{amsmath}
\usepackage{amssymb}	
\usepackage{natbib}
\usepackage{tabularx}
\usepackage{mathtools}
\usepackage{mathrsfs}
\usepackage{graphics}
\usepackage{multirow}
\usepackage{ar}

\newcommand{\kms}{\,km\,s$^{-1}$} 
\newcommand{\ntwoh}{N$_{2}$H$^{+}$}

\newcommand{\tone}{$J=1\rightarrow0$}

\newcommand{\vel}{km\,s$^{-1}$\,pc$^{-1}$}

\newcommand{\solar}{M$_{\odot}$}
\newcommand{\irdc}{${\rm G}035.39-00.33$}
\newcommand{\farc}{\overset{\prime\prime}{.}}

\usepackage[T1]{fontenc}
\usepackage{ae,aecompl}

\usepackage{mathptmx}
\usepackage{txfonts}

\title[The fragmentation of G035.39--00.33]{Investigating the structure and fragmentation of a highly filamentary IRDC\thanks{Based  on observations   carried  out   with  the   IRAM  Plateau   de  Bure Interferometer. IRAM  is supported by INSU/CNRS   (France),  MPG   (Germany)  and   IGN   (Spain).   }}

\author[J. D. Henshaw et al.]{J. D. Henshaw$^{1,2}$ \thanks{Contact e-mail: j.d.henshaw@ljmu.ac.uk},
P. Caselli$^{3}$, F. Fontani$^{4}$,  I. Jim\'{e}nez-Serra$^{5,6}$, J. C. Tan$^{7}$, 
\newauthor S. N. Longmore$^{1}$, J. E. Pineda$^{3}$, R. J. Parker$^{1}$, and A. T. Barnes$^{1}$ \\
$^{1}$ Astrophysics Research Institute, Liverpool John Moores University, Liverpool, L3 5RF, UK\\
$^{2}$ School of Physics and Astronomy, University of Leeds, Leeds LS2 9JT, UK\\
$^{3}$ Max-Planck Institute for Extraterrestrial Physics, Giessenbachstrasse 1, 85748 Garching, Germany\\
$^{4}$ INAF-Osservatorio Astrofisico di Arcetri, Largo E. Fermi 5, 50125, Firenze, Italy\\
$^{5}$ School of Physics and Astronomy, Queen Mary University of London, Mile End Road, London E1 4NS\\
$^{6}$ University College London, 132 Hampstead Road, London, NW1 2PS, UK\\
$^{7}$ Departments of Astronomy \& Physics, University of Florida, Gainesville, FL, 32611, USA
}

\date{Last updated XXX; in original form YYY}

\pubyear{2016}

\begin{document}
\label{firstpage}
\pagerange{\pageref{firstpage}--\pageref{lastpage}}
\maketitle

\begin{abstract}
We present $3.7$\,arcsec ($\sim0.05$\,pc) resolution 3.2\,mm dust continuum observations from the Institut  de Radioastronomie  Millim\'etrique Plateau de Bure Interferometer, with the aim of studying the structure and fragmentation of the filamentary infrared dark cloud (IRDC) \irdc. The continuum emission is segmented into a series of 13 quasi-regularly spaced ($\lambda_{\rm obs}\sim0.18$\,pc) cores, following the major axis of the IRDC. We compare the spatial distribution of the cores with that predicted by theoretical work describing the fragmentation of hydrodynamic fluid cylinders, finding a significant (factor of $\gtrsim8$) discrepancy between the two. Our observations are consistent with the picture emerging from kinematic studies of molecular clouds suggesting that the cores are harboured within a complex network of independent sub-filaments. This result emphasizes the importance of considering the underlying physical structure, and potentially, dynamically important magnetic fields, in any fragmentation analysis. The identified cores exhibit a range in (peak) beam-averaged column density ($3.6~\times~10^{23}\,{\rm cm^{-2}}<~N_{\rm H,c}<8.0~\times10^{23}\,{\rm cm^{-2}}$ ), mass ($8.1\,{\rm M_{\odot}}<M_{\rm c}<26.1\,{\rm M_{\odot}}$), and number density ($6.1\times10^{5}\,{\rm cm^{-3}}<n_{\rm H, c, eq}<14.7\times10^{5}\,{\rm cm^{-3}}$). Two of these cores, dark in the mid-infrared, centrally-concentrated, monolithic (with no traceable substructure at our PdBI resolution), and with estimated masses of the order $\sim20-25$\,\solar, are good candidates for the progenitors of intermediate-to-high-mass stars. Virial parameters span a range $0.2<\alpha_{\rm vir}<1.3$. Without additional support, possibly from dynamically important magnetic fields with strengths of the order of $230\,\mu{\rm G}<B<670\,\mu{\rm G}$, the cores are susceptible to gravitational collapse. These results may imply a multilayered fragmentation process, which incorporates the formation of sub-filaments, embedded cores, and the possibility of further fragmentation.

\end{abstract}




\section{Introduction}\label{Section:introduction}

Although a basic mechanism for the specific case of isolated low-mass star formation has been investigated over several decades (e.g. \citealp{shu_1987}), a  more generalized model, one that incorporates a consistent description for the formation of high-mass ($>8$\,\solar) stars, is still lacking. An important step in developing a holistic understanding of the star formation process is identifying and categorising the initial phases of high-mass star formation. Ultimately this requires detailed knowledge of their host molecular clouds.  

Discovered as silhouettes against the bright Galactic mid-infrared background, infrared dark clouds (hereafter, IRDCs) were quickly identified as having the potential to aid our understanding of the star formation process \citep{perault_1996, egan_1998}. Initial studies set about categorizing their physical properties, finding broad ranges in size ($\sim1-10$\,pc, with rare examples exceeding 50\,pc e.g. `Nessie'; \citealp{jackson_2010}), mass ($10^{2}-10^{5}$\,\solar), and column density ($\sim10^{22}-10^{25}\,{\rm cm}^{-2}$) (e.g. \citealp{carey_1998, egan_1998, rathborne_2006, simon_2006b}). Subsequent investigations categorising their temperatures ($\lesssim25\,{\rm K}$; \citealp{pillai_2006, peretto_2010, ragan_2011, fontani_2012, chira_2013}), chemistry (e.g. \citealp{sakai_2008, gibson_2009, izaskun_2010, vasyunina_2011, sanhueza_2012, sanhueza_2013, pon_2015, lackington_2016}), and kinematics (e.g. \citealp{devine_2011, henshaw_2013, henshaw_2014, peretto_2013, peretto_2014, tackenberg_2014, dirienzo_2015, schneider_2015, pon_2015b}) have ensued. Although the broad range in characteristics dictates that not all IRDCs will form high-mass stars \citep{kauffmann_2010}, identifying massive and relatively quiescent molecular clouds (those that are yet to be globally influenced by feedback effects from massive young stellar objects) is a crucially important step in understanding the initial conditions for high-mass star formation. 

The focus of this paper is \irdc, a massive ($\sim2\times10^{4}$\,\solar; see table 1 of \citealp{kainulainen_2013}) and filamentary IRDC, with a kinematic distance of 2.9\,kpc \citep{simon_2006b}. Originally part of a sample selected from the \citet{rathborne_2006} study by \citet{butler_2009}, \irdc \ was selected for further investigation due to its high-contrast against the Galactic mid-infrared background, and because it exhibits extended regions with no obvious tracers of star formation activity (4.5, 8, and 24\,\micron \ emission; \citealp{carey_2009, chambers_2009}). Since 2010, \irdc \ has been the focal point of a dedicated research effort whose aim is to provide a detailed case study of the physical structure, chemistry, and dynamics of a single IRDC.

The first results of this case study were presented by \citet[Paper I]{izaskun_2010}, who identified faint, but widespread, SiO emission throughout \irdc. Although a currently undetected population of low-mass protostars may account for this emission, \citet{izaskun_2010} discussed the possibility that such a signature may represent a ``fossil record'' of either the cloud formation process or a cloud merger. 

The potential to use \irdc \ as a laboratory for studying the early phases of the star formation process is supported by both \citet[Paper II]{hernandez_2011} and \citet[Paper VII]{barnes_2016}, who report widespread depletion of CO and widespread emission from deuterated species (in this case, ${\rm N_{2}D^{+}}$), respectively. These two observations emphasize the presence of cold ($<20\,{\rm K}$) and dense gas, yet to be globally affected by stellar feedback, where CO molecules have frozen on to the surface of dust grains leading to an enhancement in the abundance of deuterated nitrogen-bearing molecules. Comparing the observed abundance of deuterated species with that predicted by chemical models \citep{kong_2015}, \citet{barnes_2016} estimate the age of the cloud to be $\sim3$\,Myr old. This may imply, therefore, that although star formation within the cloud remains within an early evolutionary phase, the cloud itself is \emph{dynamically} old. Having existed for 5-10 local free-fall times, \irdc \ may have had sufficient time to settle into a state of near virial equilibrium, as concluded by \citet[Paper III]{hernandez_2012a}. 

The results of \citet{izaskun_2010} implied that \irdc \ comprises multiple sub-clouds. This was investigated by both \citet[Paper IV]{henshaw_2013} and \citet[Paper V]{izaskun_2014}, who performed systematic studies of the kinematics and structure of \irdc. On the largest scales, at least three line-of-sight kinematic features are present, with each exhibiting a unique velocity and density structure \citep{izaskun_2014}. This was confirmed by \citet[Paper VI]{henshaw_2014}, who performed the first high angular resolution ($\sim5$\,arcsec) study of the dense gas kinematics throughout the cloud, using observations of the \tone \ transition of \ntwoh \ taken with the Plateau de Bure Interferometer (hereafter, PdBI). It was revealed that the IRDC comprises a complex network of morphologically distinct molecular filaments. Moreover, \citet{henshaw_2014} found evidence to suggest that the kinematics of the gas are locally influenced by the presence of dense, and in some cases, starless, continuum sources. 

In this paper (VIII), we revisit the PdBI 3.2\,mm continuum emission data, which was first presented by \citet{henshaw_2014} for qualitative comparison with the \ntwoh \ (1-0) molecular line kinematics. Our primary aim is to investigate the structure, fragmentation process, and star formation potential of \irdc \ via quantitative analysis of the dust continuum emission. Details of the observations can be found in Section~\ref{Section:observations}. In Section~\ref{Section:results}, we discuss the method used to systematically identify structure within the continuum data and discuss this in the context of the complex kinematics of \irdc. Section~\ref{Section:analysis} contains our quantitative analysis of the continuum data. We begin with a discussion on the spatial distribution of the identified continuum cores and how this compares to predictions from theoretical work describing the fragmentation of fluid cylinders, before turning our attention to the cores themselves. In Section~\ref{Section:discussion_sf}, we discuss the implications of our findings in the context of star formation throughout \irdc. Finally, in Section~\ref{Section:conclusions} we summarize our findings and suggest possible avenues for future research.

\section{Observations}\label{Section:observations}

The  3.2\,mm  continuum  observations  were  carried  out  using  the Institut  de Radioastronomie  Millim\'etrique (IRAM) PdBI, France. A 6-field mosaic was used to cover the inner area  of IRDC \irdc \ (the dotted circles in Fig.~\ref{Figure:msd_map} depict the primary beam of the PdBI at 3.2\,mm $\sim54$\,arcsec). The final map size is $\sim40\,{\rm arcsec}\times150$\,arcsec (corresponding to $\sim0.6\,{\rm pc}\times2.1\,{\rm pc}$). 

Observations were carried out over six days in 2011 May, June and October, in the C and D configurations (using six and five antennas, respectively) offering baselines between 19\,m and 176\,m. Emission on scales larger than $\sim1.2(\lambda/D)\sim42$\,arcsec ($\sim0.6$\,pc), where $D=19$\,m, is filtered out by the interferometer. The 3.2\,mm continuum data was {\sc clean}ed using the Hogbom algorithm with natural weighting. This results in a synthesized beam of $\{\theta_{\rm maj},\,\theta_{\rm min}\}=\{4.3\,{\rm arcsec},\,3.1\,{\rm arcsec}\}=\{0.06\,{\rm pc},\,0.04\,{\rm pc}\}$, with a position angle of 18\fdg3. Line-free channels give a total bandwidth of $\sim$\,3\,GHz. The map noise level, $\sigma_{\rm rms}$, estimated from emission free regions, is $\sim0.07$\,mJy\,beam$^{-1}$. The reference position used to determine the relative offset positions used throughout this paper is $\alpha\,({\rm J}2000)=18^{\rm h}57^{\rm m}08^{\rm s}0$, $\delta\,({\rm J}2000)=02\degr10\arcmin30\farc0$. We refer the reader to \citet{henshaw_2014} for more details on the observations.

In addition to the 3.2\,mm continuum data, complementary PdBI \ntwoh \ (1-0) observations, first presented in \citet{henshaw_2014}, are also utilized throughout this work. These data have been combined with existing IRAM\,30\,m observations to incorporate missing short spacing information into the interferometric map. Following post-processing, the spatial and spectral resolution of the \ntwoh \ (1$-$0) data are $\sim5$\,arcsec and $0.14$\,\kms, respectively.

\section{Observational Results}\label{Section:results}

\subsection{Structure identification using continuum data}\label{Section:results_cont}

\begin{figure}
\begin{center}
\includegraphics[trim = 40mm 35mm 15mm 50mm, clip, width = 0.48\textwidth]{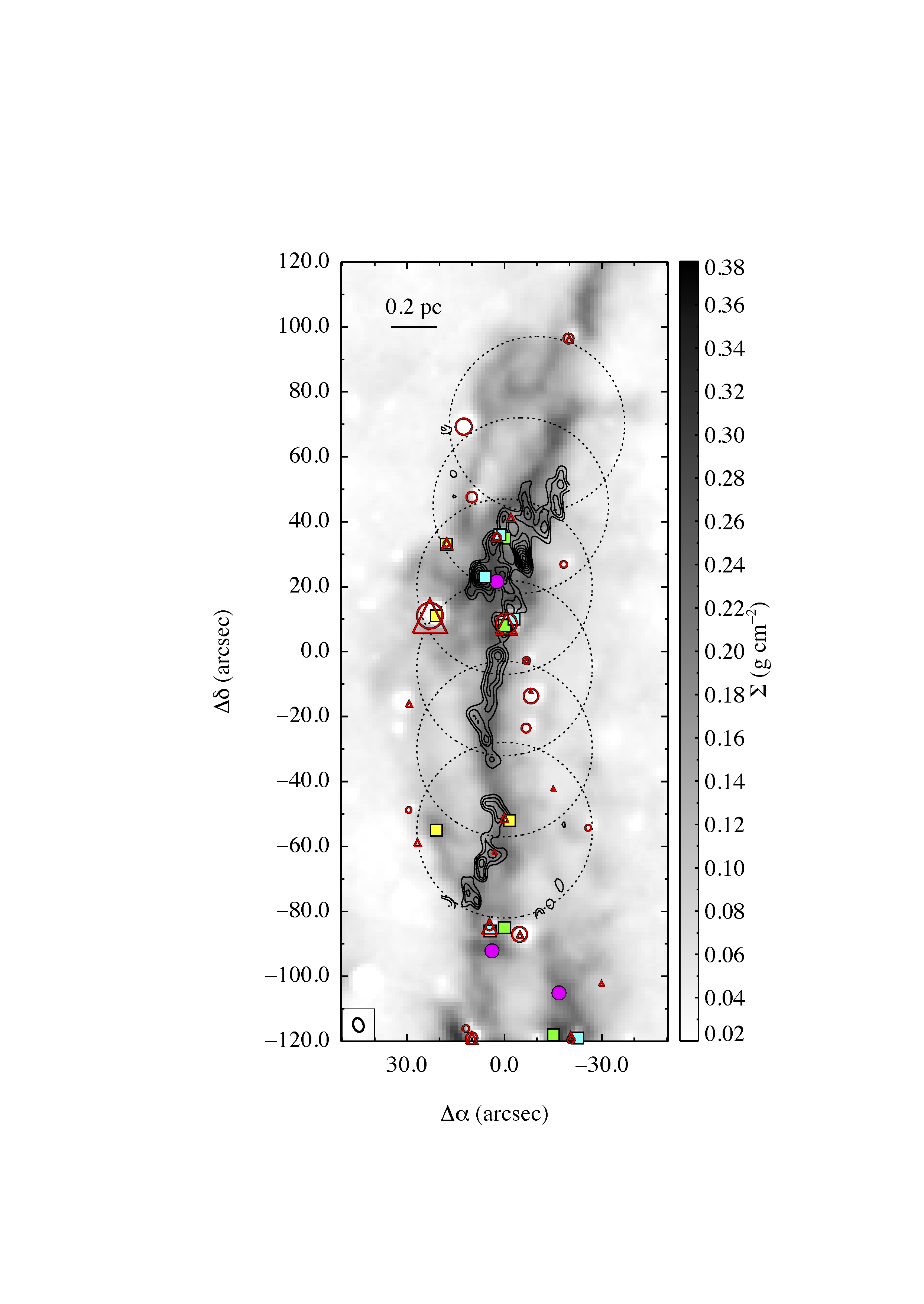}
\end{center}
\caption{The combined mid- and near-infrared extinction-derived mass surface density map of \citet[grey-scale]{kainulainen_2013} overlaid with 3.2\,mm continuum contours. Contour levels start at 3\,$\sigma_{\rm rms}$ and increase in steps of 2\,$\sigma_{\rm rms}$ (where $\sigma_{\rm rms}\approx0.07$\,mJy\,beam$^{-1}$). Large dotted circles indicate the 6-field mosaic obtained with the PdBI. Filled magenta circles indicate the locations of high-mass cores reported by \citet{butler_2012}. Filled cyan and yellow squares refer to the high-mass ($>20$\solar) and  low-mass ($<20$\solar) dense cores identified by \citet{nguyen_2011} from \emph{Herschel} observations. Open red circles and red triangles refer to the 8 and 24\,\micron\ emission, respectively \citep{carey_2009, izaskun_2010}. The symbol sizes are scaled by the source flux. Filled green squares highlight the location of extended 4.5\,\micron\  emission \citep{chambers_2009}.  }
\label{Figure:msd_map}
\end{figure}
\begin{figure}
\begin{center}
\includegraphics[trim = 40mm 35mm 25mm 50mm, clip, width = 0.45\textwidth]{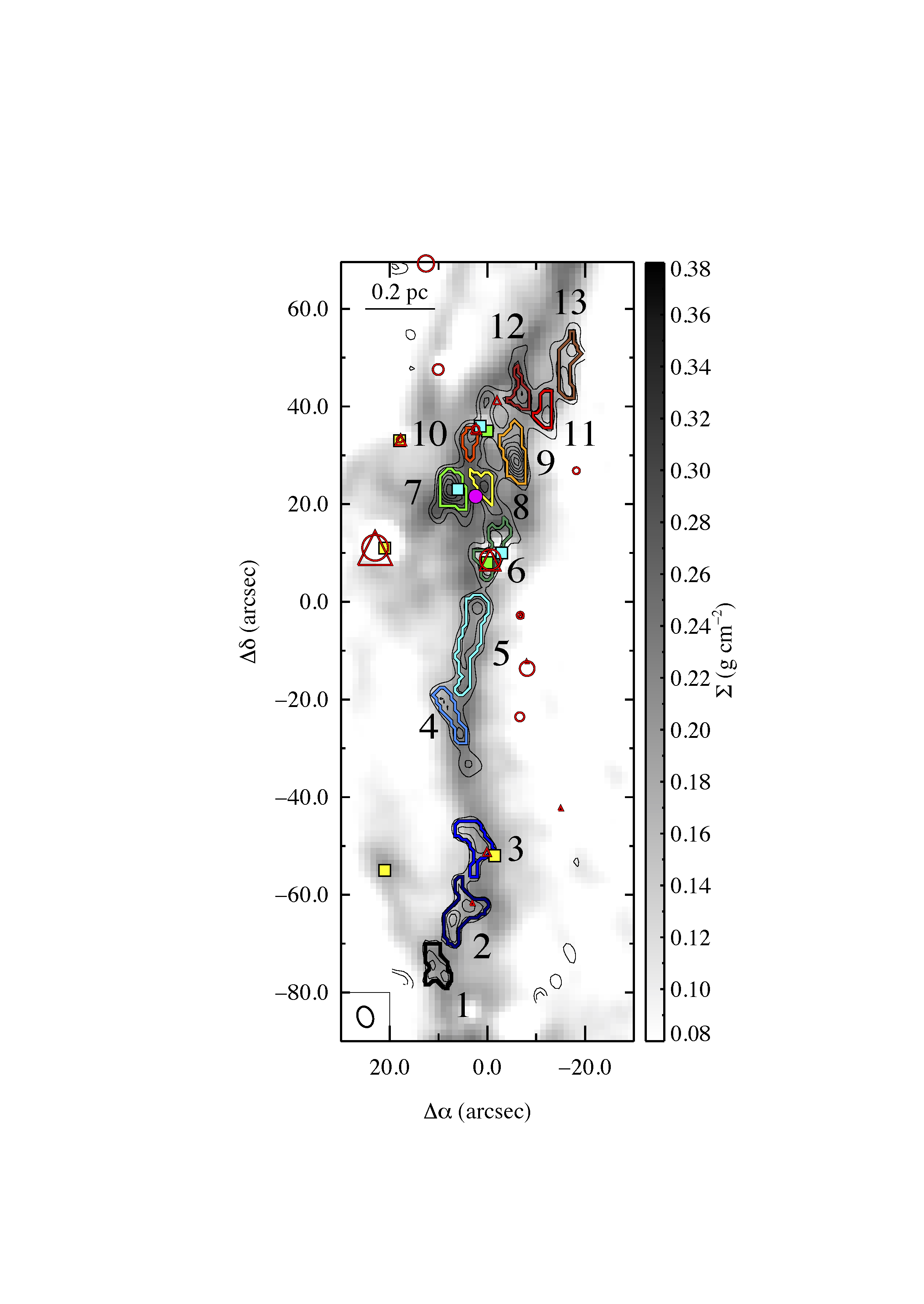}
\end{center}
\caption{Highlighting the projected location of the dendrogram leaves discussed in \S~\ref{Section:results}. Each leaf is denoted with an ID number and a coloured contour. Information relating to each leaf can be found in Table~\ref{Table:basic_leaf_info}. The background image is the mass surface density map of \citet{kainulainen_2013} and all symbols are defined in Fig.~\ref{Figure:msd_map}.}
\label{Figure:cont_dendro}
\end{figure}

Fig.~\ref{Figure:msd_map} shows the spatial extent of the PdBI 6-field mosaic (dotted circles) overlaid on the combined mid- and near-infrared extinction-derived mass surface density map of \irdc \ \citep{kainulainen_2013}. The black contours highlight the 3.2\,mm continuum emission. The locations of extended 4.5 (extended green objects; \citealp{cyganowski_2008}), 8 and 24\,\micron \ emission, indicating locations of embedded star formation, appear as green squares, red circles, and red triangles, respectively \citep{carey_2009,chambers_2009, izaskun_2010}. Qualitatively, the continuum emission appears closely related to regions of high mass surface density ($0.15\,{\rm g\,cm^{-2}}<\Sigma<0.32\,{\rm g\,cm^{-2}}$). However, there are notable exceptions to this. For instance, there is a lack of continuum emission towards $\{\Delta\alpha,\,\Delta\delta\}=\{-20.0\,{\rm arcsec},\,70.0\,{\rm arcsec}\}$. Such discrepancies may be due to a lack of sensitivity (our $3\sigma_{\rm rms}$ column density sensitivity is $N_{\rm H}\gtrsim10^{23}{\rm cm^{-2}}$; see \S~\ref{Section:analysis_phys}) or the result of missing short spacings in our interferometric map. 

\begin{table*}
	\caption{Dendrogram leaves: basic information. } \vspace{0.2cm}
	
	\centering  
	\tabcolsep=0.5cm \normalsize{
	\begin{tabular}{ l  c  c  c  c  c  c  c  c }
	\hline
	ID & 
	$\Delta\alpha^{a}$ & 
	$\Delta\delta^{a}$ & 
	$R_{\rm min}^{b}$ & 
	$R_{\rm maj}^{b}$ & 
	$\langle R\rangle^{b}$ & 
	\AR$^{c}$ & 
	$N_{\rm pix}A_{\rm pix}^{d}$ & 
	$R_{\rm eq}^{e}$  
	\\ [0.5ex]
	
 	& 
	(${\rm arcsec}$) & 
	(${\rm arcsec}$) & 
	(${\rm arcsec}$) & 
	(${\rm arcsec}$) &  
	(${\rm arcsec}$) & 
	& 
	(${\rm arcsec}$$^{2}$) & 
	(${\rm arcsec}$)   
	\\ [0.5ex]
	
	\hline 
 1 &        8.9 &      -76.9 &       2.03 &       3.80 &       2.78 &       1.87 &      43.90 &       3.74      \\ [0.5ex]
 2 &        7.4 &      -64.7 &       2.80 &       5.11 &       3.78 &       1.82 &      68.16 &       4.66      \\ [0.5ex]
 3 &        1.3 &      -50.3 &       2.52 &       4.56 &       3.39 &       1.81 &      53.72 &       4.14      \\ [0.5ex]
 4 &        5.9 &      -26.7 &       1.34 &       5.50 &       2.71 &       4.11 &      39.28 &       3.54      \\ [0.5ex]
 5 &        2.1 &       -0.9 &       1.81 &       8.56 &       3.93 &       4.73 &      84.91 &       5.20      \\ [0.5ex]
 6 &       -0.2 &        7.5 &       2.02 &       5.22 &       3.25 &       2.58 &      51.98 &       4.07      \\ [0.5ex]
 7 &        7.4 &       22.7 &       2.23 &       3.26 &       2.69 &       1.46 &      45.63 &       3.81      \\ [0.5ex]
 8 &        0.5 &       23.5 &       1.62 &       3.05 &       2.22 &       1.88 &      25.99 &       2.88      \\ [0.5ex]
 9 &       -6.3 &       28.8 &       1.91 &       4.54 &       2.94 &       2.38 &      57.18 &       4.27      \\ [0.5ex]
10 &        2.8 &       34.1 &       1.38 &       2.87 &       1.99 &       2.08 &      21.95 &       2.64      \\ [0.5ex]
11 &      -12.4 &       38.7 &       1.53 &       3.31 &       2.25 &       2.16 &      27.15 &       2.94      \\ [0.5ex]
12 &       -7.1 &       42.5 &       1.66 &       3.57 &       2.43 &       2.16 &      27.15 &       2.94      \\ [0.5ex]
13 &      -16.9 &       43.2 &       1.60 &       5.70 &       3.02 &       3.55 &      45.63 &       3.81      \\ [0.5ex]
	\hline
	\end{tabular}
\vspace{0.2cm}

\begin{minipage}{0.98\textwidth}\footnotesize{
$^{a}$ Offset location of peak leaf emission.\\
$^{b}$ Semi-minor, $R_{\rm min}$, semi-major, $R_{\rm maj}$ axes, and the geometric mean. \\
$^{c}$ Leaf aspect ratio; $\AR\equiv R_{\rm maj}/R_{\rm min}$. \\
$^{d}$ Leaf area. \\
$^{e}$ Equivalent radius; $R_{\rm eq}\equiv(N_{\rm pix}A_{\rm pix}/\pi)^{1/2}$.
}
\end{minipage}
}
\label{Table:basic_leaf_info}
\end{table*}

The continuum emission is highly structured. There are a number of prominent emission peaks arranged along the major axis of the IRDC. To investigate this further, we use dendrograms \citep{rosolowski_2008}. Specifically, our analysis makes use of {\sc astrodendro}, a {\sc python} package used to compute dendrograms of astronomical data.\footnote{For more information see: \url{http://www.dendrograms.org}} As well as providing a systematic approach to structure identification, dendrograms are also less sensitive to variation in the input parameters in comparison to alternative methods \citep{pineda_2009}. Additionally, dendrograms can be used to identify hierarchical structure, which is desirable in complex regions. The following parameters are used in computing the dendrogram: ${\rm min\_value}=~3\sigma_{\rm rms}$ (the minimum intensity considered in the analysis); ${\rm min\_delta}=~2\sigma_{\rm rms}$ (the minimum spacing between isocontours; using ${\rm min\_delta}=~1\sigma_{\rm rms}$ has no effect on the identified structure); ${\rm min\_npix}=26$ (the minimum number of pixels contained within a structure). The angular resolution of our observations is used to determine ${\rm min\_npix}=\frac{2\pi}{8{\rm ln}(2)}\frac{\theta_{\rm maj}\theta_{\rm min}}{A_{\rm pix}}$, where $A_{\rm pix}$ is the area of a single ($0.76\,{\rm arcsec}\times0.76\,{\rm arcsec}$) pixel. 

Fig.~\ref{Figure:cont_dendro} shows the result of the dendrogram analysis. Here we highlight the location of the dendrogram `leaves', the highest level of the dendrogram hierarchy, representing the smallest structures identified. A total of 13 leaves are identified. Each is denoted by an ID number designated in order of increasing offset declination. Equivalent radii of the leaves range between $0.04\,{\rm pc}~<~R_{\rm eq}~<~0.07\,{\rm pc}$ (with a median value of $R_{\rm eq}=0.05$\,pc; note that $\sqrt{\theta_{\rm maj}\theta_{\rm min}}\sim3.7\,{\rm arcsec}$ which is $\sim0.05$\,pc), whereby $R_{\rm eq}\equiv(N_{\rm pix}A_{\rm pix}/\pi)^{1/2}$, and $N_{\rm pix}$ is the number of pixels associated with a given leaf. The equivalent radius refers to that of a circle which covers an area equivalent to $N_{\rm pix}A_{\rm pix}$. These radii are, on average, 30 per cent larger than the geometric mean radii computed from the intensity-weighted second moment output from {\sc astrodendro} (see Table~\ref{Table:basic_leaf_info} and discussion by \citealp{rosolowsky_2006}). Selecting the equivalent radius therefore represents a conservative approach to estimating the physical properties of the leaves (e.g. number densities; \S~\ref{Section:analysis_phys}). The median aspect ratio of the dendrogram leaves is $2.2$. However, leaves \#4 and \#5 appear to be more filamentary, with aspect ratios $>4$ (although it is questionable as to whether these leaves represent single structures; see \S~\ref{Section:results_line}). There is some suggestion that certain leaves exhibit substructure, evidenced through either secondary continuum peaks or irregular boundaries (e.g. leaves \#2, \#6, \#13). This substructure is rejected by the algorithm, following the insertion of physically motivated input parameters selected to reflect the limitations of our observations. Only when ${\rm min\_npix}=18$ (i.e. when the pixel threshold is reduced below the number of pixels contained within one beam) does the number of structures identified deviate from that presented here (leaf~\#13 splits into two). Even when ${\rm min\_npix}$ is reduced by a factor of $>2$, 11 out of the original 13 identified leaves remain unaffected. This gives us confidence that our results are robust, and relatively insensitive to the input parameters of the dendrogram algorithm. The offset locations of peak emission, semi-minor and semi-major axes ($R_{\rm min}$, $R_{\rm maj}$) and their geometric mean, projected aspect ratios ($\AR\equiv R_{\rm maj}/R_{\rm min}$), areas ($N_{\rm pix}A_{\rm pix}$), and equivalent radii ($R_{\rm eq}$) of the leaves are listed in Table~\ref{Table:basic_leaf_info}. 

\begin{figure*}
\begin{center}
\includegraphics[trim = 0mm 0mm 0mm 0mm, clip, width = 0.33\textwidth]{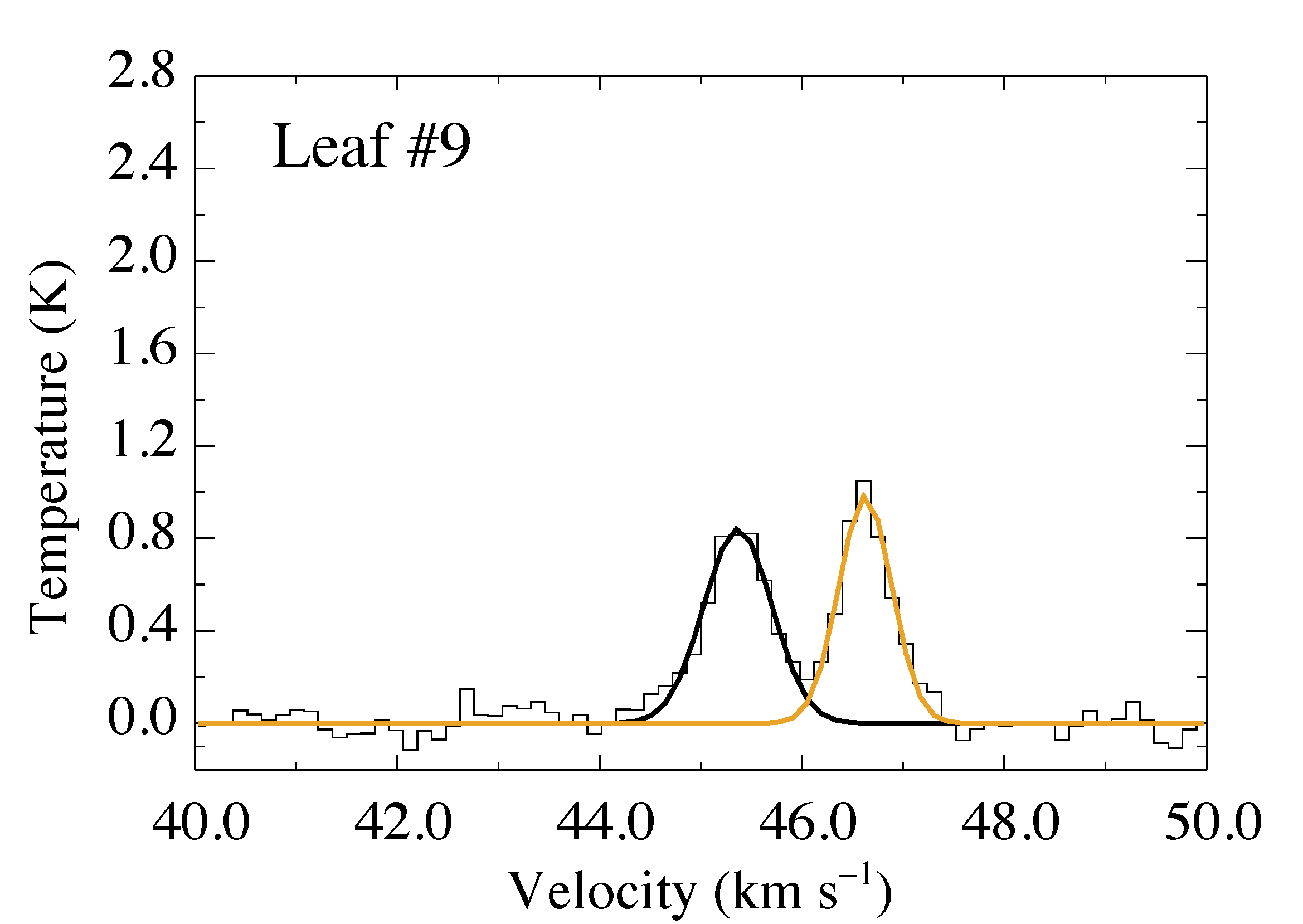}
\includegraphics[trim = 0mm 0mm 0mm 0mm, clip, width = 0.33\textwidth]{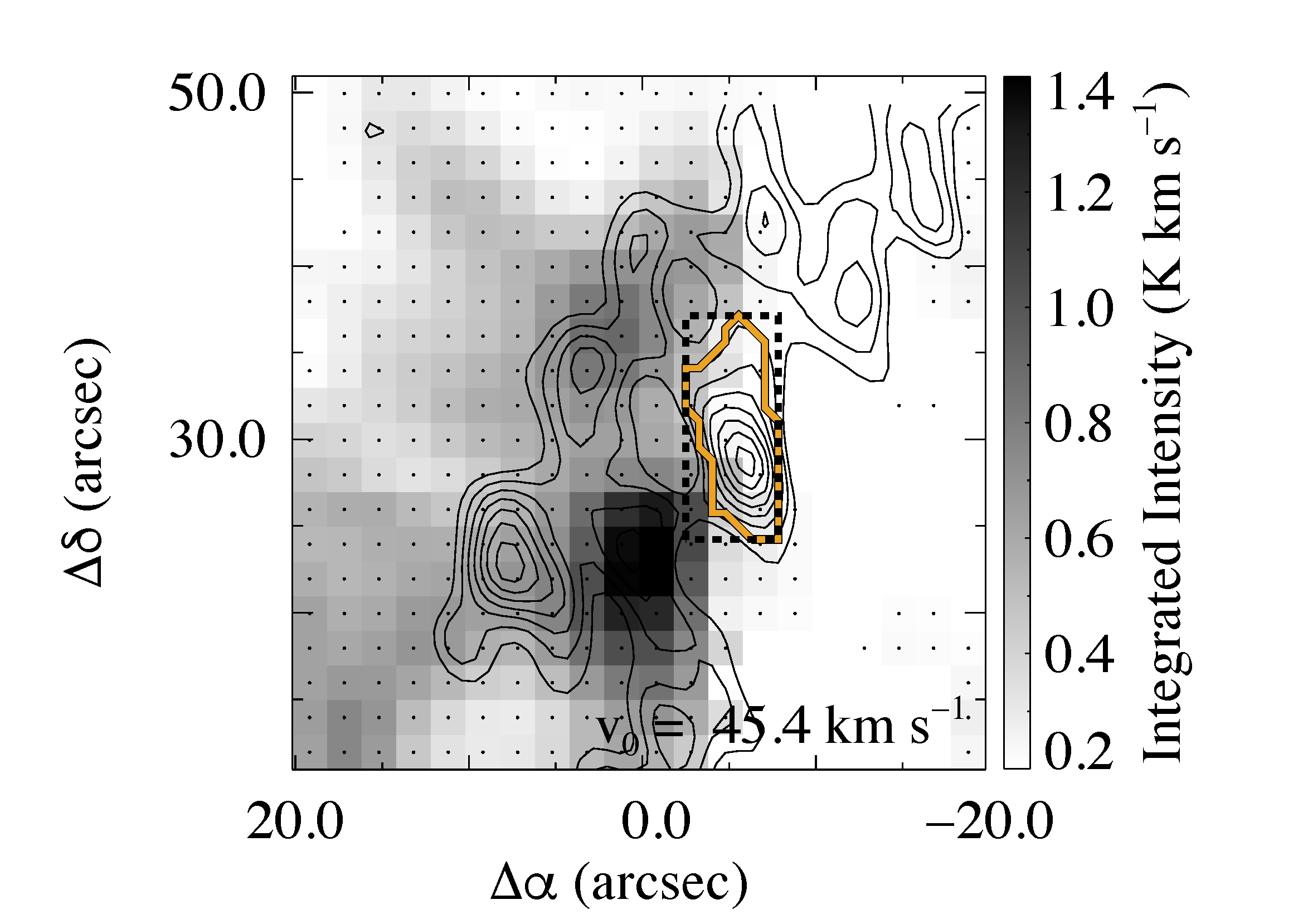}
\includegraphics[trim = 0mm 0mm 0mm 0mm, clip, width = 0.33\textwidth]{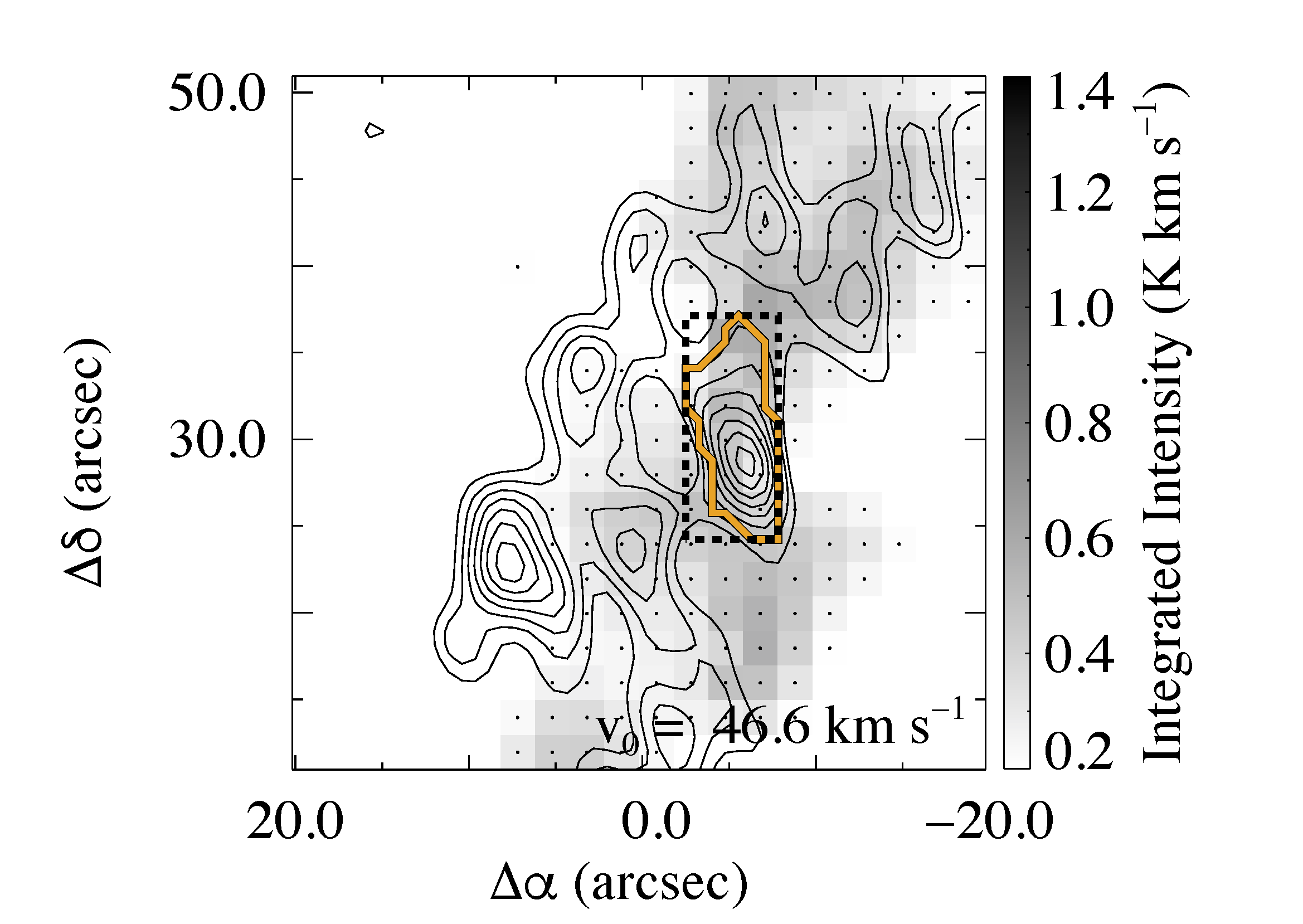}
\includegraphics[trim = 0mm 0mm 0mm 0mm, clip, width = 0.33\textwidth]{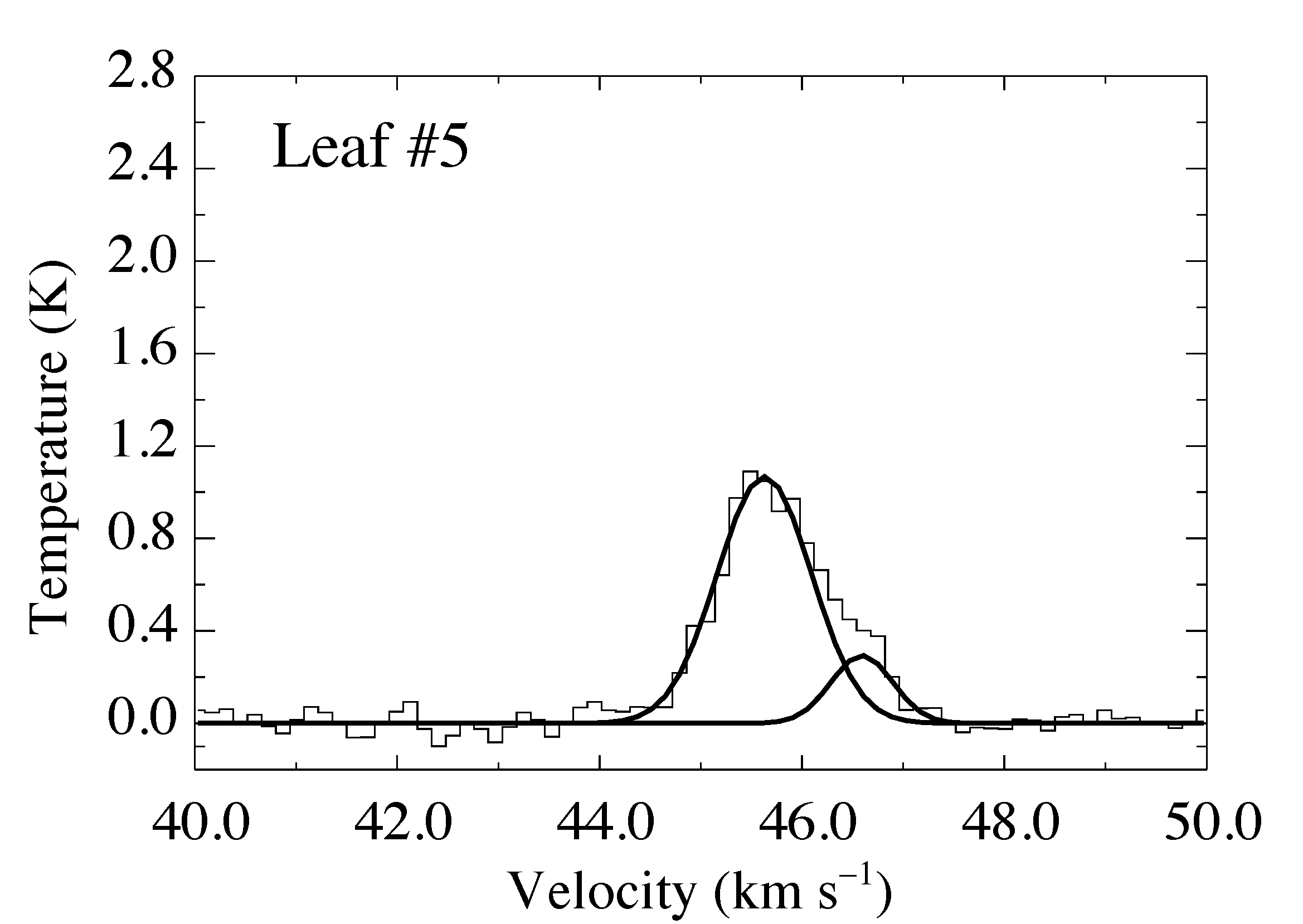}
\includegraphics[trim = 0mm 0mm 0mm 0mm, clip, width = 0.33\textwidth]{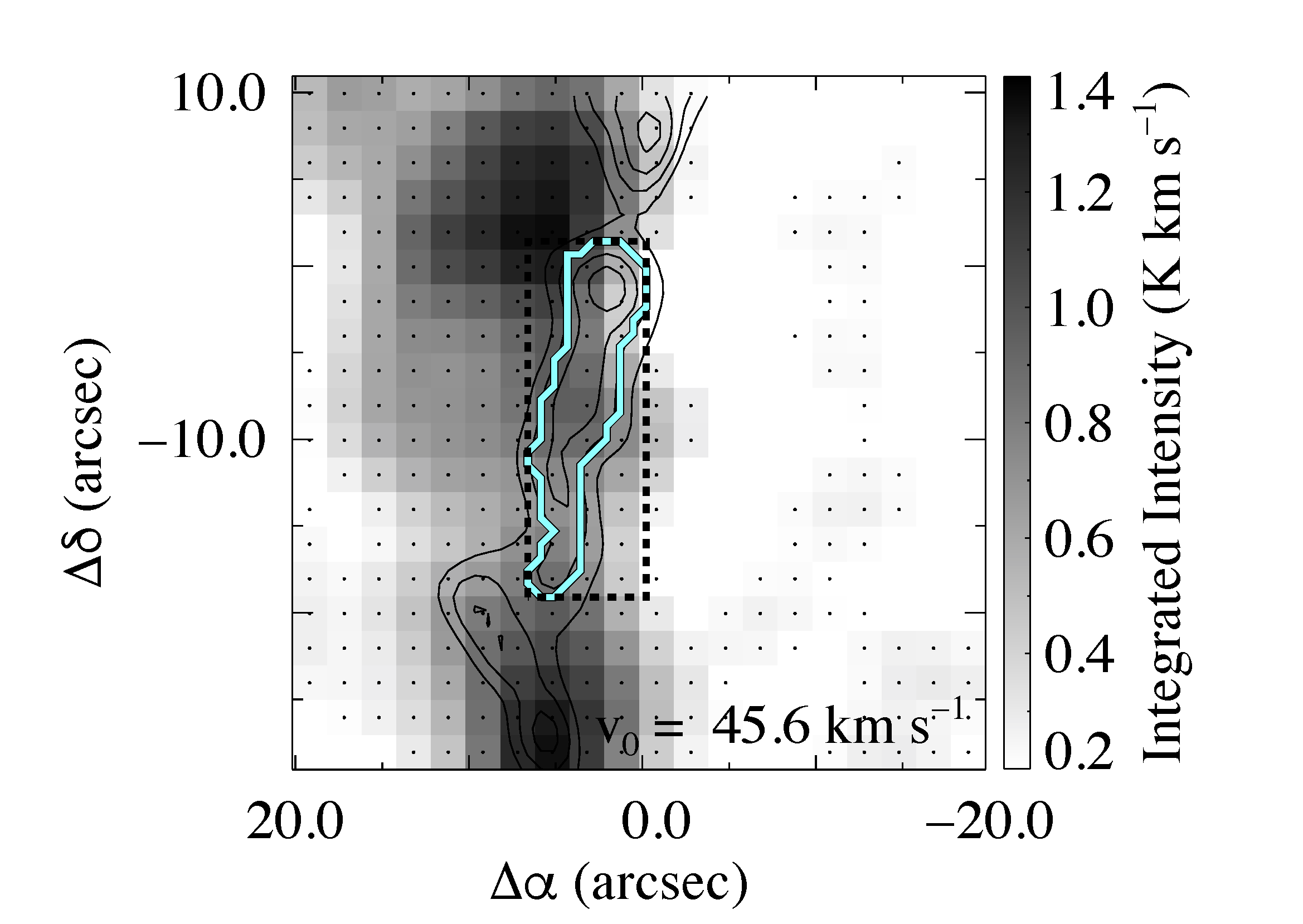}
\includegraphics[trim = 0mm 0mm 0mm 0mm, clip, width = 0.33\textwidth]{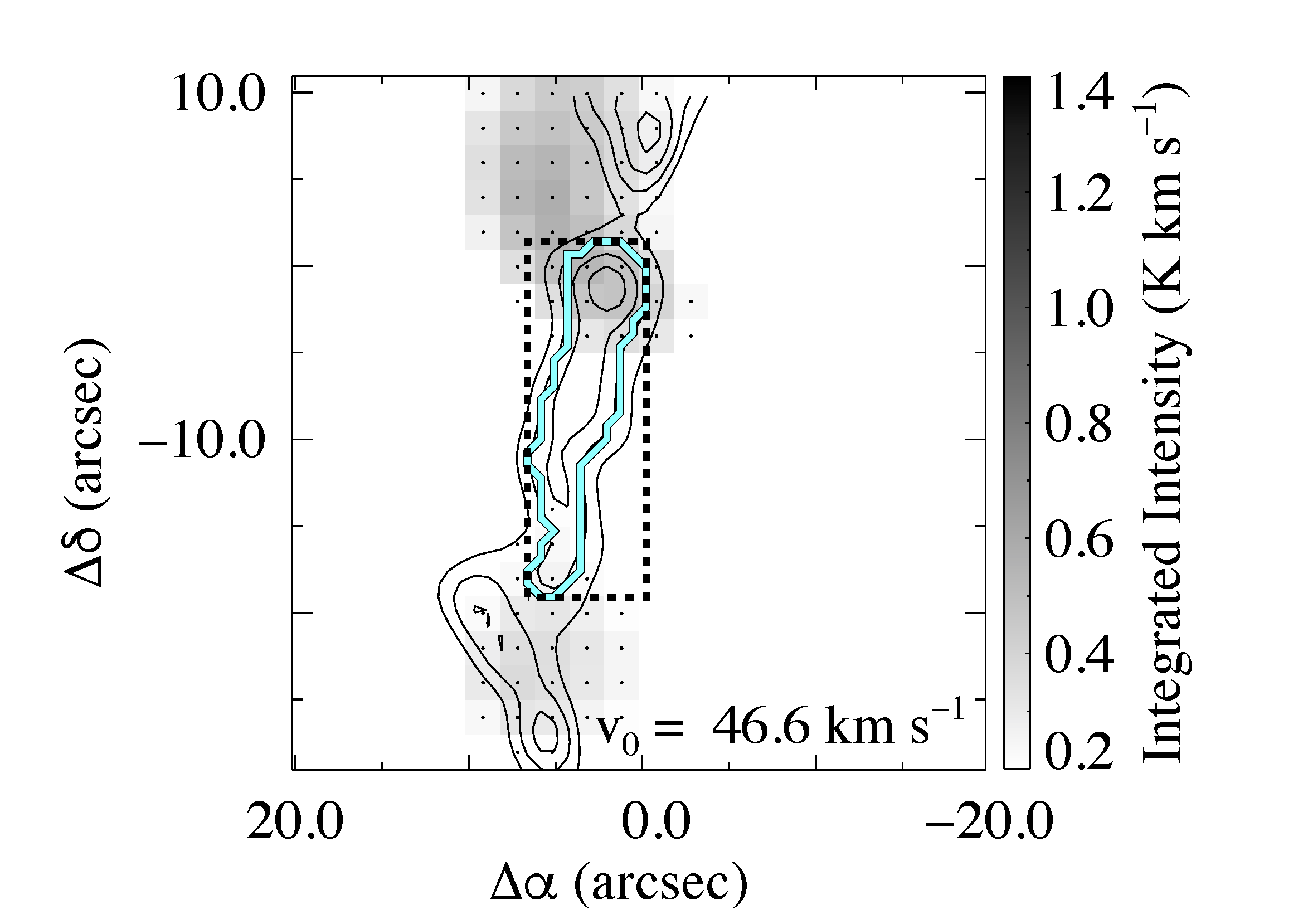}
\end{center}
\caption{Top: the spectral and spatial distribution of \ntwoh \ (1-0) emission associated with leaf~\#9. The left-hand panel is a spatially-averaged spectrum, showing the isolated ($F_{1},\,F~=~0,\,1\rightarrow1,\,2$) hyperfine component of \ntwoh \ (1-0). The spectrum has been extracted from the black dashed box seen in the centre and right-hand panels. Two spectral components are clearly evident. The solid black and orange Gaussian profiles represent the best-fitting model solution to the data (the orange line refers to the component most closely associated with the leaf extracted from the 3.2\,mm continuum data). The centre and right-hand panels display the spatial distribution of each emission feature. The plots are shown on an equivalent scale for ease of comparison, however, this has led to saturation in both central panels. The black contours are equivalent to those in Fig.~\ref{Figure:msd_map} and the orange contour corresponds to the boundary of leaf~\#9. Bottom: each panel is equivalent to those shown in the top panels but for leaf~\#5. Note how the northern and southern portions of the dendrogram leaf appear to be attributed to two independent velocity components which overlap spatially. The black Gaussian profiles reflect the fact that the continuum flux accredited to leaf~\#5 cannot be attributed to a single structure (see text for details regarding the treatment of such cases).}
\label{Figure:spec_mom}
\end{figure*}

\subsection{Comparison with molecular line observations}\label{Section:results_line}

To complement the structure-finding algorithm employed in \S~\ref{Section:results_cont}, we examine the molecular line kinematics associated with each dendrogram leaf. This investigation focuses on the isolated ($F_{1},\,F~=~0,\,1\rightarrow1,\,2$) hyperfine component of \ntwoh \ $(1-0)$, using the PdBI data first presented by \citet{henshaw_2014}.\footnote{The total optical depth of the \ntwoh \ (1-0) line can exceed $\tau=1$ in \irdc \ \citep{henshaw_2014}. However, the relative contribution made by the isolated component to the total line flux is small ($\sim11$ per cent), and it is separated from the main group by $\sim8$\,\kms \ (i.e. typically greater than the observed line width). This feature therefore often has $\tau<1$, and can be used as a reliable tracer of the line-of-sight kinematics of IRDCs.} 

To examine the kinematics, we first generate a spatially-averaged spectrum from all spectra contained within the boundary defining the maximum (projected) physical extent of each leaf. \citet{henshaw_2014} find that the \ntwoh \ (1-0) emission observed throughout \irdc \ can be attributed to a complex network of filamentary structures. Since the velocity separation between these sub-clouds is $<1$\,\kms \ (comparable to their mean FWHM line-widths) and velocity gradients of magnitude $1.5-2.5$\,\vel \ are observed throughout each, analysing the \ntwoh \ (1-0) data using dendrograms is not trivial. Gaussian profiles are therefore fitted to the isolated component of all features observed within each spatially-averaged spectrum.\footnote{This analysis was performed using {\sc scouse} (Semi-automated multi-COmponent Universal Spectral-line fitting Engine; \citealp{henshaw_2016}), which is available for download here: \url{https://github.com/jdhenshaw/SCOUSE}} The line emission is then integrated over the velocity range, $[v_{0,{\rm i}}~-~(\Delta v_{{\rm i}}/2)]~<~v_{0,{\rm i}}~<~[v_{0,{\rm i}}~+~(\Delta v_{{\rm i}}/2)]$, where $v_{0,{\rm i}}$ and $\Delta v_{\rm i}$ are the centroid velocity and full width at half maximum (FWHM) line-width of the i$^{\rm th}$ velocity component, respectively. This enables us to examine the spatial distribution of \ntwoh \ emission associated with each identified velocity component, and relate this to the relevant continuum emission peak. 

Out of the 13 spatially-averaged spectra extracted from within the leaf boundaries, 8 exhibit multiple velocity components. Typically, where multiple spectral features are evident, the emission from one will dominate over the other(s). This allows us, albeit simplistically, to link the continuum sources to a particular kinematic feature. A good example of this is leaf~\#9, shown in the top panels of Fig.~\ref{Figure:spec_mom}. The left panel is the spatially-averaged spectrum extracted from the dashed box shown in the centre and right-hand panels. While the continuum emission appears monolithic, two velocity components are evident in the \ntwoh \ spectrum. The centre and right panels show the spatial distribution of the \ntwoh \ emission associated with each velocity component. The region covered by the integrated emission in both instances is greater than that covered by the leaf alone, indicating that the \ntwoh $(1-0)$ emission is extended and not exclusively associated with the continuum peak for leaf~\#9. Although emission from the velocity component identified at $v_{\rm 0,1}=45.4$\,\kms \ is spatially coincident with the leaf (centre panel), the component at $v_{\rm 0,2}=46.6$\,\kms \ dominates (in terms of spatial coverage; right panel). We therefore speculate that the majority of flux attributed to leaf~\#9 is associated with the high(er)-velocity component. 

An exception to this is presented in the bottom panels of Fig.~\ref{Figure:spec_mom}. In contrast to leaf~\#9, the spatial distribution of each \ntwoh \ emission feature associated with leaf~\#5 (see the bottom-left panel) leads us to question whether this leaf represents a single structure with complex internal kinematics, or if this leaf, in fact, comprises two independent structures that overlap in projection. The southern portion of the leaf appears to be associated with the component identified at $v_{\rm 0}=45.6$\,\kms \ (centre panel). However, the northern portion, including the peak in 3.2\,mm continuum emission, is associated with the higher velocity component at $v_{\rm 0}=46.6$\,\kms (right-hand panel). As the top and bottom panels of Fig.~\ref{Figure:spec_mom} are continuous in declination, we can see that the northern tip of leaf~\#5 appears to be associated with a coherent structure that extends to, and beyond, leaf~\#9 (F3; \citealp{henshaw_2014}). 

We stress that the inclusion of velocity information does not completely alleviate problems associated with projection effects (see e.g. \citealp{beaumont_2013}). This becomes particularly pertinent in an environment such as \irdc, which exhibits complex morphological structure and kinematics (e.g. \citealp{henshaw_2013,henshaw_2014,  izaskun_2014}). However, the above analysis does highlight the importance of demonstrating caution when identifying structure in two-dimensional data. Repeating the above analysis for all identified leaves, we use the kinematic information as a rough guide to determine which leaves are to be analysed further in \S~\ref{Section:analysis}. The leaves are split into four categories: (1) leaves which exhibit single velocity components are retained throughout all analysis (Leaves~\#1, \#2, \#3, \#10, \#12); (2) leaves which show multiple velocity components, but where one of these dominates (in terms of either the magnitude of, or spatial coverage of, the integrated \ntwoh \ emission), are retained, and the dynamical properties of the leaves (see \S~\ref{Section:analysis_virial}) are estimated using only the kinematic properties of the dominant component (Leaves~\#7, \#9, \#11); (3) leaves which exhibit multiple velocity components, but where the continuum emission cannot be unambiguously attributed to a single velocity component, are retained, and the dynamical properties of the leaves are estimated using the kinematic properties of all components (Leaf~\#8); (4) leaves which exhibit multiple velocity components, but where different portions of the continuum emission within the leaf boundary may be associated with different velocity components, are rejected from all analysis that relies on intrinsic geometrical assumptions and/or assumes there is no underlying substructure (Leaves~\#5, \#6, \#13 and possibly \#4). We refer the reader to Appendix~\ref{App:kinematics} for a more complete description of each dendrogram leaf and its associated kinematics.

\section{Climbing the structure tree: investigating the structure and fragmentation of G035.39--00.33}\label{Section:analysis}

\subsection{Investigating the fragmentation of a filament}\label{Section:discussion_filament_frag}

We begin our analysis at the foot of the structure tree, focusing on the spatial distribution of continuum emission throughout \irdc. There are several examples in the literature of filamentary molecular clouds that exhibit a quasi-regular spacing of ``cores'' (e.g. \citealp{jackson_2010, miettinen_2012, busquet_2013, kainulainen_2013a, takahashi_2013, lu_2014, wang_2014, beuther_2015, ragan_2015, contreras_2016, teixeira_2016}). This regularity is often discussed in the context of predictions from theoretical work describing the fragmentation of fluid cylinders due to gravitational or magnetohydrodynamic-driven instabilities (e.g. \citealp{chandrasekhar_1953, nagasawa_1987, inutsuka_1992, nakamura_1993, nakamura_1995, tomisaka_1995}). In this theoretical framework, the characteristic spacing between fragments is defined by the wavelength of the fastest growing unstable mode of the fluid instability. 

\begin{figure}
\begin{center}
\includegraphics[trim = 15mm 15mm 30mm 15mm, clip, width = 0.48\textwidth]{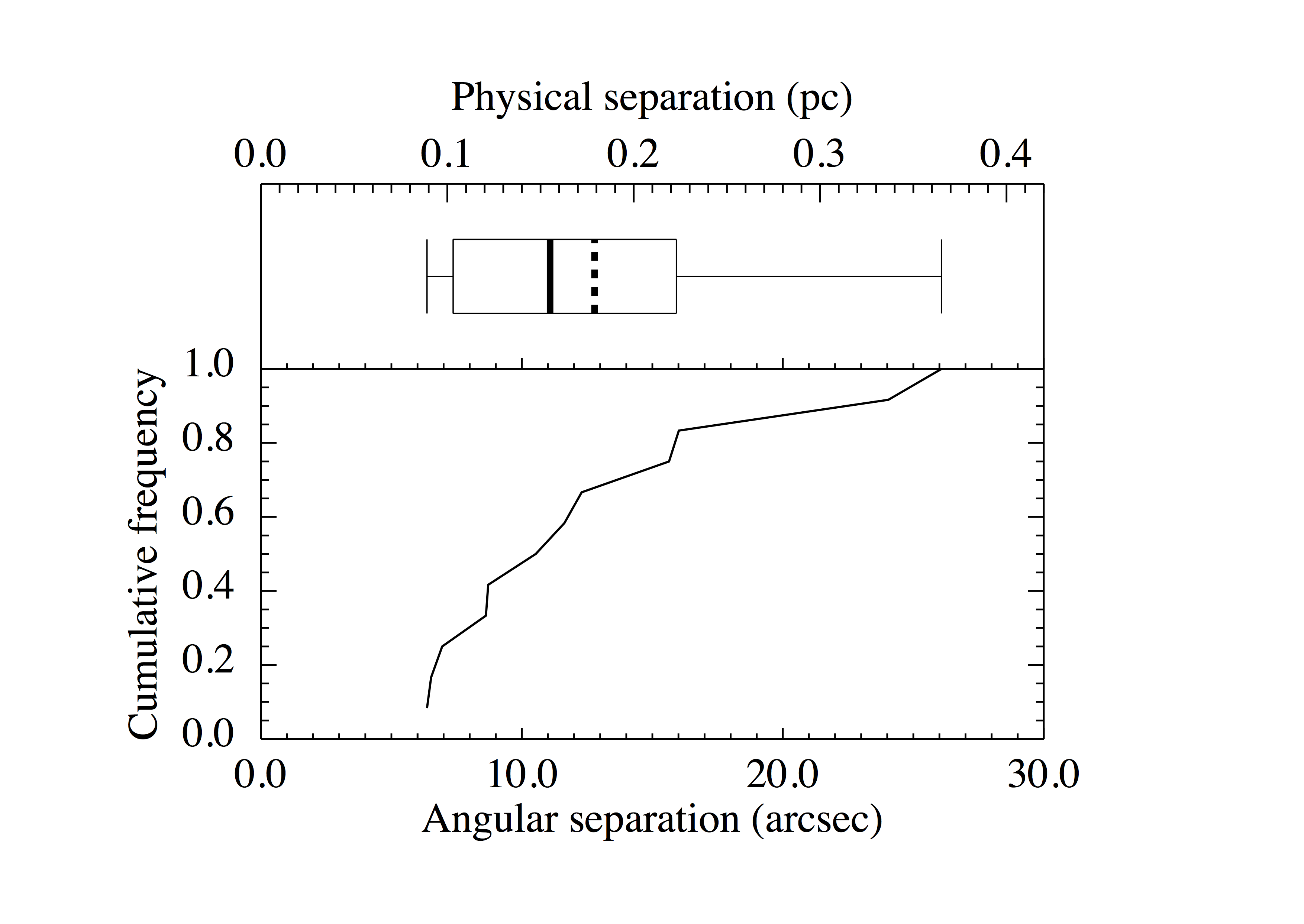}
\end{center}
\caption{Top: a box plot of the angular leaf separation. This highlights the range in separation, the interquartile range (the box itself), the median separation ($\sim11.1$\,arcsec; thick vertical line), and the mean separation ($\sim12.8$\,arcsec; vertical dashed line). Bottom: the corresponding cumulative distribution. }
\label{Figure:core_sep}
\end{figure}

The 3.2\,mm dust continuum emission associated with \irdc \ is distributed along the major axis of the filamentary IRDC (see Fig.~\ref{Figure:msd_map}). To quantify the spatial separation between the dendrogram leaves identified in \S~\ref{Section:results_cont} we use the minimum spanning tree (MST) method \citep{Prim_57}. An MST is a graph theory construct that identifies the shortest possible total path-length between a set of points where there are no closed loops. MSTs are frequently used to quantify the relative spatial distributions of both stars and gas in simulated and observed star-forming regions \citep[e.g.][]{Cartwright_04,Allison_09, gutermuth_2009, Lomax_11,Parker_15}. MSTs have the advantage over other methods that they are not biased by the inherent geometry of the region. The reference point for each dendrogram leaf is taken as the location of peak emission (see Table~\ref{Table:basic_leaf_info}), and these points are then used to construct the MST. Fig.~\ref{Figure:core_sep} includes a box plot and the corresponding cumulative distribution function of the MST lengths. The mean angular separation between dendrogram leaves according to the MST is $\sim12.8$\,arcsec (with $\sim$50\,per cent of all values falling within a factor of $\sim2$ of the mean), which corresponds to a projected physical distance of $\lambda_{\rm obs}\sim0.18$\,pc.\footnote{Alternatively, simply measuring peak-to-peak gives a mean projected core separation of $\sim0.16$\,pc.}

For an infinitely-long, isothermal, self-gravitating gas filament with $R_{\rm f}\gg H$ (where $R_{\rm f}$ and $H$ refer to the filament's radius and isothermal scaleheight, respectively), the spacing between fragments is given by \citep{ostriker_1964, nagasawa_1987}
\begin{equation}
\lambda_{\rm frag}\approx22H=\frac{22c_{\rm s}}{(4\pi G \rho_{\rm f})^{1/2}},
\label{Eq:length_scale_fil}
\end{equation}
where $c_{\rm s}=[(k_{\rm B}T_{\rm f})/(\mu_{\rm p}m_{\rm H})]^{1/2}$ is the isothermal sound speed of gas at a temperature, $T_{\rm f}$ ($c_{\rm s}\approx0.23$\,\kms \ at 15\,K; see below), $\mu_{\rm p}=2.33$ (for molecular gas at the typical interstellar abundance of H, He, and metals), and $m_{\rm H}$ is the mass of atomic hydrogen. Finally in Equation~\ref{Eq:length_scale_fil}, $G$ and $\rho_{\rm f}$, refer to the gravitational constant and filament mass density, respectively. 

Using the mass surface density map presented in Fig.~\ref{Figure:msd_map}, \citet{hernandez_2012a} estimate the filament number density, $n_{\rm H,f}$, of \irdc \ (see their Table\,1). The average value of $n_{\rm H,f}$, over the region mapped with the PdBI, is $n_{\rm H,f}\sim0.2\times10^{5}\,{\rm cm^{-3}}$ (note that this assumes that the filament is inclined by 30\degr, with respect to the plane of the sky, 0\degr). \citet{nguyen_2011} estimate the dust temperature throughout \irdc, by fitting pixel-by-pixel modified blackbody spectral energy distributions derived from \emph{Herschel} observations (excluding the 70\,\micron \ emission). They find dust temperatures ranging between 13 and 16\,K, with the lowest temperatures observed towards the centre of \irdc. However, the angular resolution of the \emph{Herschel} temperature map is 37~arcsec. These observations are therefore sensitive to more diffuse (and warm) emission than traced by our high-angular resolution PdBI observations (flux contributions from scales $>42$~arcsec are filtered out by the interferometer; see \S~\ref{Section:observations}). Thermodynamic coupling between the gas and dust in molecular clouds is valid for densities $\sim10^{5}$\,cm$^{-3}$ (e.g. \citealp{goldsmith_2001,glover_2012}), which is greater than the value estimated by \citet{hernandez_2012a}. However, in the absence of both complementary gas temperature and high-angular resolution dust temperature estimates, we assume that the gas and dust temperatures are approximately equal, and adopt a fiducial value of $T_{\rm f}=15$\,K. 

It is convenient to rewrite Equation~\ref{Eq:length_scale_fil} as
\begin{equation}
\lambda_{\rm frag,f}\approx1.2\bigg(\frac{T_{\rm f}}{15\,{\rm K}}\bigg)^{1/2}\bigg(\frac{n_{\rm H,f}}{10^{4}\,{\rm cm^{-3}}}\bigg)^{-1/2}\,{\rm pc},
\label{Eq:length_scale_therm}
\end{equation}
where we express the filament mass density in terms of the number density of hydrogen nuclei ($n_{\rm H,f}$), $\rho_{\rm f}\approx\mu_{\rm H}m_{\rm H}n_{\rm H,f}$ (with $\mu_{\rm H}=1.4$). Setting $\lambda_{\rm frag}=\lambda_{\rm obs}=0.18$\,pc in Equation~\ref{Eq:length_scale_therm}, we find that a density of $n_{\rm H,f}\sim4.4\times10^{5}\,{\rm cm^{-3}}$ would be required to reproduce the observed core spacing.  

One aspect not factored into the above analysis is the effect of inclination. The true core spacing throughout a filament inclined at an angle, $i$, with respect to the plane of the sky ($i=0\degr$) is $\lambda_{\rm obs,i}=\lambda_{\rm obs}/{\rm cos(}i{\rm )}$. \citet{hernandez_2012a} assume an inclination angle of 30\degr \ when determining the number density of the filament. Following this assumption results in an ``inclination-corrected'' leaf spacing of $\lambda_{\rm obs,i}=0.21$\,pc. Setting $\lambda_{\rm frag}=\lambda_{\rm obs,i}$ in Equation~\ref{Eq:length_scale_therm}, we find that a density of $n_{\rm H,f}\sim3.3\times10^{5}\,{\rm cm^{-3}}$, is required to reproduce the inclination-corrected spacing.

In reality, random turbulent motions may also contribute to the total gas pressure. Throughout \irdc, the non-thermal contribution to the velocity dispersion, $\sigma_{\rm NT}$, typically dominates over the estimated isothermal sound speed (by factors of 1.5-2 at the spatial resolution of the PdBI; \citealp{henshaw_2014}). Assuming that $\sigma_{\rm NT}$ is dominated by turbulence\footnote{This should serve as an upper limit to the level of turbulent support. With only the velocity dispersion as a gauge, it is difficult to distinguish between non-thermal motions that act to provide support to a cloud (i.e. random turbulence) and those generated by gravitational collapse (e.g. \citealp{vazquez-semadeni_2009, ballesteros-paredes_2011, traficante_2015}). We are also unable to quantify the contribution to $\sigma_{\rm NT}$ from ordered velocity gradients and/or unresolved structure that exists on scales smaller than our PdBI beam.}  and that this acts to support the filament against gravitational collapse, we can replace the sound speed in Equation~\ref{Eq:length_scale_fil} with the total (including both thermal and non-thermal motions) velocity dispersion, $\sigma_{v}$ \citep{fiege_2000}. Following \citet{fuller_1992}
\begin{equation}
\sigma_{v}=\sqrt{\frac{\Delta v^{2}_{\rm i}}{8{\rm ln}(2)}+k_{\rm B}T_{\rm kin}\bigg(\frac{1}{\mu_{\rm p}m_{\rm H}}-\frac{1}{m_{\rm obs}}\bigg)},
\label{Eq:disp_eff}
\end{equation}
where $m_{\rm obs}$ is the mass of the observed molecule (for \ntwoh, $m_{\rm obs}=29m_{\rm H}$). In this case, Equation~\ref{Eq:length_scale_fil} simplifies to
\begin{equation}
\lambda_{\rm frag, f}\approx5.1\bigg(\frac{\sigma_{v}}{1\,{\rm km\,s^{-1}}}\bigg)\bigg(\frac{n_{\rm H,f}}{10^{4}\,{\rm cm^{-3}}}\bigg)^{-1/2}\,{\rm pc}.
\label{Eq:length_scale_turb}
\end{equation}
If we take $\sigma_{v}\sim0.4\,{\rm km\,s^{-1}}$ (the mean value deduced from line-fitting; \citealp{henshaw_2014}), a density of $n_{\rm H, f}\sim12.9\times10^{5}\,{\rm cm^{-3}}$ would be required to reproduce $\lambda_{\rm obs}$. Alternatively, accounting for inclination, a density of $n_{\rm H,f}\sim9.4\times10^{5}\,{\rm cm^{-3}}$ would be required to reproduce $\lambda_{\rm obs,i}$.

Evidently there exists a discrepancy between the observed core separation and that predicted by this particular model. When considering thermal motions only, the density required to reproduce the observed spacing is $\sim$ an order of magnitude greater than the value estimated by \citet{hernandez_2012a}. Including turbulent gas motions only exacerbates the problem, requiring densities that are, in fact, similar to the typical density of the leaves (see \S~\ref{Section:analysis_phys}). Putting this another way, for a fixed density of $n_{\rm f}=0.2\times10^{5}$\,cm$^{-3}$, the theoretical fragment spacing is a factor of $\sim8$ greater than the observed value when considering both thermal and turbulent support ($\sim5$, when considering only thermal support). This is similar to the conclusion of \citet{pillai_2011}, who find that, in order to explain the fragment spacing in massive star forming regions G$29.96~-~0.02$ and G$35.20-1.74$, gas densities similar to their estimated core densities are required. Below we discuss two alternate scenarios which may explain the discrepancy in \irdc. First of all, we discuss the possibility that the presence of dynamically important magnetic fields may influence the fragmentation length-scale (a scenario that was recently explored by \citealp{contreras_2016}). Secondly, we discuss how the geometric interpretation inherent to the above discussion, i.e. that \irdc \ is well-represented as a single cylindrical filament, may not be applicable. 

\begin{figure*}
\begin{center}
\includegraphics[trim = 12mm 10mm 0mm 30mm, clip, width = 0.48\textwidth]{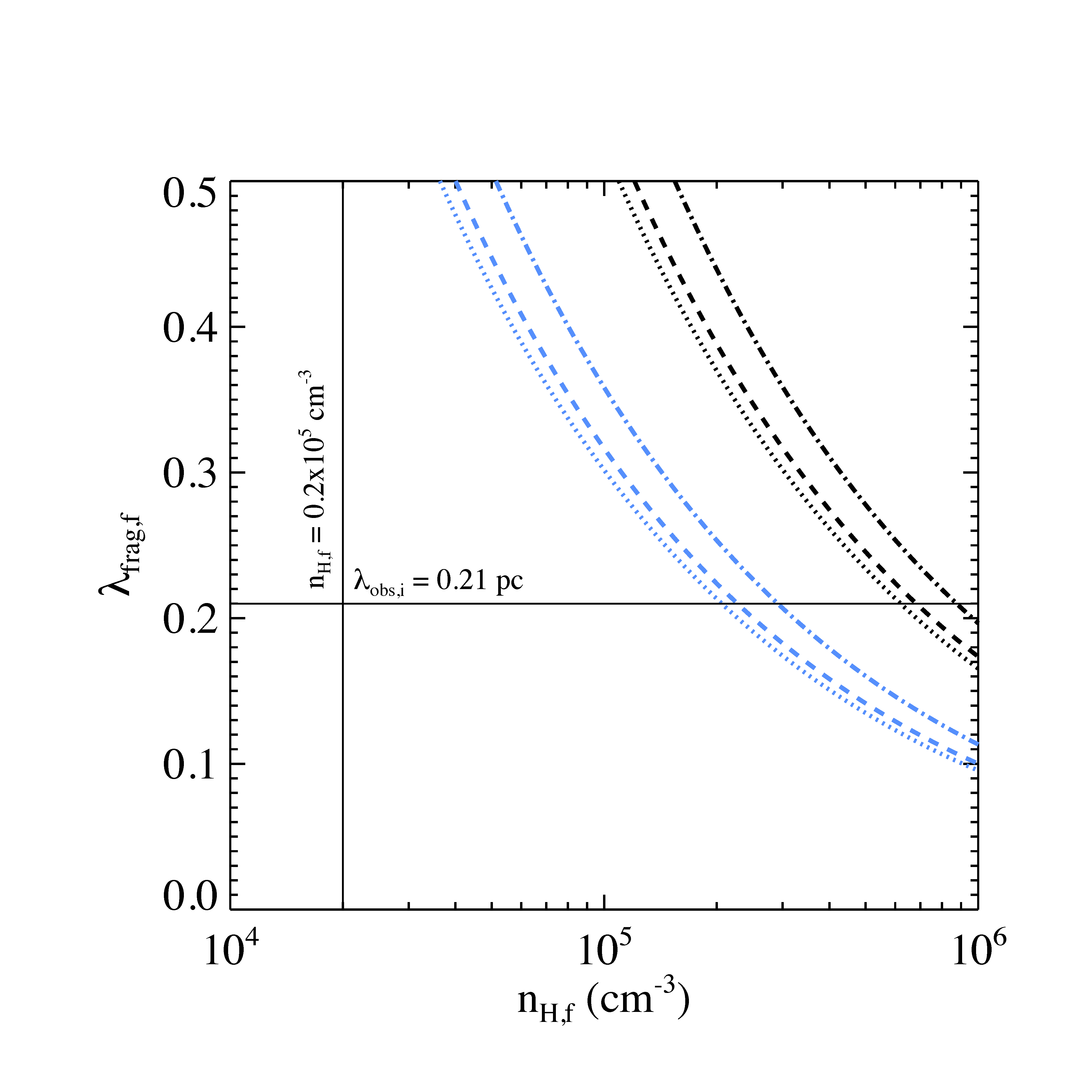}
\includegraphics[trim = 12mm 10mm 0mm 30mm, clip, width = 0.48\textwidth]{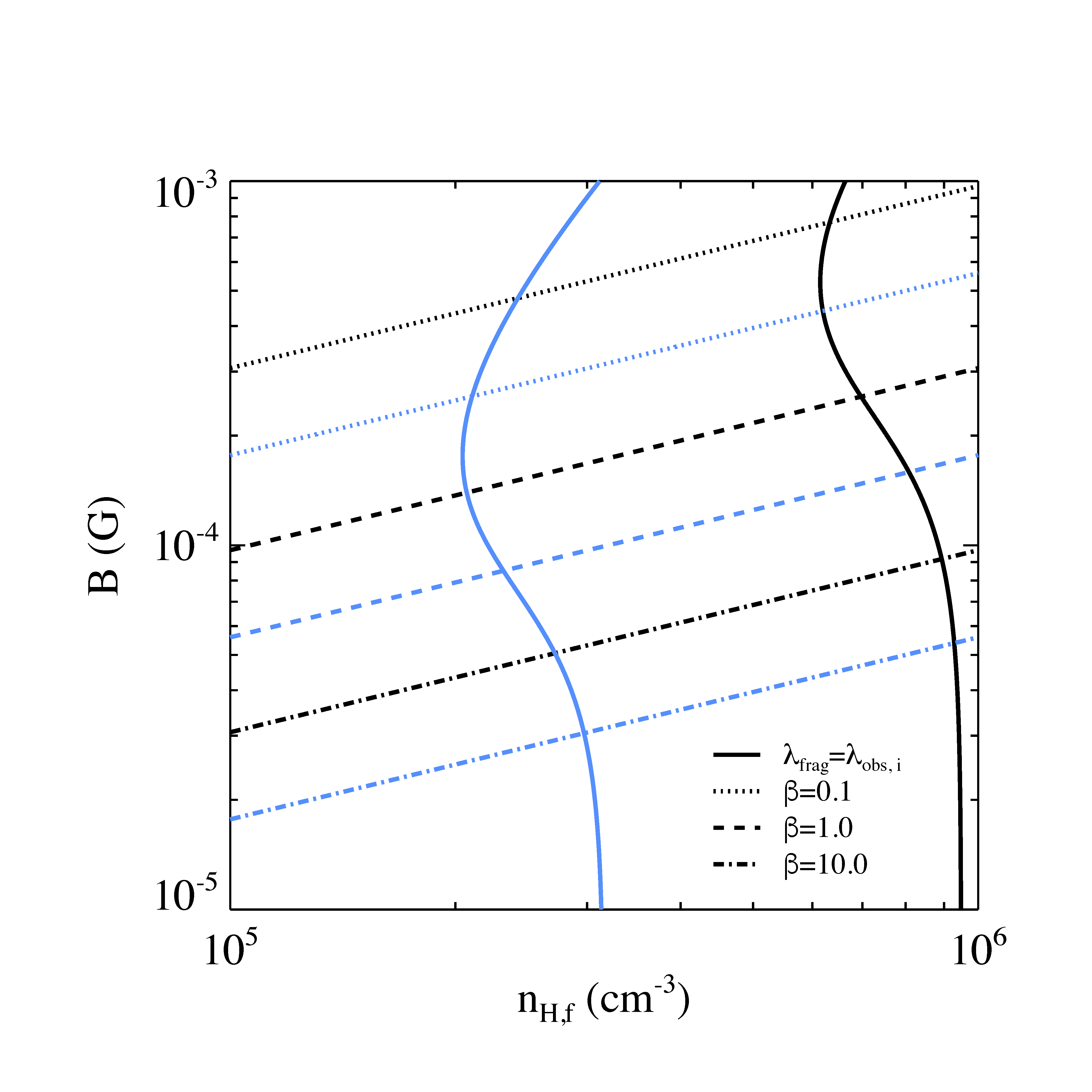}
\end{center}
\caption{Left: this figure shows how the wavelength of the fastest growing mode of the magnetohydrodynamic fluid instability discussed in \S~\ref{Section:discussion_filament_frag} changes as a function of density according to Equation~\ref{Eq:length_scale_mag} \citep{nakamura_1993,nakamura_1995} for both the thermal (blue) and thermal+non-thermal (black) cases. The dotted, dashed, and dot-dashed refer to where $\beta=0.1,1.0,10.0$. For dynamically important magnetic fields (where $\beta=0.1$; dotted lines), the wavelength of the fastest growing mode is less than in the regime where magnetic fields are unimportant ($\beta=10$; dot-dashed lines). The solid vertical and horizontal lines indicate the observed filament density ($n_{\rm H,f}=0.2\times10^{5}$\,cm$^{-3}$; \citealp{hernandez_2012a}) and inclination-corrected leaf spacing, respectively. Right: this figure shows how the filament density required for $\lambda_{\rm frag,f}=~\lambda_{\rm obs, i}$ changes as a function of magnetic field strength. The solid line(s) indicate where $\lambda_{\rm frag}=\lambda_{\rm obs,i}$. The dotted, dashed, and dot-dashed refer to the loci of $\beta=0.1,1.0,10.0$. For dynamically unimportant magnetic fields ($\beta=10$; dot-dashed lines), the density required for $\lambda_{\rm frag,f}~=~\lambda_{\rm obs,i}$ (solid lines) reverts to that derived using Equation~\ref{Eq:length_scale_fil}. This figure demonstrates that even with a dynamically important magnetic field, the reduction in the wavelength of the fastest growing mode of the instability is insufficient to explain the observed leaf spacing. }
\label{Figure:length_scale}
\end{figure*}

It is possible that the discrepancy between the observed and predicted spatial distribution of continuum emission is a result of the fact that Equation~\ref{Eq:length_scale_fil} does not account for the effects of magnetic fields. \citet[see also \citealp{nakamura_1993} and \citealp{hanawa_1993}]{nakamura_1995}, studying the fragmentation of filamentary clouds with longitudinal magnetic fields, find that the wavelength of the fastest growing perturbation is given by 
\begin{equation}
\lambda_{\rm frag, f}\approx\frac{8.73H}{[1+(1/\beta)]^{1/3}-0.6}
\label{Eq:length_scale_mag}
\end{equation}
where $\beta=(8\pi\rho_{\rm f} c^{2}_{\rm s})/B^{2}$, $B$ is the magnetic flux density, and
\begin{equation}
H=\frac{c_{\rm s}}{(4\pi G\rho_{\rm f})^{1/2}}[1+(1/\beta)]^{1/2}, 
\end{equation}
($c_{\rm s}$ and $\rho_{\rm f}$ are defined in Equation~\ref{Eq:length_scale_fil}).

The left-hand panel of Fig.~\ref{Figure:length_scale} demonstrates how $\lambda_{\rm frag,f}$ varies as a function of density according to Equation~\ref{Eq:length_scale_mag}, for both the thermal (blue) and thermal+non-thermal (black) cases. The dotted, dashed, and dot-dashed refer to where $\beta=0.1, 1.0, 10.0$. The solid vertical and horizontal lines indicate the observed filament density ($n_{\rm H,f}=0.2\times10^{5}$\,cm$^{-3}$; \citealp{hernandez_2012a}) and inclination-corrected leaf spacing ($\sim0.21$\,pc), respectively. The right-hand panel of Fig.~\ref{Figure:length_scale} shows how the density required for $\lambda_{\rm frag,f}~=~\lambda_{\rm obs,i}$ changes as a function of the magnetic field strength according to Equation~\ref{Eq:length_scale_mag}. Solid lines indicate the locus where $\lambda_{\rm frag,f}=\lambda_{\rm obs, i}$, once again, for both the thermal (blue) and thermal+non-thermal (black) cases. 

The left-hand panel demonstrates how the wavelength of the most unstable perturbation is shorter when the ratio of the magnetic pressure to the gas pressure is higher (i.e. when $\beta$ is small), for a fixed density \citep{nakamura_1993}. Note however, that the effect is small. Conversely, the right-hand panel shows how the density required for $\lambda_{\rm frag,f}~=~\lambda_{\rm obs,i}$ can be reduced if the magnetic field is dynamically important ($0.1<\beta<1.0$). As can be inferred from the right-hand panel, the density required for $\lambda_{\rm frag,f}~=~\lambda_{\rm obs,i}$ in the case of thermal+non-thermal (thermal) fragmentation is $6.2\times10^{5}\,{\rm cm^{-3}}$ ($2.1\times10^{5}\,{\rm cm^{-3}}$) for a magnetic field strength of $470\,\mu{\rm G}$ ($140\,\mu{\rm G}$), compared with $9.4\times10^{5}\,{\rm cm^{-3}}$ ($3.3\times10^{5}\,{\rm cm^{-3}}$) for a dynamically unimportant magnetic field (see above). This figure demonstrates that the decrease in $\lambda_{\rm frag, f}$ according to Equation~\ref{Eq:length_scale_mag}, due to the presence of a dynamically important longitudinal magnetic field, on its own, cannot account for the observed discrepancy. 

An alternative possibility is that the underlying assumption of the above model, that \irdc \ can be described simplistically as a single cylindrical filament, may be a poor one. We stress that this is not necessarily mutually exclusive from a scenario which includes dynamically important magnetic fields (the above analysis only accounts for a very specific configuration of magnetic field). However, in this example, the discrepancy may relate to the structure of the cloud itself. Although low-angular resolution dust continuum maps may hint towards a relatively ``simplistic'' cloud morphology, the reality is anything but simple. \citet{henshaw_2014} argue that \irdc \ is organized into a serpentine network of morphologically distinct molecular sub-filaments. Each sub-filament displays not only unique kinematic properties (in terms of $\sigma_{v}$ and a complex pattern of velocity gradients), but also its own density structure, as demonstrated by \citet[albeit these densities are derived from CO observations with $\sim20$\,arcsec resolution]{izaskun_2014}.

The presence of multiple line-of-sight structures, if confirmed, influences our investigation in two key ways. First, if, as suggested by \citet{henshaw_2014}, the leaves are associated with otherwise independent molecular filaments then we may underestimate the true spacing. The density required to reproduce the true spacing may therefore be lower than that stated above (assuming the velocity dispersion remains constant; Equation~\ref{Eq:length_scale_turb}). Secondly, the relevant properties in Equation~\ref{Eq:length_scale_turb} ($n_{\rm H,f}$, inclination, $\sigma_{v}$) should be those relating to the individual sub-filaments. The fiducial value of the density assumed above ($n_{\rm H, f}=0.2\times10^{5}\,{\rm cm^{-3}}$) is derived from the mass surface density map (Fig.~\ref{Figure:msd_map}), assuming cylindrical geometry with a radius of $R_{\rm f}\approx30$\,arcsec or $\sim0.4$\,pc at a distance of 2900\,pc \citep{hernandez_2012a}, which may not reflect the central density of the sub-filaments. From Equation~\ref{Eq:length_scale_turb}, a factor of 10 increase in the density would give $\lambda_{\rm frag}\sim0.45\,{\rm pc}\sim2\lambda_{\rm obs,i}$. 

\subsection{Estimating the physical properties of the dendrogram leaves}\label{Section:analysis_phys}

\subsubsection{Initial considerations}

Having focused on the distribution of continuum emission throughout \irdc, we now turn our attention to analysing the leaf properties. It is worth noting that the combination of missing continuum flux in our interferometric map and complex kinematic structure (discussed in Sections~\ref{Section:results_line} and \ref{Section:discussion_filament_frag}, and more fully in \citealp{henshaw_2014}), makes it difficult to unambiguously apportion flux to any given continuum source. We therefore employ two different approaches to estimating the physical properties of the dendrogram leaves. 

Both methods make an underlying assumption regarding the translation of a region of emission, defined by an isosurface, into physical three-dimensional space (we refer the reader to \citealp{rosolowski_2008} for a more complete description of the philosophy behind these methods). The first approach, which is most conservative, makes the assumption that each leaf represents a discrete object superimposed on a background of flux, $F^{\rm bg}_{\nu}$ (which is subtracted from each leaf pixel prior to the estimation of physical properties). This has been used in recent studies as a way of accounting for the fact that emission from more diffuse or larger-scale structures can contaminate the flux of small-scale structures (e.g. \citealp{ragan_2013, pineda_2015}). The second approach assumes that there is no background contribution, and that \emph{all} of the flux within the leaf boundary is attributed to that structure. The reality probably lies somewhere in between, and these two methods provide lower and upper bounds to the source flux, respectively. We present results from both approaches throughout the following analysis and denote the background-subtracted properties with `b' (see Table~\ref{Table:derived_leaf_info}).

It is also prudent, prior to the determination of physical properties, to estimate the contribution to the continuum flux from free-free emission originating from embedded radio sources. To estimate the free-free contribution at 93\,GHz we inspect images from The Coordinated Radio and Infrared Survey for High-mass Star Formation (CORNISH) survey \citep{hoare_2012, purcell_2013}. We identify one $5\sigma$ source ($\sim$2\,mJy at 5\,GHz) at a position $\alpha\,({\rm J}2000)=18^{\rm h}57^{\rm m}08^{\rm s}37$, $\delta\,({\rm J}2000)=02\degr10\arcmin32\farc71$, corresponding to an offset location of $\{\Delta\alpha,\,\Delta\delta\}=\{5.8\,{\rm arcsec},\,2.2\,{\rm arcsec}\}$ in our PdBI map. We note however, that due to artefacts in the CORNISH images, reliable source detections are limited to $\geq7\sigma$. Since the aforementioned $5\sigma$ source does not coincide with one of the 3.2\,mm continuum peaks, nor is there evidence for 8, 24, 70\,\micron \ emission at this location (see Fig.~\ref{Figure:msd_map} and \citealp{nguyen_2011}), it is possible that this is an image artefact in the CORNISH map. In the optically thin regime, free-free emission has a frequency dependence of $S_{\nu}\propto\nu^{-0.1}$. Based on the rms noise of the CORNISH images (0.37\,mJy at 5\,GHz), we estimate an upper limit to the free-free contribution of 0.28\,mJy at 93\,GHz. Since no other detections are made, we expect this to be a fairly generous upper limit and anticipate that the contribution to our PdBI continuum flux from free-free emission is small. 

\subsubsection{Estimating the physical properties}

\begin{table*}
	\caption{Dendrogram leaves: physical properties (see \S~\ref{Section:analysis_phys}). Leaves rejected from the analysis (see \S~\ref{Section:results_line} and Appendix~\ref{App:kinematics}) are clearly marked. } \vspace{0.2cm}
	
	\centering  
	\tabcolsep=0.2cm \normalsize{
	\begin{tabular}{ l  c  c  c  c  c  c  c  c  c  c  c  c  c  c }
	\hline
	ID & 
	$\Delta\alpha$ & 
	$\Delta\delta$ & 
	\multicolumn{2}{ c }{Flux density$^{a}$} & 
	\multicolumn{2}{ c }{Integrated flux$^{b}$} & 
	\multicolumn{2}{ c }{Column density$^{c}$} & 
	\multicolumn{2}{ c }{Mass$^{d}$} & 
	\multicolumn{2}{ c }{Number density$^{e}$} & 
	\multicolumn{2}{ c }{Free-fall time$^{f}$} 
	\\ [0.5ex]

	&  
	&  
	& 
	\multicolumn{2}{ c }{$\times10^{-3}$} & 
	\multicolumn{2}{ c }{$\times10^{-3}$} & 
	\multicolumn{2}{ c }{$\times10^{23}$} & 
	\multicolumn{2}{ c }{} & 
	\multicolumn{2}{ c }{$\times10^{5}$} & 
	\multicolumn{2}{ c }{$\times10^{4}$} 
	\\ [0.5ex]
	
	& 
	(arcsec) & 
	(arcsec) & 
	\multicolumn{2}{ c }{(Jy beam$^{-1}$)} &
	\multicolumn{2}{ c }{(Jy)} & 
	\multicolumn{2}{ c }{(cm$^{-2}$)} & 
	\multicolumn{2}{ c }{(\solar)} & 
	\multicolumn{2}{ c }{(cm$^{-3}$)} & 
	\multicolumn{2}{ c }{(yr)} 
	\\ [0.5ex]
	
	\hline
	
	&  
	&  
	& 
	$F^{\rm peak}_{\nu}$ & 
	$F^{\rm bg}_{\nu}$ & 
	$S_{\nu}$ & 
	$S^{\rm b}_{\nu}$ & 
	$N_{\rm H, c}$ & 
	$N^{\rm b}_{\rm H, c}$ & 
	$M_{\rm c}$ & 
	$M^{\rm b}_{\rm c}$ & 
	$n_{\rm H, c, eq}$ & 
	$n^{\rm b}_{\rm H,c,eq}$ & 
	$t_{\rm ff, c}$ & 
	$t^{\rm b}_{\rm ff,c}$ 
	\\ [0.5ex]

	\hline 
	
 1 &        8.9 &      -76.9 &  0.57 &  0.21 &  1.03 &  0.41 &  3.67 &  2.31 & 10.68 &  4.30 &  6.77 &  2.73 &  5.29 &  8.33      \\ [0.5ex]
 2 &        7.4 &      -64.7 &  0.72 &  0.25 &  1.83 &  0.73 &  4.61 &  3.03 & 19.07 &  7.60 &  6.24 &  2.49 &  5.50 &  8.72      \\ [0.5ex]
 3 &        1.3 &      -50.3 &  0.62 &  0.25 &  1.42 &  0.55 &  3.94 &  2.35 & 14.82 &  5.72 &  6.93 &  2.68 &  5.22 &  8.40      \\ [0.5ex]
 4 &        5.9 &      -26.7 &  0.56 &  0.31 &  1.05 &  0.26 &  3.56 &  1.60 & 10.90 &  2.71 &  $-$ &  $-$ &  $-$ &  $-$      \\ [0.5ex]
 5 &        2.1 &       -0.9 &  0.73 &  0.31 &  2.50 &  0.80 &  4.65 &  2.69 & 26.07 &  8.35 &  $-$ &  $-$ &  $-$ &  $-$      \\ [0.5ex]
 6 &       -0.2 &        7.5 &  0.68 &  0.35 &  1.60 &  0.41 &  4.36 &  2.12 & 16.63 &  4.25 &  $-$ &  $-$ &  $-$ &  $-$      \\ [0.5ex]
 7 &        7.4 &       22.7 &  1.14 &  0.42 &  2.10 &  0.84 &  7.29 &  4.59 & 21.83 &  8.70 & 13.05 &  5.20 &  3.81 &  6.03      \\ [0.5ex]
 8 &        0.5 &       23.5 &  0.70 &  0.44 &  0.90 &  0.15 &  4.45 &  1.64 &  9.38 &  1.60 & 13.05 &  2.22 &  3.81 &  9.22      \\ [0.5ex]
 9 &       -6.3 &       28.8 &  1.24 &  0.32 &  2.35 &  1.16 &  7.96 &  5.94 & 24.42 & 12.10 & 10.40 &  5.16 &  4.26 &  6.06      \\ [0.5ex]
10 &        2.8 &       34.1 &  0.73 &  0.43 &  0.79 &  0.17 &  4.69 &  1.93 &  8.23 &  1.75 & 14.74 &  3.13 &  3.58 &  7.77      \\ [0.5ex]
11 &      -12.4 &       38.7 &  0.59 &  0.36 &  0.78 &  0.15 &  3.75 &  1.46 &  8.14 &  1.52 & 10.60 &  1.99 &  4.22 &  9.76      \\ [0.5ex]
12 &       -7.1 &       42.5 &  0.64 &  0.36 &  0.79 &  0.15 &  4.12 &  1.80 &  8.26 &  1.55 & 10.76 &  2.03 &  4.19 &  9.66      \\ [0.5ex]
13 &      -16.9 &       43.2 &  0.58 &  0.35 &  1.35 &  0.30 &  3.70 &  1.46 & 14.01 &  3.12 &  $-$ &  $-$ &  $-$ &  $-$      \\ [0.5ex]

	\hline
	\end{tabular}
	\vspace{0.2cm}

\begin{minipage}{1.0\textwidth}\footnotesize{
$^{a}$ Peak ($F^{\rm peak}_{\nu}$) and background ($F^{\rm bg}_{\nu}$) flux density of each leaf. Uncertainty: $\sigma F_{\nu}\sim0.07$\,mJy\,beam$^{-1}$. \\
$^{b}$ Integrated flux density of each leaf before ($S_{\nu}$) and after ($S^{\rm b}_{\nu}$) background subtraction. Uncertainties: $\langle\sigma S_{\nu}\rangle\sim0.02$\,mJy, $\langle\sigma S^{\rm b}_{\nu}\rangle\sim0.03$\,mJy.\\
$^{c}$ Beam-averaged column density at peak flux density before ($N_{\rm H, c}$) and after ($N^{\rm b}_{\rm H, c}$) background subtraction. Uncertainty: $\sigma N_{\rm H, c}\sim50$ per cent. \\
$^{d}$ Leaf mass before ($M_{\rm c}$) and after ($M^{\rm b}_{\rm c}$) background subtraction. Uncertainty: $\sigma M_{\rm c}\sim60$ per cent. \\
$^{e}$ Leaf number density before ($n_{\rm H, c, eq}$) and after ($n^{\rm b}_{\rm H, c, eq}$) background subtraction. Uncertainty: $\sigma n_{\rm H, c, eq}\sim75$ per cent.\\
$^{f}$ Leaf free-fall time before ($t_{\rm ff, c}$) and after ($t^{\rm b}_{\rm ff, c}$) background subtraction. Uncertainty: $\sigma t_{\rm ff, c}\sim40$ per cent.\\
}
\end{minipage}
}
\label{Table:derived_leaf_info}
\end{table*}

Assuming optically thin dust continuum emission, the beam-averaged column density at the location of peak emission, $N_{\rm H,c}$ (where the subscript `c' distinguishes core/leaf properties from the filament properties, `f', discussed in \S~\ref{Section:discussion_filament_frag}) is estimated for each dendrogram leaf using
\begin{equation}
N_{\rm H, c}=\frac{F^{\rm peak}_{\rm \nu}R_{\rm gd}}{\Omega_{\rm A}\mu_{\rm H}m_{\rm H}\kappa_{\nu}B_{\rm \nu}(T_{\rm d})},
\end{equation}
where $F^{\rm peak}_{\rm \nu}$ is the peak flux density of the leaf (in Jy beam$^{-1}$), $R_{\rm gd}$ is the total (gas plus dust)-to-(refractory-component-)dust-mass ratio, $\Omega_{\rm A}~=~[(\pi/4{\rm ln}2)\theta_{\rm maj}\theta_{\rm min}]$ is the beam solid angle ($\theta_{\rm maj}$ and $\theta_{\rm min}$ are the major and minor axes of the synthesized beam, respectively; see \S~\ref{Section:observations}), $\kappa_{\nu}$ is the dust opacity per unit mass at a frequency $\nu$, and $B_{\nu}(T_{\rm d})$ is the Planck function at a dust temperature, $T_{\rm d}$. 

The dust opacity per unit mass is determined from $\kappa_{\nu}=~\kappa_{0}(\nu/\nu_0)^{\beta}$, assuming a dust emissivity index, $\beta$, where $\kappa_{0}$ is based on the moderately coagulated thin ice mantle dust model of \citet{ossenkopf_1994} at a frequency, $\nu_{0}$. At a frequency of $\sim93$\,GHz we adopt a value of $\kappa_{\nu}\approx0.186$\,cm$^{2}$g$^{-1}$ (extrapolating from $\kappa_{0}=0.899$\,cm$^{2}$g$^{-1}$ at $\nu_{0}=230$\,GHz with $\beta=1.75$; e.g. \citealp{battersby_2011}). From \citet[table\,23.1]{draine_2011}, the \emph{hydrogen}-to-(refractory-component-)dust-mass ratio is $R_{\rm gd, H}\sim100$. Therefore we adopt a value of $R_{\rm gd}=141$ for the \emph{total} (gas plus dust)-to-(refractory-component-)dust-mass ratio (assuming a typical interstellar composition of H, He, and metals). As discussed in \S~\ref{Section:discussion_filament_frag}, there are currently no gas or dust temperature estimates for \irdc \ at a resolution equivalent to those studied here. For the leaf analysis we assume that $T_{\rm c}<T_{\rm f}$, that $T_{\rm d}=T_{\rm c}$, and $T_{\rm d}=13\,{\rm K}$ (at the lower end of the range derived by \citealp{nguyen_2011}). The corresponding column density sensitivity derived from our $3\sigma_{\rm rms}$ flux level of 0.21 mJy\,beam$^{-1}$ is $N_{\rm H}\sim1.3\times10^{23}$\,cm$^{-2}$.

The derived column densities range from $3.6~\times~10^{23}\,{\rm cm^{-2}}<~N_{\rm H,c}<8.0~\times10^{23}\,{\rm cm^{-2}}$ ($1.5\times10^{23}\,{\rm cm^{-2}}<N_{\rm H,c}^{\rm b}<5.9\times10^{23}\,{\rm cm^{-2}}$), with a mean value of $N_{\rm H,c}\sim4.7\times10^{23}\,{\rm cm^{-2}}$ ($N^{\rm b}_{\rm H,c}\sim2.5\times10^{23}\,{\rm cm^{-2}}$). Given the uncertainties in the flux calibration ($\sim10$ per cent), dust opacity per unit mass ($\sim30$ per cent; accounting for different degrees of coagulation in the \citealp{ossenkopf_1994} models), total (gas plus dust)-to-dust mass ratio ($\sim30$ per cent), and temperature ($\pm3$\,K), the uncertainty in the derived beam-averaged column density is $\sim50$ per cent (after summing in quadrature). These values, as well as those estimated below, can be found in Table~\ref{Table:derived_leaf_info}.

The mass of each leaf is estimated using 
\begin{equation}
M_{\rm c}=\frac{d^{2}S_{\nu}R_{\rm gd}}{\kappa_{\nu}B_{\nu}(T_{\rm d})}, 
\label{Eq:mass_calc}
\end{equation}
where $d$ is the distance to the source ($\sim2.9$\,kpc) and $S_{\nu}$ is the integrated leaf flux (in Jy). The resultant leaf masses range from $8.1\,{\rm M_{\odot}}<M_{\rm c}<26.1\,{\rm M_{\odot}}$ \ ($1.5\,{\rm M_{\odot}}<M_{\rm c}^{\rm b}<12.1\,{\rm M_{\odot}}$). Note however, that the masses of leaves~\#4 ($M_{\rm c}=10.9$\,\solar), \#5 (26.1\,\solar), \#6 (16.6\,\solar), and \#13 (14.0\,\solar) cannot be unambiguously attributed to single structures (see \S~\ref{Section:results_line} and Appendix~\ref{App:kinematics}). These leaves are therefore rejected from any further analysis which is reliant on a geometrical assumption. Combining the uncertainties in $R_{\rm gd}$, $k_{\nu}$, $T_{\rm d}$, with those  in the integrated flux density (typically $\sim2$ per cent) and the distance measurement ($\sim15$ per cent; \citealp{simon_2006b}), the uncertainty in the derived mass is expected $\sim60$ per cent. 

For comparison, we also estimate the mass from the mass surface density map of \citet{kainulainen_2013a}. We extract this mass estimate, $M_{\rm MIREX}$, from within the boundary defining the maximum (projected) physical extent of each leaf (see Fig.~\ref{Figure:spec_mom} and those in Appendix~\ref{App:kinematics}). We find $4.8\,{\rm M_{\odot}}<M_{\rm MIREX}<26.8\,{\rm M_{\odot}}$. Comparing the masses of directly, we find \textbf{$0.4<M_{\rm MIREX}/M_{\rm c}<1.1$}, with an average value of $\langle M_{\rm MIREX}/M_{\rm c}\rangle\sim0.84$. Due to the lack of short spacings in our interferometric map, we caution against drawing firm conclusions from this comparison. However, the fact that the estimates agree (within the uncertainties) gives us confidence in the reliability of our continuum derived masses. Finally, comparing the total mass of (all) the leaves with the mass of the inner filament estimated by \citet{hernandez_2012a}, we find that the leaves make up $\sim10$ per cent of the total mass of the region.

The equivalent particle number density at the surface of a leaf with radius, $R_{\rm eq}$, and mass, $M_{\rm c}$, can be estimated using
\begin{equation}
n_{\rm H, c, eq}=\frac{M_{\rm c}}{\frac{4}{3}\pi R_{\rm eq}^{3}\mu_{\rm H}m_{\rm H}}.
\end{equation}
The range in particle number density is $6.1\times10^{5}\,{\rm cm^{-3}}<n_{\rm H, c, eq}<14.7\times10^{5}\,{\rm cm^{-3}}$ ($1.9\times10^{5}\,{\rm cm^{-3}}<n^{\rm b}_{\rm H, c, eq}<5.2\times10^{5}\,{\rm cm^{-3}}$), with a mean value of $n_{\rm H, c, eq}=9.5\times10^{5}\,{\rm cm^{-3}}$ ($n^{\rm b}_{\rm H, c, eq}=2.7\times10^{5}\,{\rm cm^{-3}}$). The corresponding range in the local free-fall time,
\begin{equation}
t_{\rm ff, c}=\bigg(\frac{3\pi}{32G\mu_{\rm H}m_{\rm H} n_{\rm H,c, eq}}\bigg)^{1/2},
\end{equation}
for each of the dendrogram leaves is $3.6\times10^{4}\,{\rm yr}<t_{\rm ff, c}<5.5\times10^{4}\,{\rm yr}$ ($6.0\times10^{4}\,{\rm yr}<t^{\rm b}_{\rm ff, c}<10.1\times10^{4}\,{\rm yr}$). Assuming a mean \emph{filament} density of $n_{\rm f}=0.2\times10^{5}\,{\rm cm^{-3}}$ (see \S~\ref{Section:discussion_filament_frag}), we find $t_{\rm ff, f}\sim2.4\times10^{5}$\,yr, which is $\sim$ an order of magnitude greater than the estimated free-fall time the embedded smaller scale structures. Given the uncertainty in the mass estimate ($\sim60$ per cent) and in the distance measurement ($\sim15$ per cent) the relative uncertainties in the number density and free-fall time are $\sim75$ and $\sim40$ per cent, respectively.

\subsection{Dynamical properties of the dendrogram leaves}\label{Section:analysis_virial}

Using the physical properties of the dendrogram leaves derived in \S~\ref{Section:analysis_phys}, we can now assess whether the leaves themselves are susceptible to gravitational collapse (and potentially further fragmentation). In the following sections, we evaluate the support provided by different mechanisms. 

\subsubsection{Thermal support}

To determine the likelihood that the dendrogram leaves will collapse we first consider the thermal Jeans mass
\begin{equation}
M_{\rm J}=\frac{\pi^{5/2}c^{3}_{\rm s}}{6(G^{3}\rho_{\rm c})^{1/2}}\sim2.2\,\bigg(\frac{T_{\rm c}}{15\,{\rm K}}\bigg)^{3/2}\bigg(\frac{n_{\rm H,c}}{10^{5}\,{\rm cm^{-3}}}\bigg)^{-1/2}\,{\rm M_{\odot}},
\end{equation}
and thermal Jeans length
\begin{equation}
\lambda_{\rm J, c}=c_{\rm s}\bigg(\frac{\pi}{G\rho_{\rm c}}\bigg)^{1/2}\sim0.11\,\bigg(\frac{T_{\rm c}}{15\,{\rm K}}\bigg)^{1/2}\bigg(\frac{n_{\rm H,c}}{10^{5}\,{\rm cm^{-3}}}\bigg)^{-1/2}\,{\rm pc}.
\end{equation}
This analysis assumes that only thermal pressure contributes to supporting the leaves. Utilizing the densities presented in Table~\ref{Table:derived_leaf_info}, we estimate a range in Jeans masses $0.5\,{\rm M_{\odot}}<M_{\rm J,c}<0.7\,{\rm M_{\odot}}$ ($0.8\,{\rm M_{\odot}}<M^{\rm b}_{\rm J,c}<1.3\,{\rm M_{\odot}}$). Comparing these values with the derived leaf masses, we find $16<M_{\rm c}/M_{\rm J,c}<45$ ($1<M^{\rm b}/M^{\rm b}_{\rm J,c}<15$). Since $M_{\rm c}/M_{\rm J,c}\gg 1$, this implies that thermal support alone cannot provide sufficient support against collapse. Without any additional support the leaves would be expected to collapse on a time-scale equivalent to the free-fall time, $\langle t_{\rm ff,c} \rangle \sim 5\times10^{4}$\,yr ($\langle t^{\rm b}_{\rm ff,c} \rangle \sim 9\times10^{4}$\,yr), and possibly fragment, with a corresponding length scale of $0.03\,{\rm pc}<\lambda_{\rm J,c}<0.04\,{\rm pc}$ ($0.04\,{\rm pc}<\lambda^{\rm b}_{\rm J,c}<0.07\,{\rm pc}$). 

\begin{table*}
	\caption{Dendrogram leaves: virial analysis. Leaves rejected from the analysis (see \S~\ref{Section:results_line} and Appendix~\ref{App:kinematics}) are not included in this table. Leaf~\#8 cannot be unambiguously linked to either of the two velocity components identified and so both entries are included (repeated values are indicated with `...').} \vspace{0.2cm}
	
	\centering  
	\tabcolsep=0.14cm \normalsize{
	\begin{tabular}{  c  c  c  c  c  c  c  c  c  c  c  c  c  c }
	\hline
	ID & 
	$\Delta\alpha$ & 
	$\Delta\delta$ & 
	$v_{\rm 0}^{a}$ & 
	$\Delta v^{a}$ & 
	$\sigma_{v}^{b}$ & 
	$R_{\rm eq}$ &
	$M_{\rm c}$ &
	\multicolumn{4}{ c }{Estimated virial ratio$^{c}$} &
	\multicolumn{2}{ c }{Model best fit$^{d}$} 
	\\ [0.5ex]
	
	&
	&
	&
	&
	&
	&
	&
	&
	$\kappa_{\rho}=0.0$ &
	$\kappa_{\rho}=1.0$ &
	$\kappa_{\rho}=1.5$ &
	$\kappa_{\rho}=2.0$ &
	&
	\\ [0.5ex]
	
 	& 
	(arcsec) & 
	(arcsec) & 
	(\kms) & 
	(\kms) & 
	(\kms) &   
	(pc) &
	(\solar) &
	$a_{\rho}=1$ &
        $a_{\rho}=10/9$ &
	$a_{\rho}=5/4$ &
	$a_{\rho}=5/3$ &
	&
	\\ [0.5ex]
	 
	\hline 
	
	&
	&
	&
	&
	&
	&
	&
	&
	$\alpha_{\rm vir}$ &
	$\alpha_{\rm vir}$ &
	$\alpha_{\rm vir}$ &
	$\alpha_{\rm vir}$ &
	$\kappa_{\rho}$ &
	$\alpha_{\rm vir}$
	\\ [0.5ex]
	
	\hline

 1 &    8.9 &  -76.9 &  45.18 (0.02) &   0.93 (0.04) &   0.45 (0.01) &  0.053 &10.68&   1.14 &   1.02 &   0.91 &   0.68 &   1.90 &   0.74      \\ [0.5ex]
 2 &    7.4 &  -64.7 &  45.40 (0.01) &   1.13 (0.02) &   0.52 (0.01) &  0.065 &19.07&   1.08 &   0.97 &   0.86 &   0.65 &   1.82 &   0.75      \\ [0.5ex]
 3 &    1.3 &  -50.3 &  45.49 (0.01) &   0.86 (0.02) &   0.42 (0.01) &  0.058 &14.82&   0.80 &   0.72 &   0.64 &   0.48 &   1.72 &   0.59      \\ [0.5ex]
 7 &    7.4 &   22.7 &  45.14 (0.02) &   0.99 (0.04) &   0.47 (0.01) &  0.054 &21.83&   0.62 &   0.56 &   0.50 &   0.37 &   1.96 &   0.39      \\ [0.5ex]
 8 &    0.5 &   23.5 &  45.17 (0.01) &   0.49 (0.02) &   0.29 (0.01) &  0.040 & 9.38&   0.43 &   0.39 &   0.34 &   0.26 &   1.84 &   0.29      \\ [0.5ex]
 $...$ &  $...$ &  $...$ &   45.84 (0.02) &   1.37 (0.03) &   0.62 (0.01) &   $...$ & $...$ &   1.89 &   1.70 &   1.51 &   1.13 &    $...$ &   1.29      \\ [0.5ex]
 9 &   -6.3 &   28.8 &  46.63 (0.01) &   0.62 (0.03) &   0.33 (0.01) &  0.060 &24.42&   0.31 &   0.28 &   0.25 &   0.19 &   2.14 &   0.16      \\ [0.5ex]
10 &    2.8 &   34.1 &  45.79 (0.01) &   0.98 (0.02) &   0.47 (0.01) &  0.037 & 8.23&   1.13 &   1.02 &   0.90 &   0.68 &   2.00 &   0.68      \\ [0.5ex]
11 &  -12.4 &   38.7 &  46.75 (0.03) &   0.58 (0.06) &   0.32 (0.02) &  0.041 & 8.14&   0.61 &   0.55 &   0.49 &   0.36 &   1.84 &   0.41      \\ [0.5ex]
12 &   -7.1 &   42.5 &  46.07 (0.01) &   1.32 (0.02) &   0.60 (0.01) &  0.041 & 8.26&   2.06 &   1.85 &   1.64 &   1.23 &   1.98 &   1.26      \\ [0.5ex]

	\hline 
\end{tabular}
	\vspace{0.2cm}

\begin{minipage}{1.0\textwidth}\footnotesize{
$^{a}$ Centroid velocity (plus uncertainty) and FWHM line-width (plus uncertainty). See \S~\ref{Section:results_line} for more details on the method.  \\
$^{b}$ Total (thermal plus non-thermal) velocity dispersion of the mean particle. Derived from Equation~\ref{Eq:disp_eff} using the FWHM line-width of the \ntwoh \ (1-0) isolated hyperfine component. \\
$^{c}$ Estimated virial ratios assuming a density profile of the form $\rho\propto r^{-\kappa_{\rho}}$ (\S~\ref{Section:analysis_thermturb}). Uncertainty: $\sigma\alpha_{\rm vir}\sim60$ per cent. \\
$^{d}$ Model best-fitting solutions to $\kappa_{\rho}$.\\
}
\end{minipage}
}
\label{Table:virial_analysis}
\end{table*}

\subsubsection{Thermal $+$ turbulent support}\label{Section:analysis_thermturb}

The relative importance of a cloud fragment's kinetic and gravitational energy can be expressed in the form of the dimensionless virial parameter, $\alpha_{\rm vir}$ \citep{bertoldi_1992}:
\begin{equation}
\alpha_{\rm vir}\,\equiv\,\frac{5\sigma_{v}^{2}R_{\rm eq}}{GM_{\rm c}}=\frac{M_{\rm vir}}{M_{\rm c}}=2a\frac{E_{\rm kin}}{|E_{\rm pot}|}.
\label{Eq:virial}
\end{equation}
where $\sigma_{v}$ incorporates contributions to the kinetic energy from thermal motions and non-thermal motions within the gas (the best-fitting solution to each leaf spectrum determined in \S~\ref{Section:results_line} provides the means to estimate $\sigma_{v}$; see Equation~\ref{Eq:disp_eff}),  $M_{\rm vir}\equiv5R_{\rm eq}\sigma_{v}^{2}/G$ is the virial mass, the parameter $a=a_{\theta}a_{\rho}\equiv5R|E_{\rm pot}|/3GM^{2}$, in Equation~\ref{Eq:virial} is the ratio of gravitational energy, $E_{\rm pot}$ (assuming negligible external tides), to that of a uniform sphere, and $E_{\rm kin}$ is the kinetic energy. 

Deviations from spherical symmetry are accounted for in $a_{\theta}$. \citet{bertoldi_1992} consider a triaxial ellipsoid with equatorial radius, $R$, and an extent in the third dimension, $2Z$, such that the aspect ratio is $y=Z/R$. They show that for ${\rm log_{10}}(Z/R)~<~|1|$, $a_{\theta}\approx1.0\pm0.2$. Consequently we ignore the effect of clump elongation in the following analysis. 

The parameter, $a_{\rho}$, measures the effect of a nonuniform density distribution. It is estimated using
\begin{align}
a_{\rho} = \frac{(1-\kappa_{\rho}/3)}{(1-2\kappa_{\rho}/5)},
\end{align}
whereby $\kappa_{\rho}$ relates to a density structure of the form, $\rho_{\rm c}(r)\propto r^{-\kappa_{\rho}}$. Ignoring the effect of both surface pressure and magnetic fields, a cloud in virial equilibrium has $|E_{\rm pot}| = 2E_{\rm kin}$. For a virialized spherical cloud with a power law density distribution, $\rho_{\rm c}\propto r^{-\kappa_{\rho}}$, $\alpha_{\rm vir}=a=1$ when $\kappa_{\rho}=0$ (i.e. uniform density) and $\alpha_{\rm vir}=a=5/3$ when $\kappa_{\rho}=2$ (i.e. a singular isothermal sphere). 

Recent high-angular resolution studies of IRDCs find $\kappa_{\rho}=1.5-2.0$ (e.g. \citealp{zhang_2009, wang_2011, butler_2012, palau_2014}). Many of our identified leaves are only marginally resolved. However, we can make a crude attempt to measure $\kappa_{\rho}$ by investigating the radial flux density profile of the dendrogram leaves (making assumptions about the leaf geometry). Assuming that the density, $\rho_{\rm c}$, and temperature, $T_{\rm c}$, scale as a power law with radius, $\rho_{\rm c}(r)\propto r^{-\kappa_{\rho}}$ and $T_{\rm c}(r)\propto r^{-\kappa_{T}}$, then the flux density of dust emission is given by $F_{\nu}\propto\int{\rho_{\rm c}T_{\rm c} ds}$, where $S$ is the length along the line-of-sight. Assuming spherical symmetry, the flux density scales as $F_{\nu}\propto r^{-(\kappa_{\rho}+\kappa_{T} - 1)}$, which simplifies to $F_{\nu}~\propto~r^{-(\kappa_{\rho} - 1)}$ (valid for $\kappa_{\rho}>1$, e.g. \citealp{ward-thompson_1994, andre_1996, longmore_2011}) in the isothermal case. Using a profile of the form (a subscript `m' denotes model parameters)
\begin{align*}
 F_{\rm \nu,m}(r)&=
  \begin{cases}
   F^{\rm peak}_{\rm \nu,m}\bigg(\frac{r}{R^{\rm peak}_{\rm eq,m}}\bigg)^{-(\kappa_{\rho}-1)}  & \text{where} \ r < R_{\rm eq} \\
   \text{const.} & \text{where} \ r > R_{\rm eq} 
  \end{cases}
\end{align*}
we generate synthetic images, characterized by a peak flux density, $F^{\rm peak}_{\rm \nu,m}$, at an equivalent radius, $R^{\rm peak}_{\rm eq,m}~=~(A_{\rm pix,m}/\pi)^{1/2}$. Where $r>R_{\rm eq}$ we set the constant value equivalent to the minimum value of $F_{\rm \nu,m}(r<R_{\rm eq})$. The synthetic images are then convolved with a Gaussian kernel, the FWHM of which is equivalent to $\langle\theta\rangle~=~(\theta_{\rm max}\theta_{\rm min})^{1/2}$, and the peak model flux density is normalized to the observed peak flux density, $F^{\rm peak}_{\nu}$, at an equivalent radius, $R^{\rm peak}_{\rm eq}$.  

\begin{figure*}
\begin{center}
\includegraphics[trim = 35mm 10mm 0mm 0mm, clip, width = 0.48\textwidth]{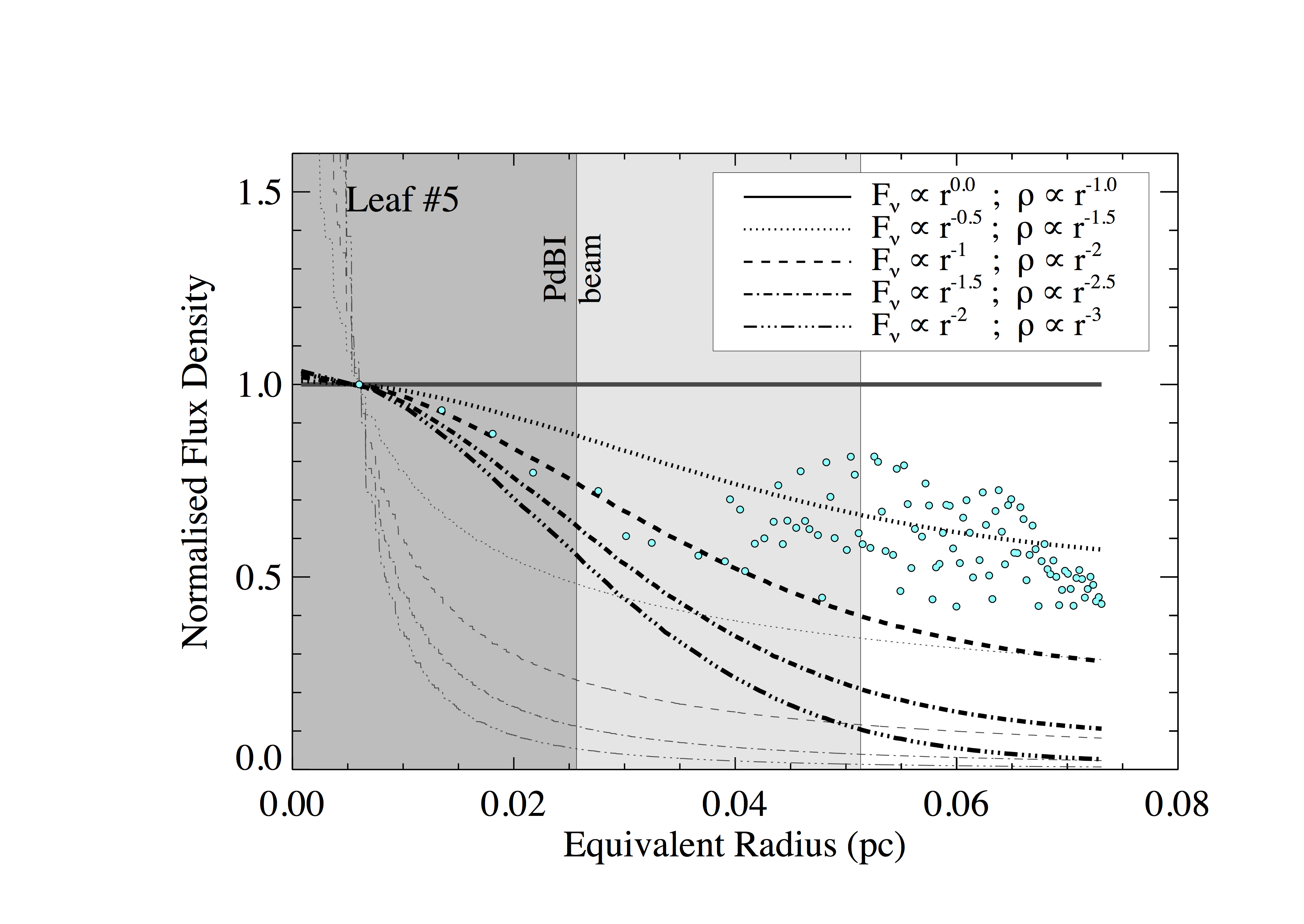}
\includegraphics[trim = 35mm 10mm 0mm 0mm, clip, width = 0.48\textwidth]{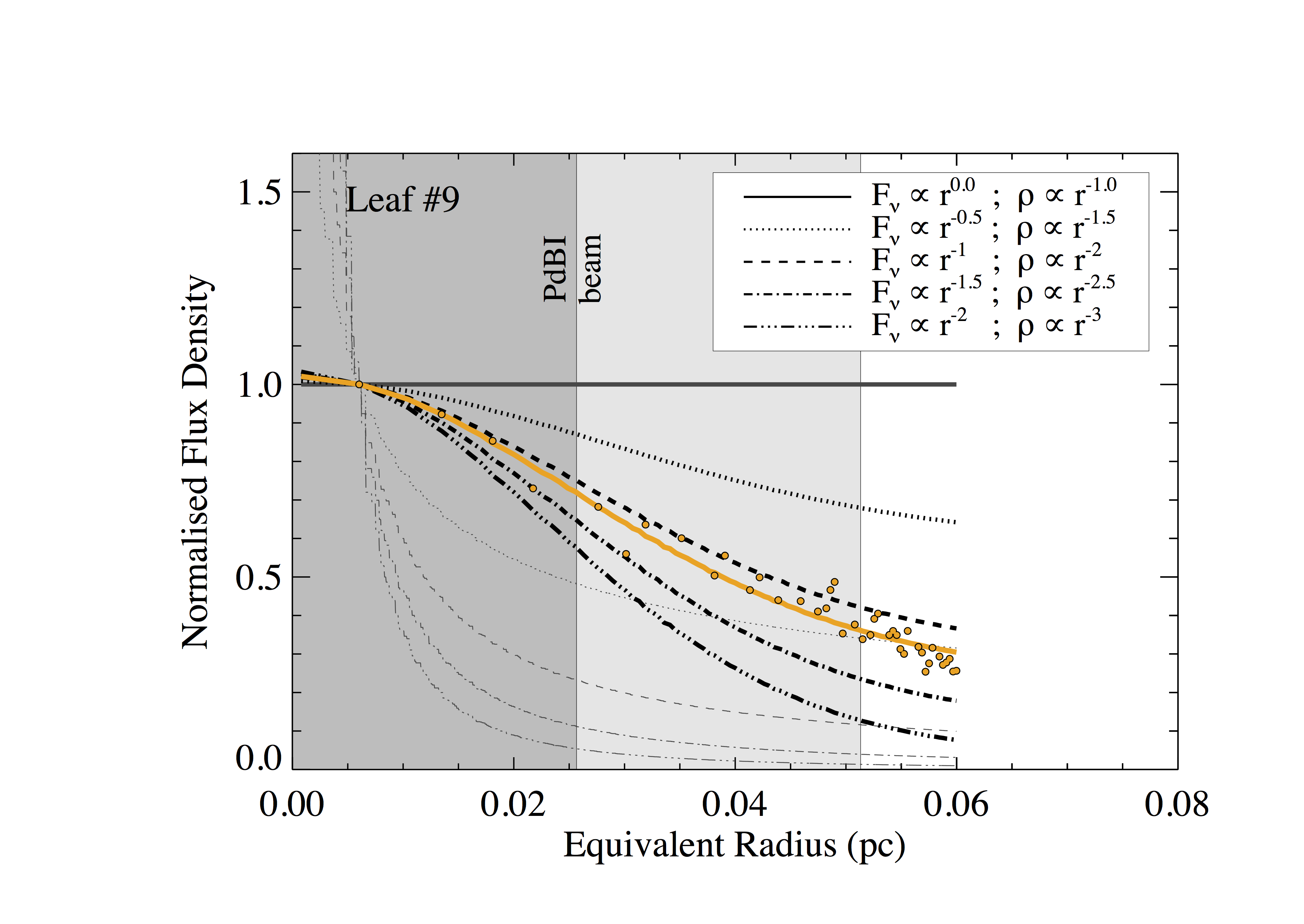}
\end{center}
\caption{Flux density as a function of equivalent radius for leaves~\#5 (left-hand panel) and \#9 (right-hand panel). Black lines are model radial flux density profiles before (light) and after (heavy) beam correction (see \S~\ref{Section:analysis_thermturb}). Each line corresponds to a density profile of the form $F_{\rm \nu}\propto r^{-\kappa_{\rho}}$. The corresponding values of $\kappa_{\rho}$ are included in the legend. The dark grey block indicates the extent of the geometric mean radius of the synthesized PdBI beam (1.83\,arcsec or 0.025\,pc at a distance of 2900\,pc). The light grey block is twice this value. The solid orange line in the right-hand panel indicate the closest-matching model solution to the observations. The lack of a corresponding model solution in the left-hand panel reflects the fact that leaf~\#5 has been rejected from our analysis. This figure supports this decision, since the additional peaks in the radial flux density profile may signify the presence of underlying substructure. Radial flux density profiles for the additional leaves can be found in Appendix~\ref{App:kinematics}. }
\label{Figure:radial_FD_profile}
\end{figure*}

Fig.~\ref{Figure:radial_FD_profile} displays the radial flux density profiles for leaves~\#5 and \#9 (as in Figure~\ref{Figure:spec_mom}). The left-hand panel displays the radial flux density profile for leaf~\#5. For reasons discussed in \S~\ref{Section:results_line}, this leaf is not included in the analysis. However, in addition to the fact that the northern and southern portions of the leaf can be attributed to different kinematic components, this image supports our decision to reject this leaf. As can be seen from Fig.~\ref{Figure:radial_FD_profile}, the flux density profile exhibits additional peaks, possibly signifying the presence of underlying substructure (evidence, in this example, for the superposition of fragments associated with different sub-filaments). The right-hand panel displays the radial flux density profile for leaf~\#9. In contrast to the profile of leaf~\#5, the flux density decreases smoothly as a function of radius. The implication is that the leaf is monolithic (at the spatial resolution of our PdBI observations). The remaining radial flux density profiles can be found in Appendix~\ref{App:kinematics}. The light and heavy lines in Fig.~\ref{Figure:radial_FD_profile} signify the synthetic radial flux density profiles before and after beam correction, respectively. The solid orange line represents our best-fitting model solution to the radial flux density profile for leaf~\#9  ($\kappa_{\rho}=2.14$). On average, we find $\kappa_{\rho}=1.9$. 

The above considerations enable us to define a critical value, $\alpha_{\rm vir, cr}\equiv a$, which serves as a gauge to assess the stability of a cloud fragment. In this simplistic formalism, cloud fragments with $\alpha_{\rm vir}<\alpha_{\rm vir,cr}$ are susceptible to gravitational collapse in the absence of additional internal support and cloud fragments with $\alpha_{\rm vir}>\alpha_{\rm vir,cr}$ may expand in the absence of pressure confinement (more detailed stability analysis, accounting for the effects of surface pressure, finds $\alpha_{\rm vir, cr}\approx2$ for isothermal, non-magnetized cloud fragments in equilibrium; \citealp{mckee_1999a, kauffmann_2013}). Normalization of $\alpha_{\rm vir}$ by $a$ gives $\alpha_{\rm vir, cr}=1$, allowing for ease of comparison between virial parameters of a cloud fragment estimated assuming different density profiles. For simplicity, we do not carry out this analysis on the background-subtracted leaves, because of complications in estimating background-subtracted velocity dispersions. For leaf masses without background-subtraction, we find $0.3<\alpha_{\rm vir}<2.1$ (with a mean value, $\langle\alpha_{\rm vir}\rangle=1.0$) and $0.2<\alpha_{\rm vir}<1.2$ ($\langle\alpha_{\rm vir}\rangle=0.6$) for $\kappa_{\rho}=0$ and $\kappa_{\rho}=2.0$, respectively. Incorporating our density profile analysis returns virial parameters spanning the range $0.2<\alpha_{\rm vir}<1.3$ (with a mean value, $\langle\alpha_{\rm vir}\rangle=0.7$). Table~\ref{Table:virial_analysis} lists the virial parameters estimated for each of the dendrogram leaves.

This analysis indicates that, when taking $\sigma_{\rm NT}$ as an upper limit on the level of turbulent support, all dendrogram leaves are consistent with being either sub-virial or approximately virial ($\alpha_{\rm vir}\lesssim 1$ to within a factor of $\gtrsim2$ uncertainty for $\kappa_{\rho}=1.5-2$). In the absence of additional support, dendrogram leaves that are strongly sub-virial should undergo fairly rapid ($\sim t_{\rm ff,c}$) global collapse. Leaves~\#7 and \#9 are consistent with this picture ($\alpha_{\rm vir}\sim0.4$ and $\sim0.2$, respectively), and other regions of massive star formation where low virial parameters have been reported (e.g. \citealp{csengeri_2011a, pillai_2011, li_2013, peretto_2013, battersby_2014, beuther_2015, lu_2015}). However, low virial parameters such as these may instead be indicative of strong magnetic support (as suggested by e.g. \citealp{tan_2013b}). 

\subsubsection{Thermal + turbulent + magnetic support}\label{Section:analysis_mag}

An additional possibility is that the leaves are supported by magnetic fields, the effects of which have been thus far neglected. Following \citet[see \citealp{pillai_2011} for a recent adaptation]{bertoldi_1992}, one can define a magnetic virial mass, $M_{B,{\rm vir}}$, which incorporates the effect of both gas and magnetic pressure in providing support to the fragment
\begin{equation}
M_{B,{\rm vir}}=\frac{5R}{G}\bigg(\sigma^{2}_{v}+\frac{\sigma^{2}_{\rm A}}{6}\bigg),
\end{equation}
and a magnetic virial ratio
\begin{equation}
\alpha_{B,{\rm vir}}=\frac{M_{B,{\rm vir}}}{M_{\rm c}}
\end{equation}
where $\sigma_{\rm A}=B/\sqrt{(4\pi\rho_{\rm c})}$ is the Alfv\'{e}n velocity. Setting $M_{B,{\rm vir}}=M_{\rm c}$, and hence $\alpha_{B,{\rm vir}}=1$, for our fragments with $\alpha_{\rm vir}<1$ (Table~\ref{Table:virial_analysis}), we can estimate values of $B$ necessary for virial equilibrium. From this analysis we find field strengths in the range $230\,\mu{\rm G}<B<670\,\mu{\rm G}$ (with a median value of $\sim520\,\mu{\rm G}$) would be required to support the dendrogram leaves. These values are consistent with the similarly derived field strengths necessary for support in other massive star forming regions (e.g. \citealp{pillai_2011, tan_2013b}).

Comparing these values with the empirically-derived median field strength, $B_{\rm med}$, versus density relation of \citet{crutcher_2010} (valid for $n_{\rm H}>300\,{\rm cm^{-3}}$),
\begin{equation}
B_{\rm med}\approx220\,\bigg(\frac{n_{\rm H}}{10^{5}\,{\rm cm^{-3}}}\bigg)^{0.65}\,\mu{\rm G},
\end{equation}
assuming a distribution that is flat from 0 to $B_{\rm max}=2B_{\rm med}$, we find $0.3<B/B_{\rm med}<0.9$. Additionally, field strengths of the order $\sim{\rm mG}$ have recently been derived using observations towards massive star forming regions (e.g. \citealp{girart_2013,frau_2014,qiu_2014, pillai_2015,pillai_2016}). This indicates that the field strengths required to provide support to our identified leaves are broadly consistent with observations of cloud fragments of comparable density. However, additional observations would be needed to quantify this further. 

\section{Implications for star formation within G035.39--00.33}\label{Section:discussion_sf}

In \S~\ref{Section:discussion_filament_frag}, we find that the spacing between continuum sources throughout \irdc \ is significantly (factor of $\sim8$) smaller than that predicted by gravitational instabilities in purely hydrodynamical fluid cylinders. At face value this appears to suggest that magnetic fields may have a significant role to play in the reconciliation of the observed and predicted spatial distribution. However, complex line-of-sight structure and the presence of sub-filaments observed throughout \irdc, may make a significant contribution to the discrepancy. 

The idea that the continuum sources may be associated with different sub-filaments is qualitatively supported by leaves that appear close to one another in projection, but show clear differences in their radial velocities. For example, leaves~\#9 and \#10 are separated by a projected distance of $\sim0.15$\,pc but their line centroid velocities differ by $\sim0.8$\,\kms \ (see Table~\ref{Table:virial_analysis}). Conversely, leaves~\#9 and \#11 have a similar spatial separation but their line centroids differ by $\sim0.1$\,\kms. In this particular example, leaves~\#9 and \#11 seem to be consistent with the mean velocity of filament F3 of \citet{henshaw_2014}, $(46.86\pm0.04)$\,\kms, whereas leaf \#11 is consistent with the mean velocity of F2b, $(46.00\pm0.05)$\,\kms \ (filament F2a has a mean centroid velocity of $[45.34\pm0.04]$\,\kms, for reference). If the continuum sources can indeed be attributed to different sub-filaments, then the assumption that \irdc \ can be described simplistically, as a single cylindrical filament, is invalid. If confirmed, the observed spacing is most likely influenced by a combination of several important factors, including, the number of sub-filaments, differences in the individual sub-filament properties (e.g. density, inclination, velocity dispersion), and the strength and orientation of the magnetic field.

The origins of the complex physical and kinematic gas structure of \irdc \ are currently unknown. Whether the observed sub-filaments are a result of the fragmentation process (cf. the `\emph{fray and fragment}' scenario proposed by \citealp{tafalla_2015}) or whether they first formed at the stagnation points of a turbulent velocity field and have been brought together by gravitational contraction on larger scales (cf. the `\emph{fray and gather}' scenario proposed by \citealp{smith_2016}), remains an open question. However, the presence of widespread emission from shocked gas tracers (e.g. SiO; \citealp{izaskun_2010}) throughout \irdc \ (and other molecular clouds, e.g. \citealp{nguyen_2013, duarte-cabral_2014}), may point towards a dynamical origin. The prevalence of sub-filaments in many observational studies (e.g. \citealp{hacar_2013, peretto_2013, alves-de-oliveira_2014, fernandez-lopez_2014, lee_2014, panopoulou_2014, peretto_2014, dirienzo_2015}) emphasizes the importance of considering the underlying physical structure in any fragmentation analysis. Putting this another way, this highlights the danger of using simple geometric models, without prior consideration of the kinematic information. 

The formation of sub-filaments, followed by the formation of cores native to those sub-filaments (and potentially, further fragmentation of those cores), may signify a multilayered fragmentation process within \irdc, similar to that proposed in other molecular clouds (e.g. \citealp{teixeira_2006, hacar_2013, kainulainen_2013a, takahashi_2013, wang_2014, beuther_2015b, tafalla_2015}). We find that the majority the cores within \irdc, including the two most massive objects (leaves~\#7 and \#9), are located towards the H6 region ($\{\Delta\alpha,\,\Delta\delta\}\sim\{3\,{\rm arcsec},\,20\,{\rm arcsec}\}$; \citealp{butler_2012}). This is also the location at which several of the sub-filaments meet (\citealp{henshaw_2013, henshaw_2014, izaskun_2014}), which is reminiscent of studies highlighting the formation of star clusters at the junctions of complex filamentary systems (e.g. \citealp{myers_2009, schneider_2012, peretto_2014}), and consistent with simulations (e.g. \citealp{dale_2011, amyers_2013, smith_2013}).

Interestingly, the analysis presented in \S~\ref{Section:analysis_phys} shows that there is only a factor of $\sim3$ difference between the highest mass (leaf~\#9; $\sim24$\,\solar) and lowest mass (leaf~\#11; $\sim8$\,\solar) cores identified in this study. This is in spite of the fact that our PdBI data are theoretically sensitive to masses that are a factor of $\sim4$ lower than this ($\sim2.0$\,\solar).\footnote{This has been conservatively estimated by integrating a uniform $4\sigma_{\rm rms}$ flux density (0.28\,mJy\,beam$^{-1}$) over 26 pixels (${\rm min\_npix}$; see \S~\ref{Section:results_cont}), and using this in Equation~\ref{Eq:mass_calc}.} The steep slope of the locally-invariant stellar initial mass function (IMF) implies that many low-mass stars form within clusters alongside high-mass stars \citep{bastian_2010, offner_2013}. Assuming that the mass distribution of pre-stellar cores, the core mass function (CMF), takes a form $dN/d({\rm log}\,m)\propto M^{-\Gamma}$, where $\Gamma=1-1.5$, for masses $M>0.5$\,\solar \ (e.g. \citealp{motte_1998}), it follows that for every 25\,\solar \ core (equivalent to the mass of the most massive leaf detected in this paper) one might expect to find $10\pm3$ cores in the range 2-8\,\solar, i.e. the mass range covering our sensitivity and the lowest mass leaf detected in the present investigation. 

In a recent study by \citet{zhang_2015}, who present ALMA observations of IRDC ${\rm G}28.34+0.06$, an apparent dearth of low-mass dense cores was also noted. \citet{zhang_2015} explain that it would be counterintuitive for stars to form first in the lower density regions surrounding massive clumps within which high-mass stars are forming in ${\rm G}28.34+0.06$, and instead favour the interpretation that low-mass cores and stars form at a later stage, after the formation of massive stars. However, this is in contrast to the work of \citet{foster_2014}, who detect a population of low-mass protostars in the IRDC ${\rm G}34.43+00.24$. This newly-identified population of low-mass stars is situated in the interclump medium of the filamentary cloud. Their presence leads the authors to suggest that the population of low-mass stars may have formed before, or perhaps coevally with, the high-mass stars. 

Close inspection of the continuum map presented in Fig.~\ref{Figure:cont_dendro}, and studying the radial flux density profiles of the dendrogram leaves (see Appendix~\ref{App:kinematics}), indicates that several of the leaves exhibit substructure. Our ability to detect lower mass cores may therefore be limited by our angular resolution. To assess this further, in \S~\ref{Section:analysis_virial} we sought to establish whether the leaves which are kinematically coherent (i.e. those leaves which can be attributed to a single velocity structure but may harbour underlying substructure in the continuum), are susceptible to collapse, and potentially further fragmentation. This analysis reveals that in the absence of additional support, possibly from magnetic fields with strengths of the order $230\,\mu{\rm G}<B<670\,\mu{\rm G}$ (determined by equating the leaf masses with a critical core mass that incorporates the effect of both gas and magnetic pressure in providing support to the fragment; \S~\ref{Section:analysis_mag}), the leaves may collapse. 

The above implies that our ascent of \irdc's structure tree is not yet complete. Future, high angular resolution (approaching the Jeans length of the individual leaves, $\lambda_{\rm J,c}\lesssim0.03\,{\rm pc}$; \S~\ref{Section:analysis_virial}) and high sensitivity continuum observations \citep{henshaw_2016c}, as well as observations of molecular lines with higher critical densities, are needed to probe further sub-fragmentation (akin to that observed in low mass cores; e.g. \citealp{pineda_2011,pineda_2015}). This will determine whether the lack of cores identified between 2-8\,\solar \  has a physical origin or if this can be explained by observational bias. Such observations will also aid in testing the predictions of hydrodynamical simulations of collapsing cloud cores, which show an increased level of fragmentation in cores with shallower ($\rho\propto r^{-1}$) density profiles compared to those with steeper ($\rho\propto r^{-2}$) profiles \citep{girichidis_2011}. Specifically, this will help to establish the fate of leaves~\#7 and \#9, which appear centrally-concentrated ($\rho\propto r^{-1.96}$ and $r^{-2.14}$, respectively; \S~\ref{Section:analysis_thermturb}) and monolithic at the resolution of our PdBI observations. With steep density profiles, estimated masses of the order $\sim20-25$\,\solar \ (\S~\ref{Section:analysis_phys}), and dark at 8\,\micron \ and 24\,\micron \ (note that leaf~\#7 has a 70\,\micron \ counterpart; \citealp{nguyen_2011}), these are currently the best candidates for progenitors of intermediate-to-high mass stars within the mapped region.

\section{Summary \& conclusions}\label{Section:conclusions}

We use high angular resolution ($\sim4$\,arcsec; $0.05$\,pc) 3.2\,mm PdBI continuum observations to perform a structural analysis of the filamentary IRDC \irdc. To date, these are the highest-angular resolution continuum observations of \irdc, surpassing previous observations by factors of $\sim2-3$ (i.e. the 70\,\micron \ \emph{Herschel} data presented by \citealp{nguyen_2011} and the 1.2\,mm observations presented by \citealp{rathborne_2006}). Our analysis leads us to conclude the following:

\begin{enumerate}
\item The continuum emission is highly structured. It is segmented into a series of 13 quasi-regularly spaced ($\lambda_{\rm obs}\sim0.18\,{\rm pc}$) cores, identified as `leaves' in the dendrogram analysis, that follow the major axis of the \irdc. 
\item Comparison between continuum and \ntwoh \ (1-0) observations suggests that some of the identified leaves may reflect a superposition of structures associated with different velocity components. Although the translation between position-position-velocity and true three dimensional space can be uncertain, this result emphasises the importance of exercising caution when attempting to classify structure in two-dimensional maps. 
\item Some leaves which appear to be kinematically coherent (i.e. they can be attributed to a single velocity component) can also exhibit structured continuum emission, which is evident in their radial flux density profiles. However, the scales at which this substructure resides is beyond the angular resolution of the observations in this study and further investigation is required.
\item There is a significant (a factor of $\sim8$) discrepancy between the spatial separation of the leaves and that predicted by theoretical work describing the fragmentation of purely hydrodynamic fluid cylinders. Consistent with the kinematic analysis of \citet{henshaw_2014}, who find evidence for the presence of sub-filaments observed throughout \irdc, this result emphasizes the importance of considering the underlying physical structure (and potentially, dynamically important magnetic fields) in any fragmentation analysis.
\item The leaves exhibit a range in column density ($3.6~\times~10^{23}\,{\rm cm^{-2}}<~N_{\rm H,c}<8.0~\times10^{23}\,{\rm cm^{-2}}$ ), mass ($8.1\,{\rm M_{\odot}}<M_{\rm c}<26.1\,{\rm M_{\odot}}$), and number density ($6.1\times10^{5}\,{\rm cm^{-3}}<n_{\rm H, c, eq}<14.7\times10^{5}\,{\rm cm^{-3}}$). 
\item We used the derived physical properties of the leaves to assess their dynamical state, and determine the likelihood that they will undergo gravitational collapse. All dendrogram leaves are consistent with being either sub-virial or approximately virial ($\alpha_{\rm vir}\lesssim 1$, within the 60 per cent uncertainty, for $\kappa_{\rho}=1.5-2$). Absolute values span a range $0.2<\alpha_{\rm vir}<1.3$. In the absence of additional support, possibly from magnetic fields with strengths of the order of $230\,\mu{\rm G}<B<670\,\mu{\rm G}$, leaves that are strongly sub-virial are susceptible to gravitational collapse, and possibly further fragmentation. Leaves~\#7 and \#9 are consistent with this picture ($\alpha_{\rm vir}\sim0.4$ and $\sim0.2$, respectively).
\item The formation of sub-filaments, followed by the formation of cores native to those sub-filaments, and the possibility of further fragmentation may imply a multilayered fragmentation process within \irdc.
\item Additional fragmentation may explain the presence of multiple peaks observed in the radial flux density profiles of several of the kinematically coherent leaves (i.e. those continuum sources that can be attributed to a single velocity component). In contrast however, leaves~\#7 and ~\#9, dark in the mid-infrared, centrally concentrated ($\rho\propto r^{-1.96}$ and $r^{-2.14}$, respectively), monolithic (with no discernible substructure at our PdBI resolution), and with estimated masses of the order of $\sim20-25$\,\solar, are good candidates for progenitors of intermediate-to-high mass stars. 
\end{enumerate}

Looking towards future investigations, higher-angular resolution dust continuum observations will assist in determining whether or not the structures identified in this work have fragmented further. Similarly, constraining the strength and orientation of the magnetic fields, as well as searching for infall motions through molecular line observations, will help to assess whether cores such as these deviate from virial equilibrium. 

\section*{Acknowledgements}
\addcontentsline{toc}{section}{Acknowledgements}
We would like to thank the anonymous referee for the constructive report which has helped to improve the paper. We would like to thank Michael Butler and Jouni Kainulainen for providing the mass surface density map used in this work. Based on observations carried out under project number V008 with the IRAM PdBI. IRAM is supported by INSU/CNRS (France), MPG (Germany) and IGN (Spain). JDH would like to thank Gary Fuller and Tom Hartquist for their constructive comments on an early version of this project, as well as Nate Bastian, Yanett Contreras, Luke Maud, Fumitaka Nakamura, and Andy Pon for helpful discussions. PC and JEP acknowledge support from European Research Council (ERC; project PALs 320620). IJ-S acknowledges the funding received from the STFC through an Ernest Rutherford Fellowship (proposal number ST/L004801/1). JCT acknowledges NASA grant ADAP10-0110. RJP acknowledges support from the Royal Astronomical Society in the form of a research fellowship.




\bibliographystyle{mnras}
\bibliography{References/references} 



\appendix
\section{Leaf descriptions}\label{App:kinematics}

In Section~\ref{Section:results_line} we highlighted the importance of demonstrating caution when using structure-finding algorithms on two-dimensional data such as continuum images. Projection effects can lead to spurious estimates of physical properties. Although the inclusion of kinematic information does not resolve all of these issues, it can help to remove some ambiguity. In this appendix, we expand on the discussion of \S~\ref{Section:results_line}, and discuss each identified continuum source in more detail.\\

\begin{figure*}
\begin{center}
\includegraphics[trim = 0mm 0mm 0mm 0mm, clip, width = 0.33\textwidth]{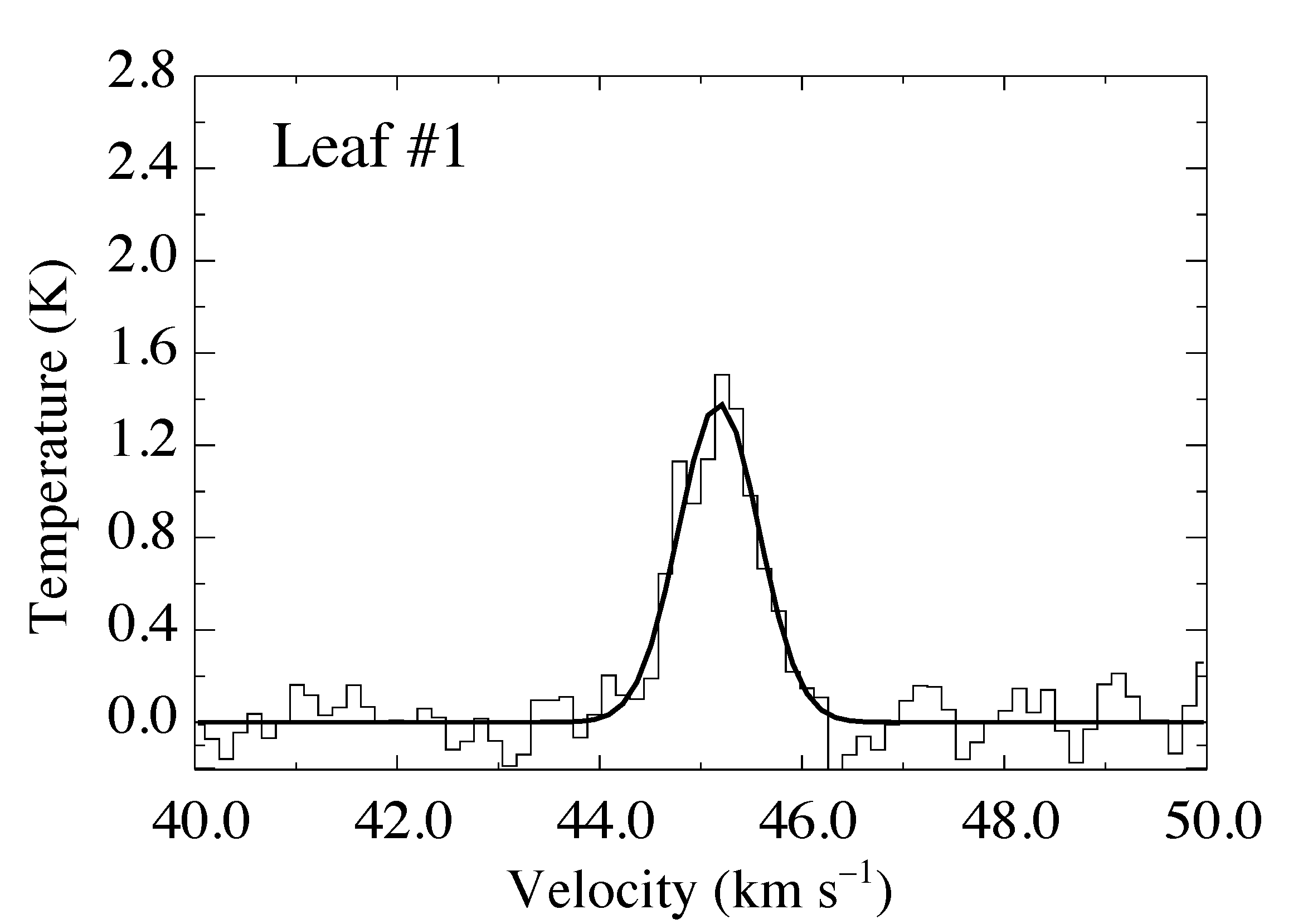}
\includegraphics[trim = 0mm 0mm 0mm 0mm, clip, width = 0.33\textwidth]{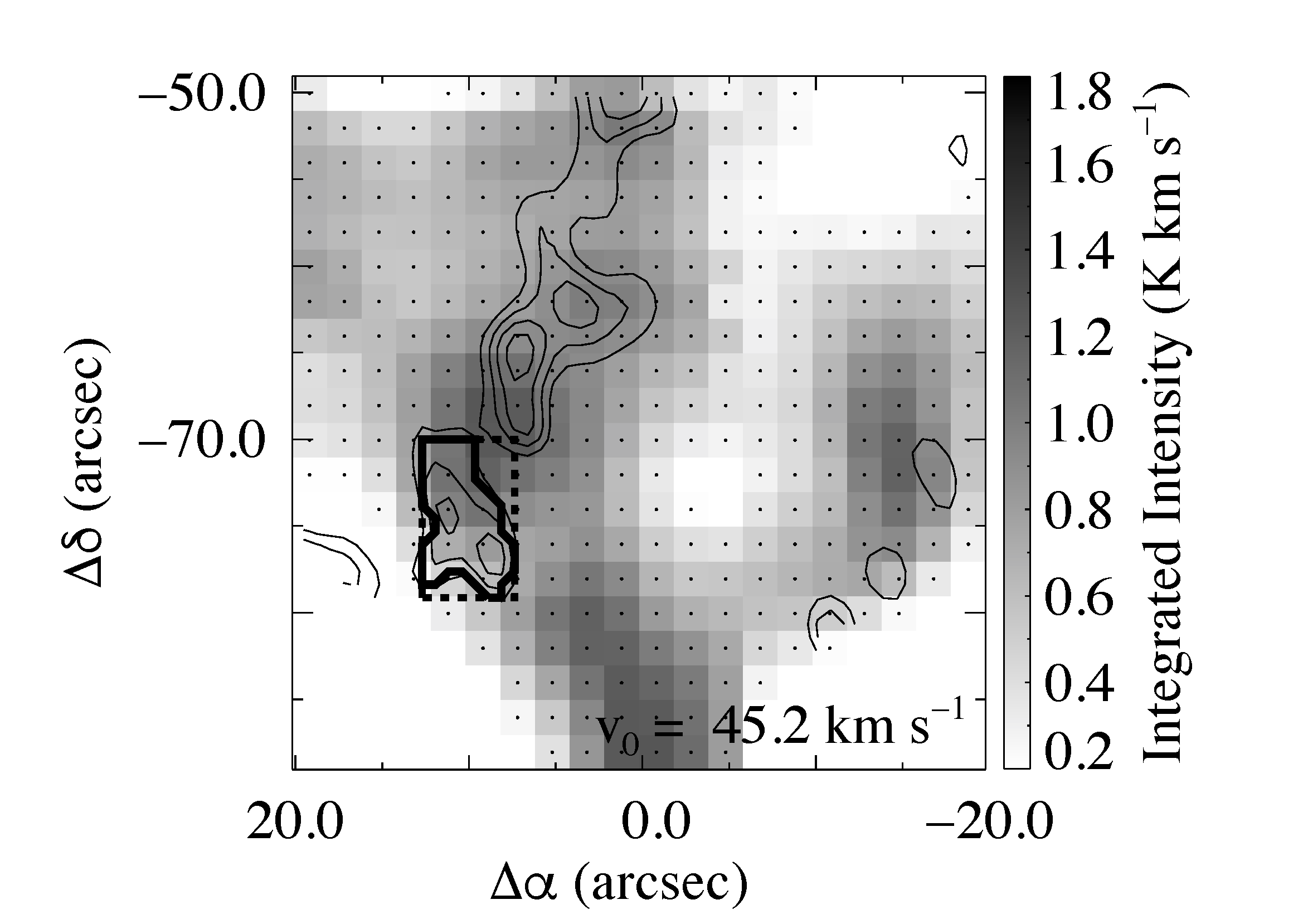}
\includegraphics[trim = 35mm 10mm 0mm 0mm, clip, width = 0.45\textwidth]{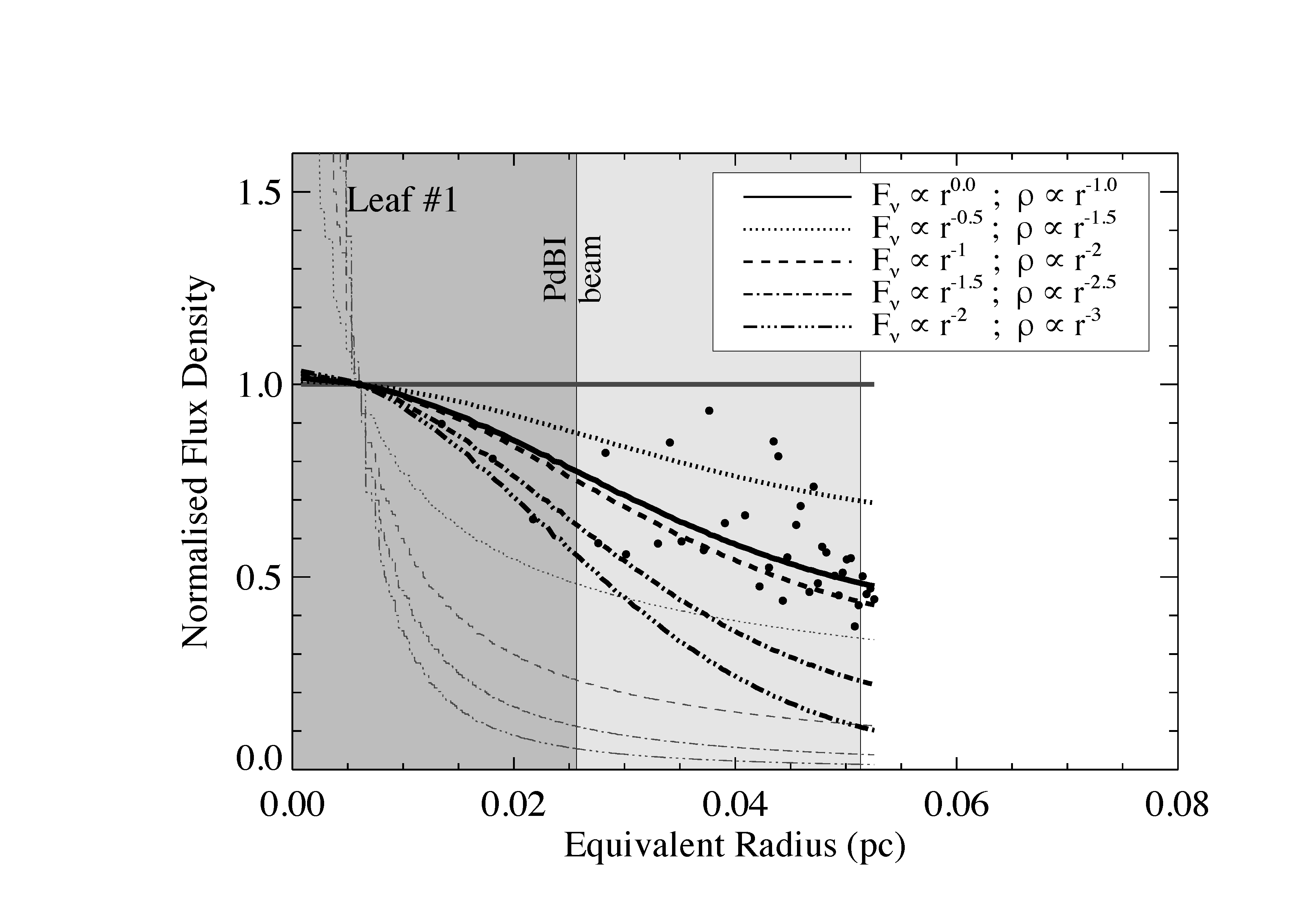}
\end{center}
\caption{The average spectrum, spatial distribution of integrated \ntwoh \ (1-0) emission, and the radial flux density profile of leaf~\#1. The top-left hand panel is a spatially-averaged spectrum, showing the isolated ($F_{1},\,F~=~0,\,1\rightarrow1,\,2$) hyperfine component of \ntwoh \ (1-0). The spectrum has been extracted from the black dashed box seen in the top-right hand panel. The line is singly peaked. The solid black Gaussian profiles represents the best-fitting model solution to the data. The right-hand panel displays the spatial distribution of the emission feature. The light black contours are equivalent to those in Fig.~\ref{Figure:msd_map} and the heavy black contour corresponds to the boundary of leaf~\#1. The bottom panel shows the flux density as a function of equivalent radius. Black lines are model radial flux density profiles before (light) and after (heavy) beam correction (see \S~\ref{Section:analysis_thermturb}). Each line corresponds to a density profile of the form $F_{\rm \nu}\propto r^{-\kappa_{\rho}}$. The corresponding values of $\kappa_{\rho}$ are included in the legend. The dark grey block indicates the extent of the geometric mean radius of the synthesized PdBI beam (1.83\,arcsec or 0.025\,pc at a distance of 2900\,pc). The light grey block is twice this value. The thick black line indicates the closest-matching model solution to the observations.}
\label{Figure:leaf1}
\end{figure*}

\noindent\textbf{Leaf~\#1:} situated at $\{\Delta\alpha,\,\Delta\delta\}=\{8.9\,{\rm arcsec},\,-76.9\,{\rm arcsec}\}$, leaf~\#1 is the southernmost continuum peak identified within the mapped region. It has an aspect ratio, $\AR=1.87$, and an equivalent radius, $R_{\rm eq}=3.74\,{\rm arcsec}$ (corresponding to an estimated physical radius of $\sim0.05$\,pc at an assumed distance of 2900\,pc). There is some suggestion that the leaf may have a secondary peak (see Figure~\ref{Figure:cont_dendro}). The radial flux density profile also appears to suggest this. However, this cannot be confirmed at the resolution of our PdBI observations. The spatially-averaged spectrum of \ntwoh, extracted from within the boundary of the leaf, is singly-peaked (see left-hand panel of Fig.~\ref{Figure:leaf1}). The centroid velocity and FWHM line-width of this spectral component are $v_{\rm 0}=45.18\,{\rm km\,s^{-1}}\pm0.02\,{\rm km\,s^{-1}}$ and $\Delta v~=~0.94\,{\rm km\,s^{-1}}\pm0.04\,{\rm km\,s^{-1}}$, respectively. \\

\begin{figure*}
\begin{center}
\includegraphics[trim = 0mm 0mm 0mm 0mm, clip, width = 0.33\textwidth]{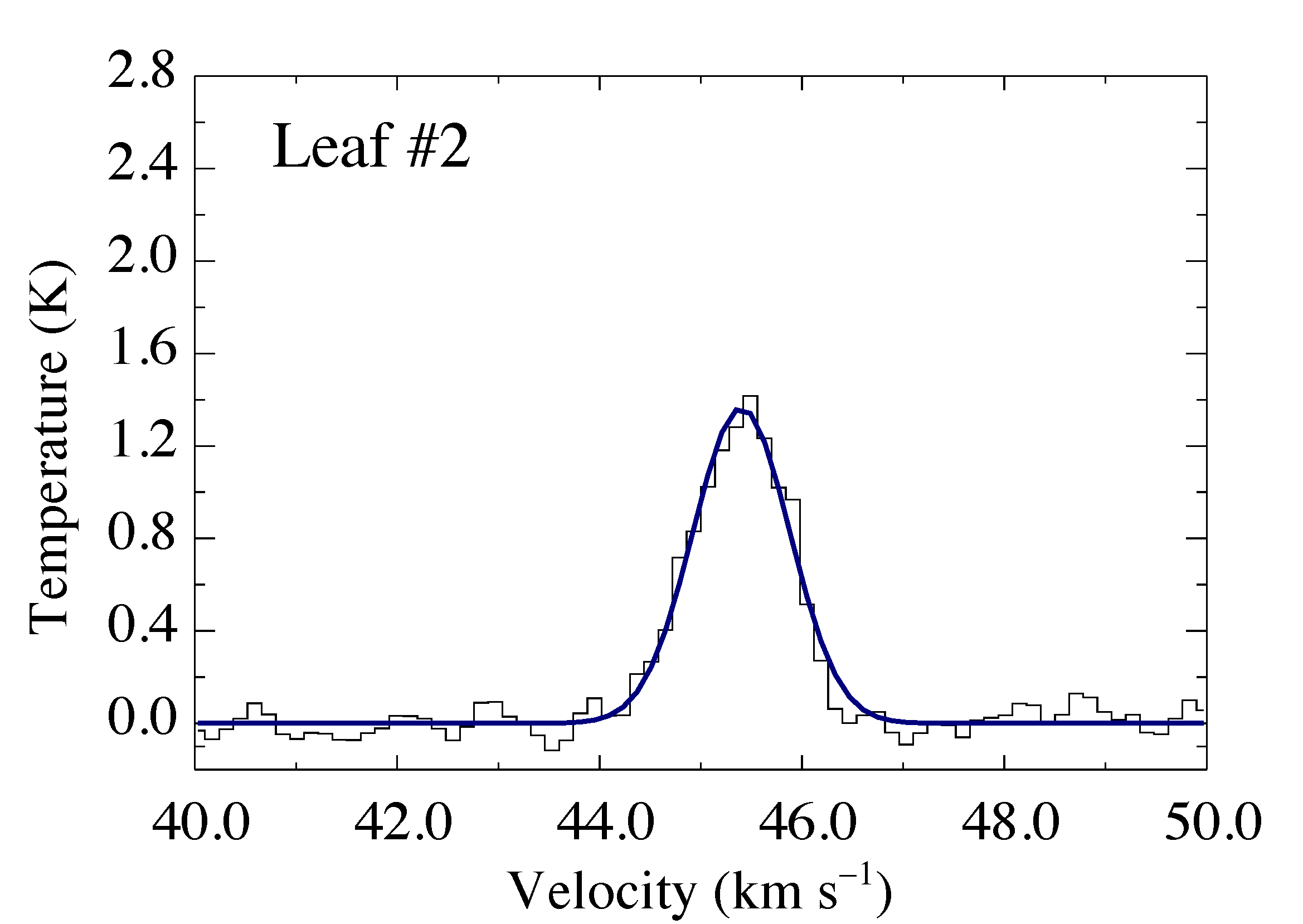}
\includegraphics[trim = 0mm 0mm 0mm 0mm, clip, width = 0.33\textwidth]{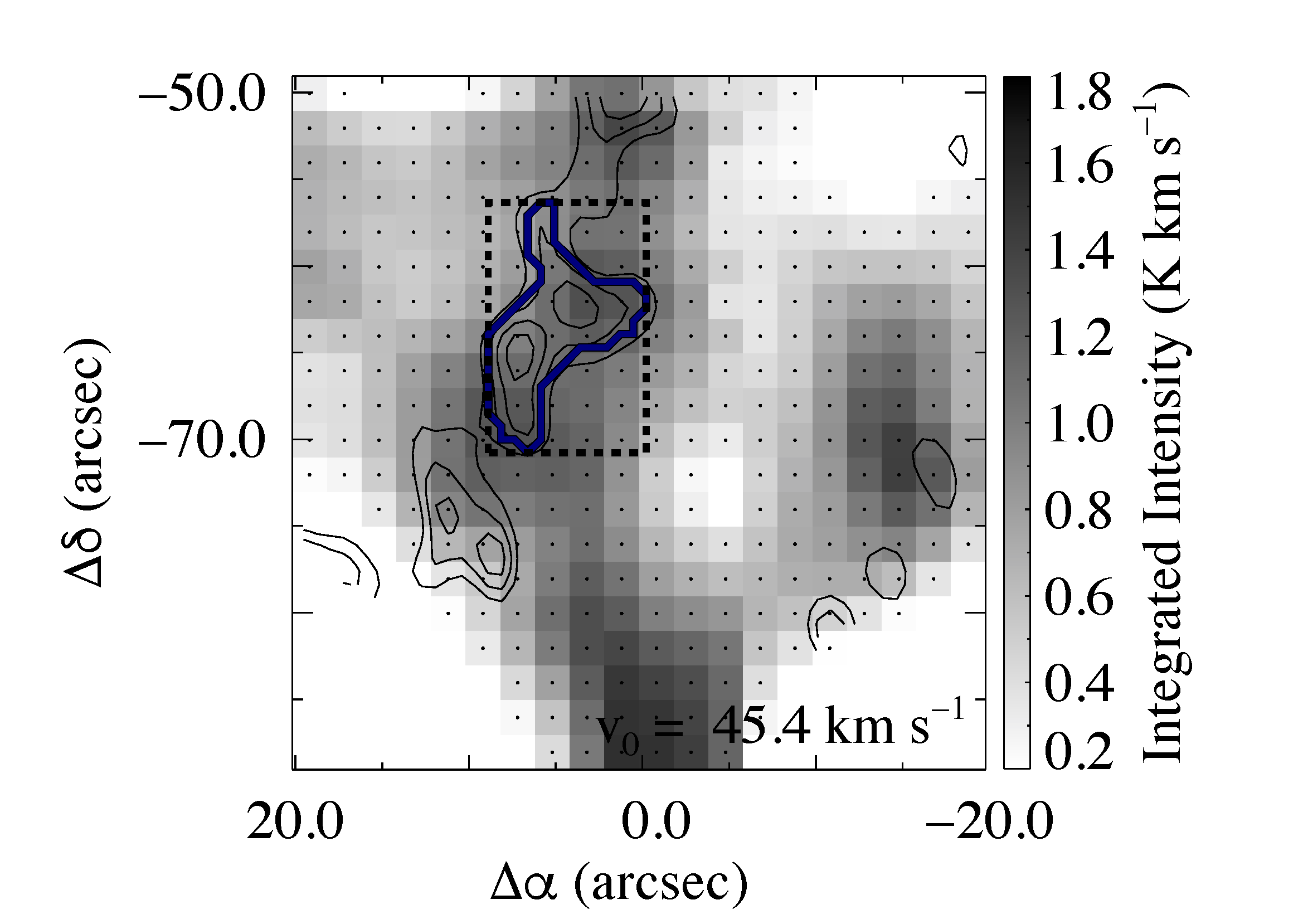}
\includegraphics[trim = 35mm 10mm 0mm 0mm, clip, width = 0.45\textwidth]{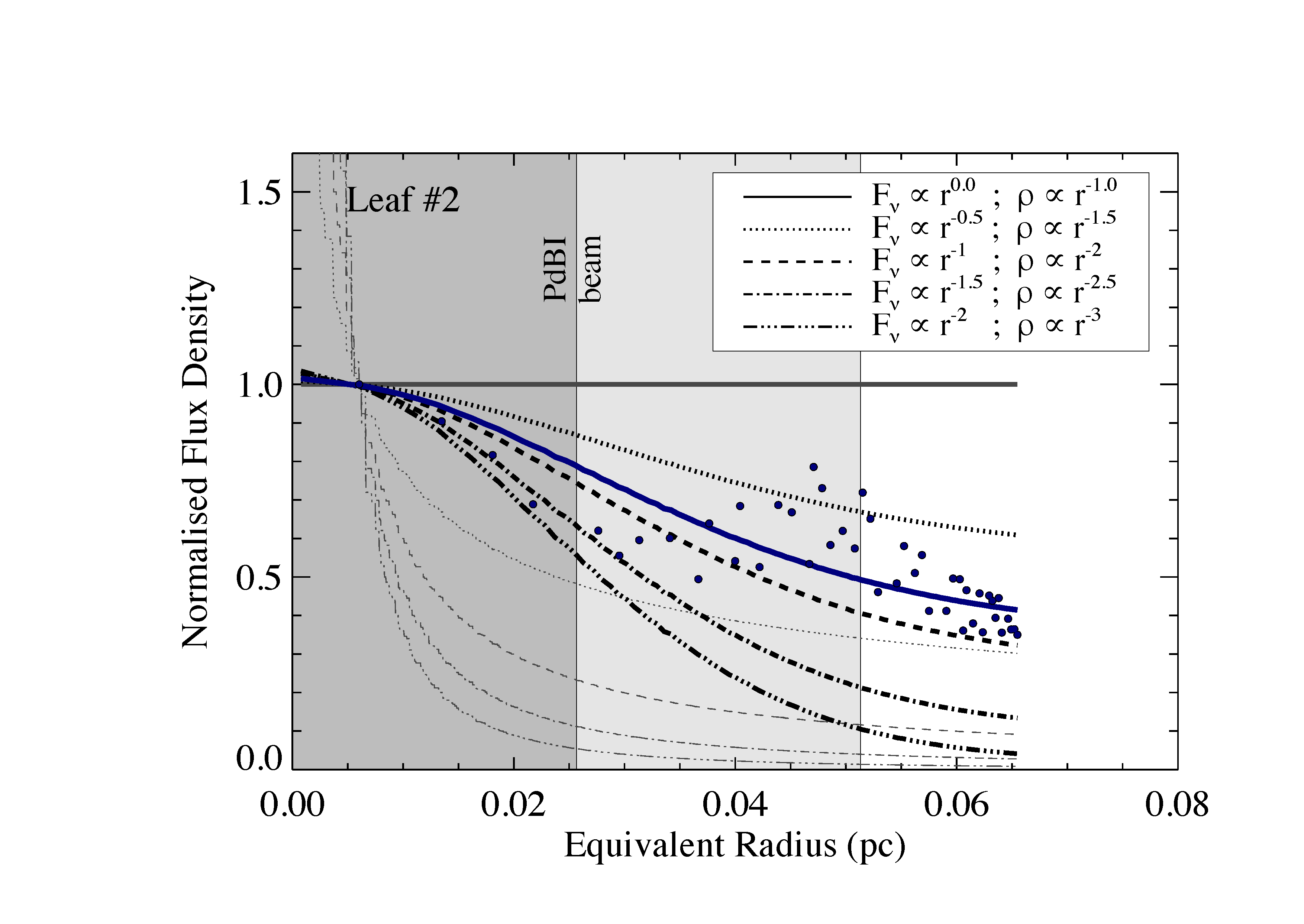}
\end{center}
\caption{The average spectrum, spatial distribution of integrated \ntwoh \ (1-0) emission, and the radial flux density profile of leaf~\#2.  }
\label{Figure:leaf2}
\end{figure*}

\noindent\textbf{Leaf~\#2:} is situated at $\{\Delta\alpha,\,\Delta\delta\}=\{7.4\,{\rm arcsec},\,-64.7\,{\rm arcsec}\}$. There is a suggestion of a secondary peak located to the north-west. The boundary of the leaf is highly irregular. The secondary peak is spatially coincident with a 24\,\micron \ source (see Figs~\ref{Figure:msd_map} and \ref{Figure:cont_dendro}). These two factors suggest that this leaf may consist of two (or more) structures (see also Fig.~\ref{Figure:leaf2}), with one (or more) of those exhibiting signatures of embedded star formation. A closer look at the \ntwoh \ emission reveals that the spectrum is singly-peaked (see left-hand panel of Fig.~\ref{Figure:leaf2}), and that the area covered by this emission feature is much larger than the leaf itself (and very similar to that for leaf~\#1). The centroid velocity and FWHM line-width of this spectral component are $v_{\rm 0}=45.40\,{\rm km\,s^{-1}}\pm0.01\,{\rm km\,s^{-1}}$ and $\Delta v~=~1.13\,{\rm km\,s^{-1}}\pm0.02\,{\rm km\,s^{-1}}$, respectively. \\

\begin{figure*}
\begin{center}
\includegraphics[trim = 0mm 0mm 0mm 0mm, clip, width = 0.33\textwidth]{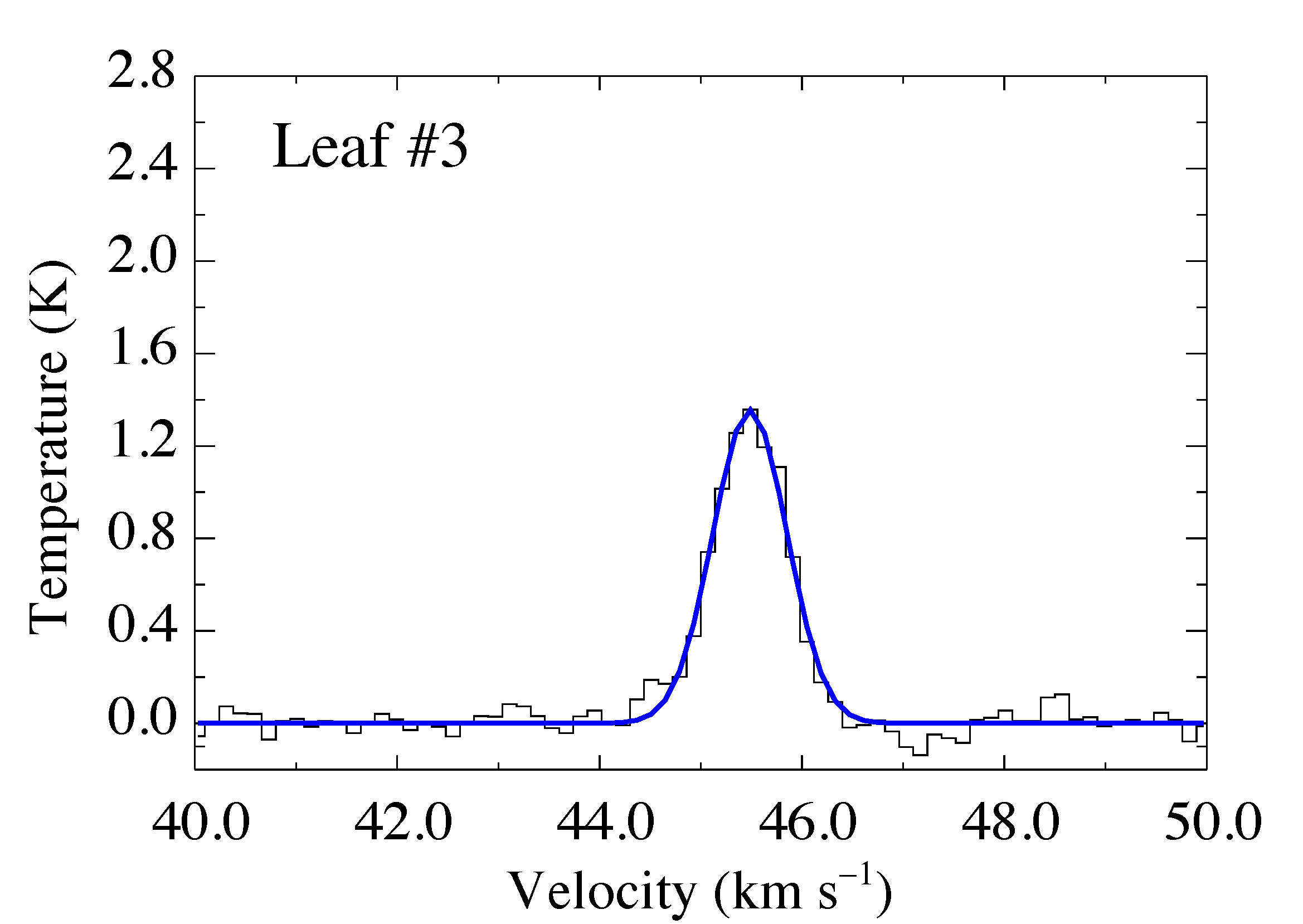}
\includegraphics[trim = 0mm 0mm 0mm 0mm, clip, width = 0.33\textwidth]{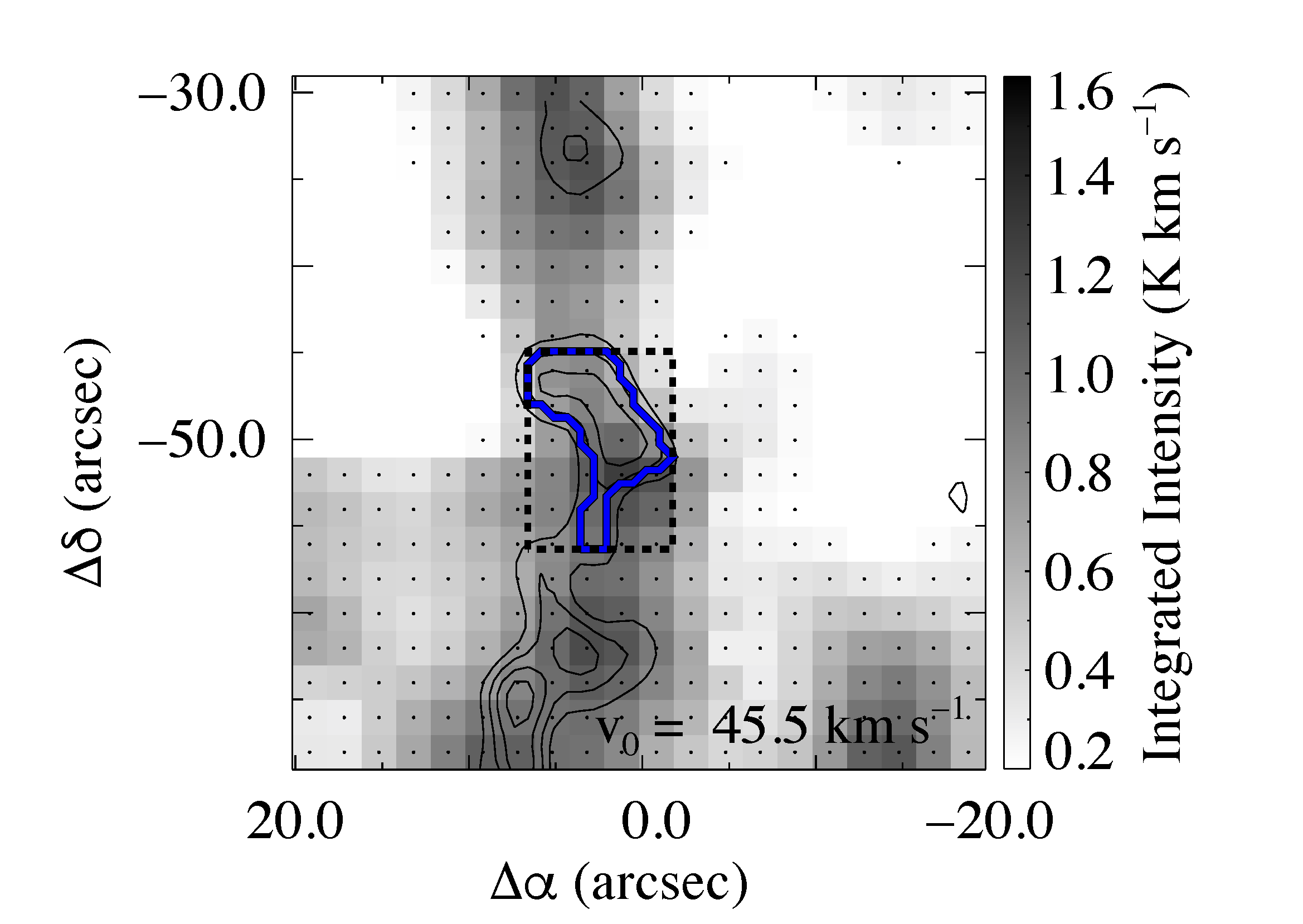}
\includegraphics[trim = 35mm 10mm 0mm 0mm, clip, width = 0.45\textwidth]{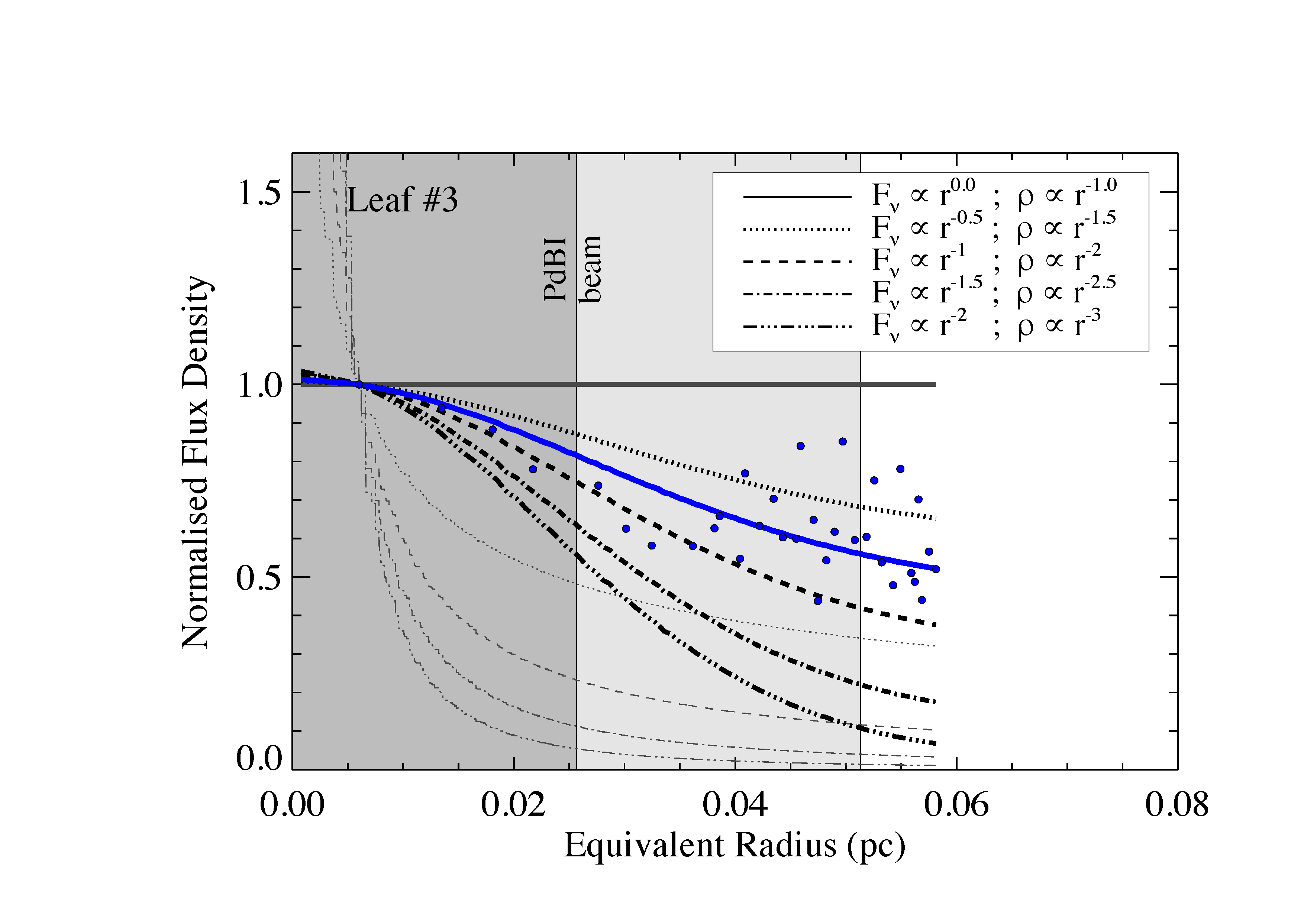}
\end{center}
\caption{The average spectrum, spatial distribution of integrated \ntwoh \ (1-0) emission, and the radial flux density profile of leaf~\#3.   }
\label{Figure:leaf3}
\end{figure*}

\noindent\textbf{Leaf~\#3:} is situated at $\{\Delta\alpha,\,\Delta\delta\}=\{1.3\,{\rm arcsec},\,-50.3\,{\rm arcsec}\}$. Similar to leaves~\#1 and \#2, leaf~\#3 has an aspect ratio $\AR=1.81$. Its major axis is aligned from north-east to south-west. There is a 24\,\micron \ emission source situated to the south-west of the leaf (Fig.~\ref{Figure:cont_dendro}). This region is also bright in 70\,\micron \ \emph{Herschel} images, and is identified as a ``low-mass dense core'' by \citet[core \#12; their table~1]{nguyen_2011}. The mass estimated from the \emph{Herschel} images is $\sim12\pm7$\,\solar, which is consistent with our 3.2\,mm continuum-derived masses, $M_{\rm c}~\sim~15$\,\solar \ and $M^{\rm b}_{\rm c}~\sim~6$\,\solar \ (within the factor of $\sim$2 uncertainty, see \S~\ref{Section:analysis}). The \ntwoh \ (1-0) emission associated with leaf~\#3 is best described with a single spectral component (see Figure~\ref{Figure:leaf3}). The centroid velocity and FWHM line-width of this spectral component are $v_{\rm 0}=45.49\,{\rm km\,s^{-1}}\pm0.01\,{\rm km\,s^{-1}}$ and $\Delta v~=~0.86\,{\rm km\,s^{-1}}\pm0.02\,{\rm km\,s^{-1}}$, respectively. As can be seen from Figs~\ref{Figure:leaf1}-\ref{Figure:leaf3} there is a velocity gradient, with the velocity increasing from the south (leaf~\#1) to the north (leaf~\#3). \\

\begin{figure*}
\begin{center}
\includegraphics[trim = 0mm 0mm 0mm 0mm, clip, width = 0.33\textwidth]{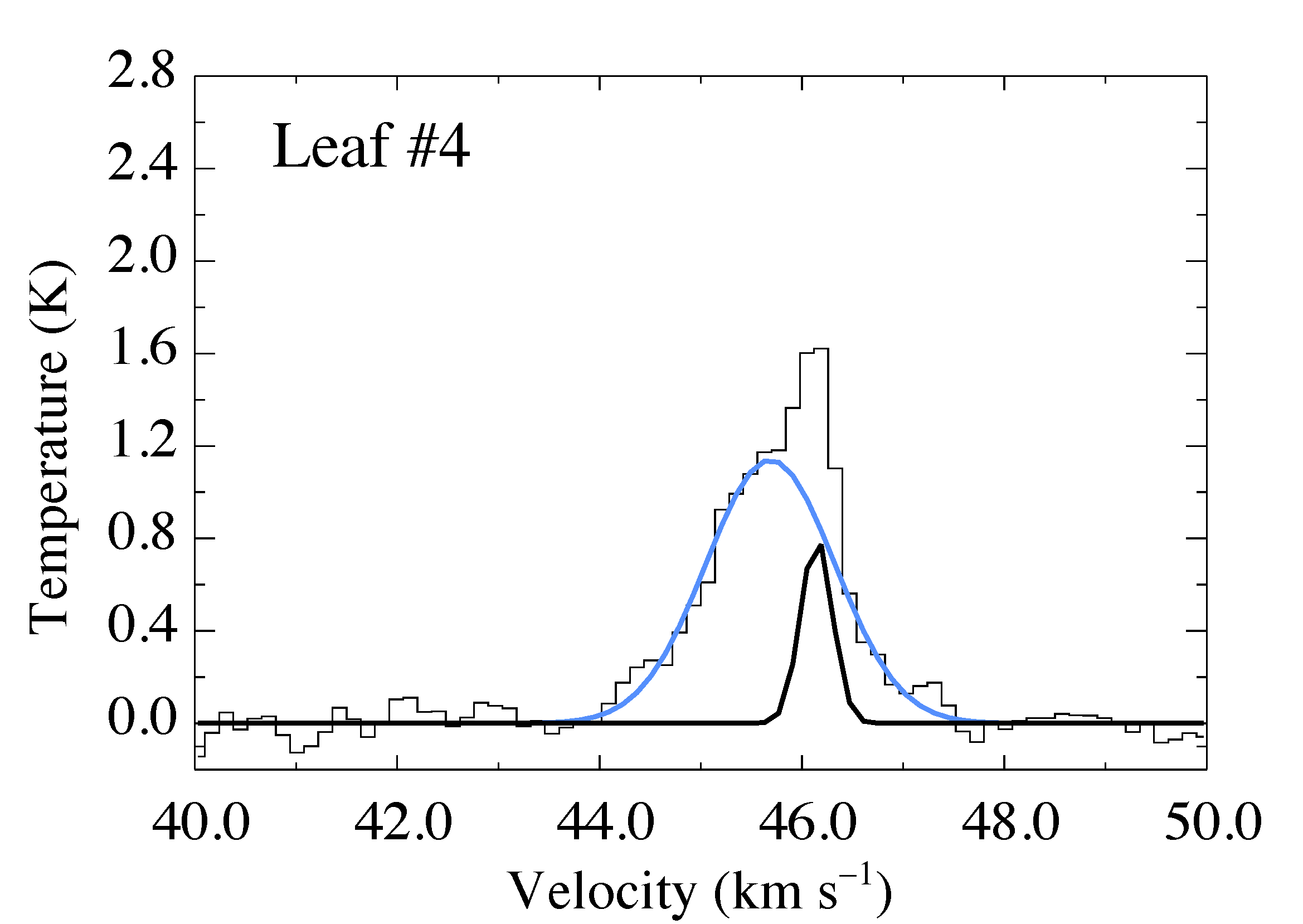}
\includegraphics[trim = 0mm 0mm 0mm 0mm, clip, width = 0.33\textwidth]{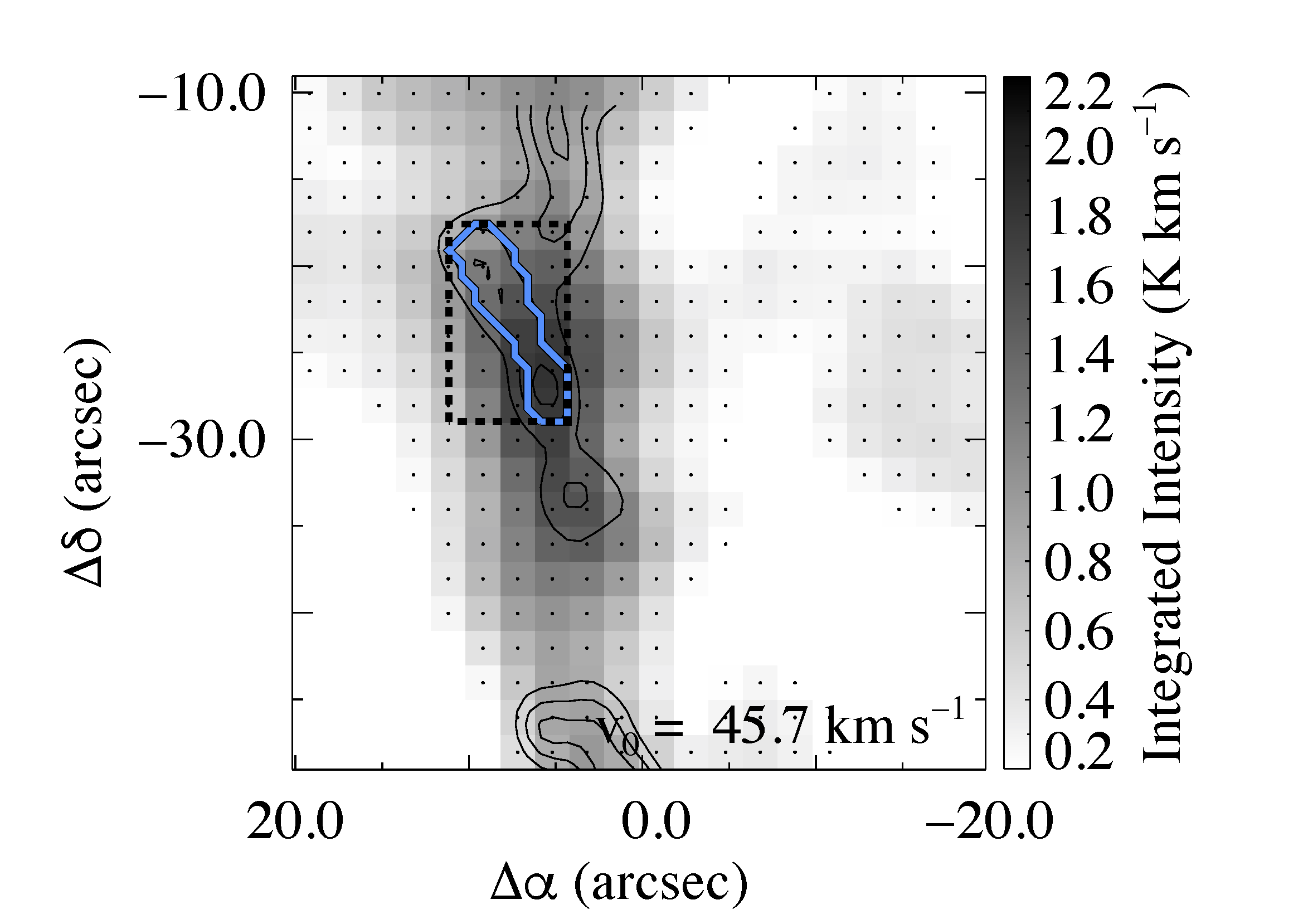}
\includegraphics[trim = 0mm 0mm 0mm 0mm, clip, width = 0.33\textwidth]{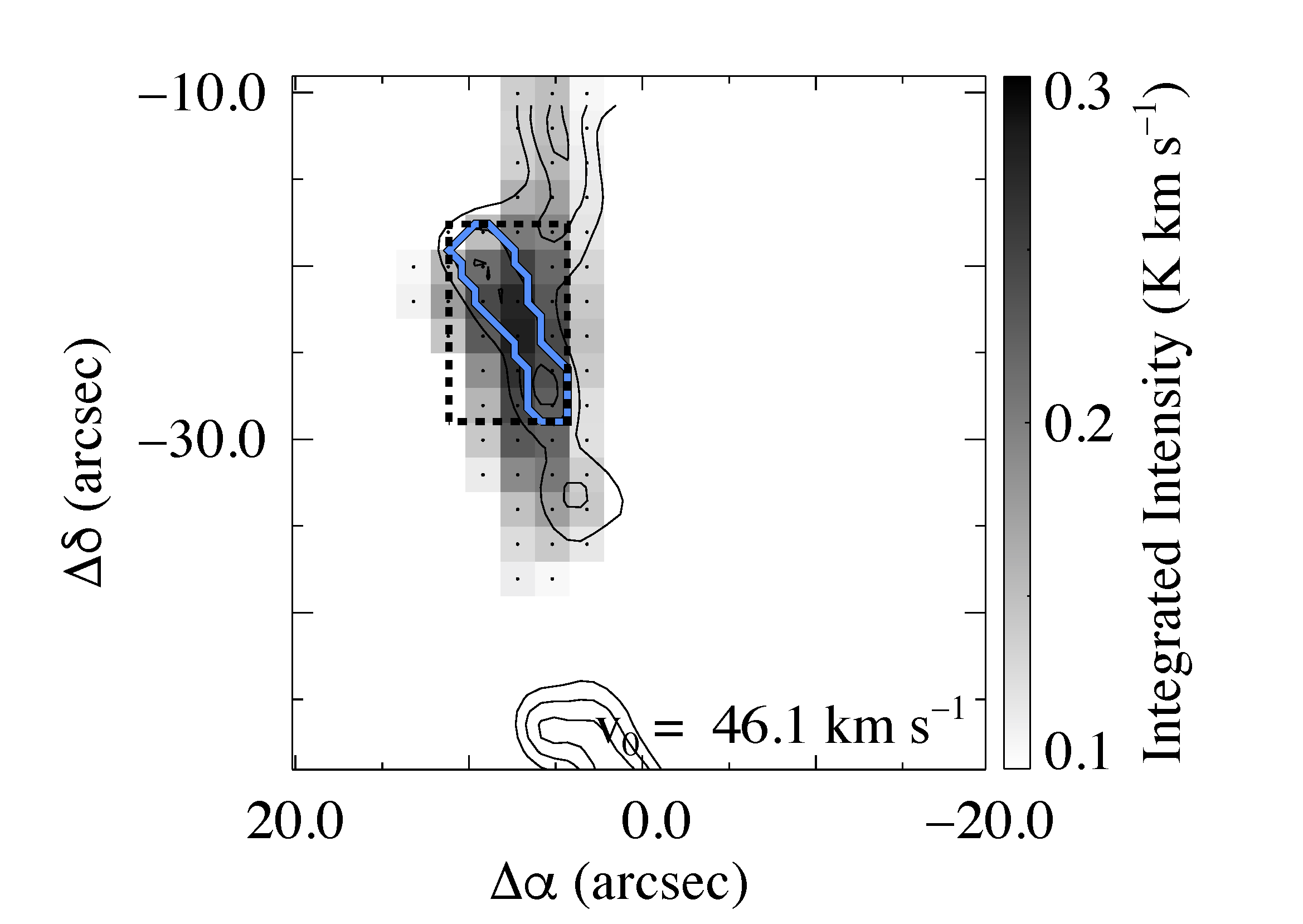}
\includegraphics[trim = 35mm 10mm 0mm 0mm, clip, width = 0.45\textwidth]{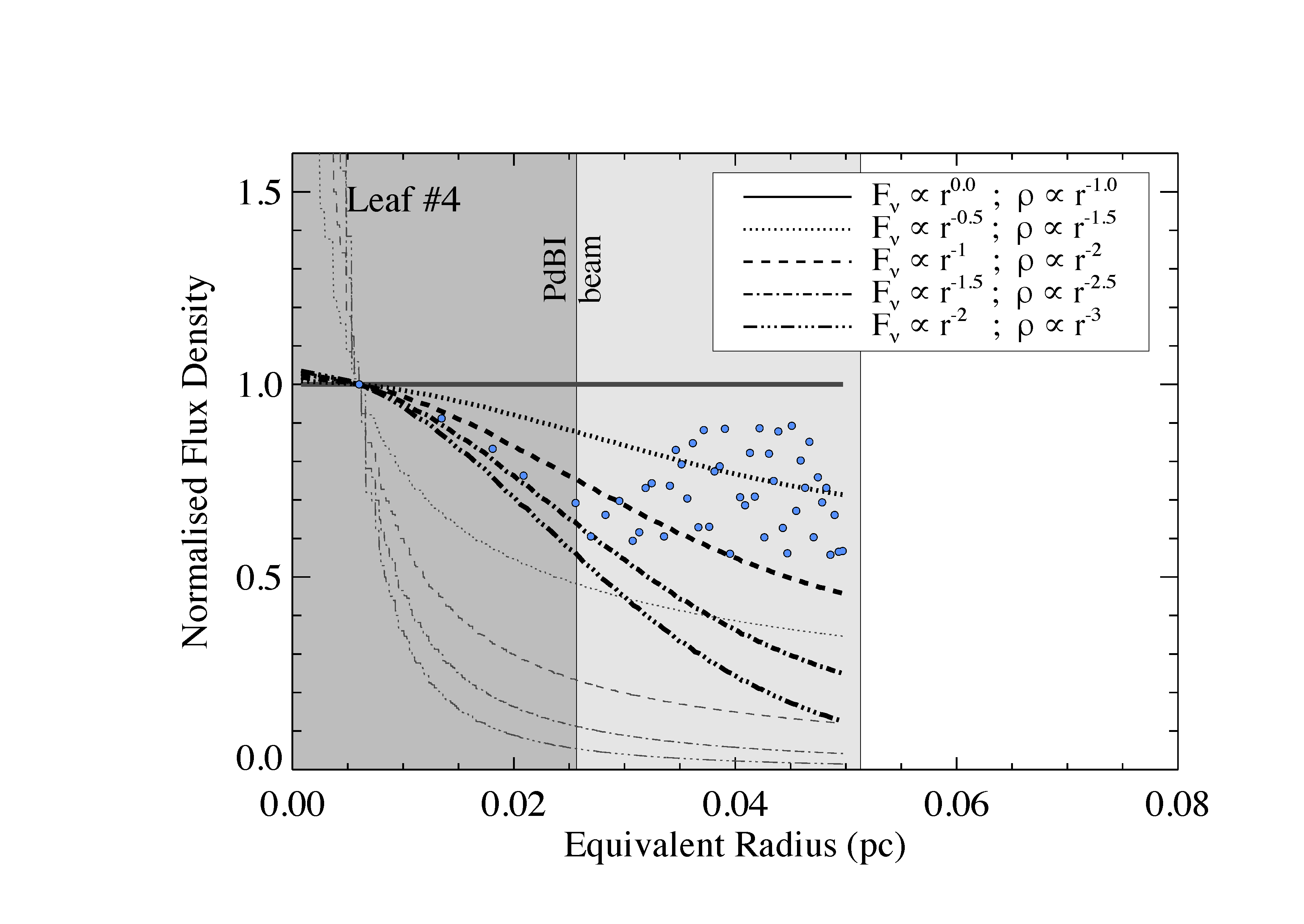}
\end{center}
\caption{The average spectrum, spatial distribution of integrated \ntwoh \ (1-0) emission, and the radial flux density profile of leaf~\#4. Note the difference in scaling, which is selected to enhance the features of both components.   }
\label{Figure:leaf4}
\end{figure*}

\noindent\textbf{Leaf~\#4:} is situated at $\{\Delta\alpha,\,\Delta\delta\}=\{5.9\,{\rm arcsec},\,-26.7\,{\rm arcsec}\}$. Unlike the three leaves described above, leaf~\#4 is elongated, with an aspect ratio, $\AR=4.11$. There is a second peak in the continuum emission to the south-west of the filamentary leaf that has not been identified during the dendrogram analysis. Higher-angular resolution observations would be needed to ascertain whether or not this represents a separate structure, or a continuation of leaf~\#4. The flux density does not decrease uniformly as a function of equivalent radius (Fig.~\ref{Figure:leaf4}), which is consistent with the filamentary nature of this continuum source. 

The spatially-averaged \ntwoh \ (1-0) spectrum taken from the boundary encompassing the leaf is shown in the left-hand panel of Fig.~\ref{Figure:leaf4}. There is a slight asymmetry in the line-profile implying the presence of two velocity components. We fitted the spectrum using both one- and two-component models, finding that the latter was more successful in reproducing the observed profile. As in \citet{henshaw_2016}, we base our judgement on: (i) the signal-to-noise level of each component (both of which are $>3$); (ii) the separation in velocity between the two observed components (which is greater than 0.5 times the FWHM of the narrowest component), which enables one to determine if the two components are distinguishable; (iii) the Akaike information criterion (\citealp{akaike_1974}), which provides a statistical method of selecting the best model from a number of choices. The centroid velocities of the two components are $v_{\rm 0,1}=45.69\,{\rm km\,s^{-1}}\pm0.03\,{\rm km\,s^{-1}}$ and $v_{\rm 0,2}=46.14\,{\rm km\,s^{-1}}\pm0.01\,{\rm km\,s^{-1}}$. The FWHM line-widths of the two components are $\Delta v_{1}~=~1.50\,{\rm km\,s^{-1}}\pm0.05\,{\rm km\,s^{-1}}$ and $\Delta v_{2}~=~0.37\,{\rm km\,s^{-1}}\pm0.04\,{\rm km\,s^{-1}}$. As can be seen from the centre and right-hand panels of Fig.~\ref{Figure:leaf4} the low(er)-velocity component dominates in terms of integrated emission (note the difference in grey-scale between the two plots). We therefore speculate that majority of the mass attributed to leaf~\#4 is associated with the low-velocity component. \\

\noindent\textbf{Leaf~\#5:} is situated at $\{\Delta\alpha,\,\Delta\delta\}=\{2.1\,{\rm arcsec},\,-0.9\,{\rm arcsec}\}$. As with leaf~\#4, it is elongated, exhibiting the greatest aspect ratio of the identified structures, $\AR=4.73$. The figures describing the \ntwoh \ emission associated with leaf~\#5 can be found in the main text (bottom panels Fig.~\ref{Figure:spec_mom}). For reasons discussed in \S~\ref{Section:results_line}, leaf~\#5 is rejected from the analysis in \S~\ref{Section:analysis} (other than the mass estimation). However, the centroid velocities and FWHM line-widths associated with the two identified components are $v_{\rm 0,1}=45.63\,{\rm km\,s^{-1}}\pm0.03\,{\rm km\,s^{-1}}$ and $v_{\rm 0,2}=46.59\,{\rm km\,s^{-1}}\pm0.08\,{\rm km\,s^{-1}}$ and $\Delta v_{1}~=~1.10\,{\rm km\,s^{-1}}\pm0.07\,{\rm km\,s^{-1}}$ and $\Delta v_{2}~=~0.72\,{\rm km\,s^{-1}}\pm0.15\,{\rm km\,s^{-1}}$, respectively. \\

\begin{figure*}
\begin{center}
\includegraphics[trim = 0mm 0mm 0mm 0mm, clip, width = 0.33\textwidth]{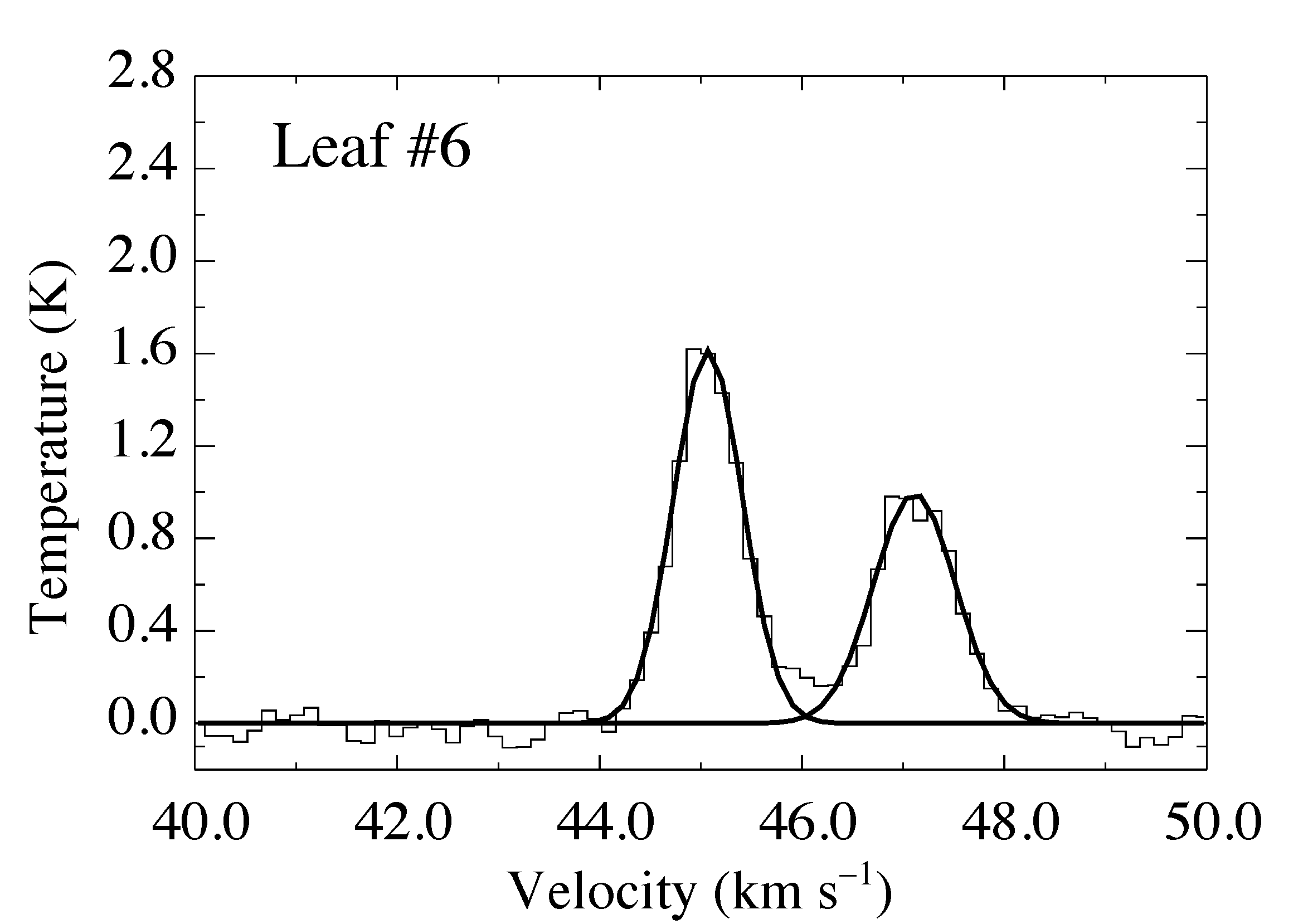}
\includegraphics[trim = 0mm 0mm 0mm 0mm, clip, width = 0.33\textwidth]{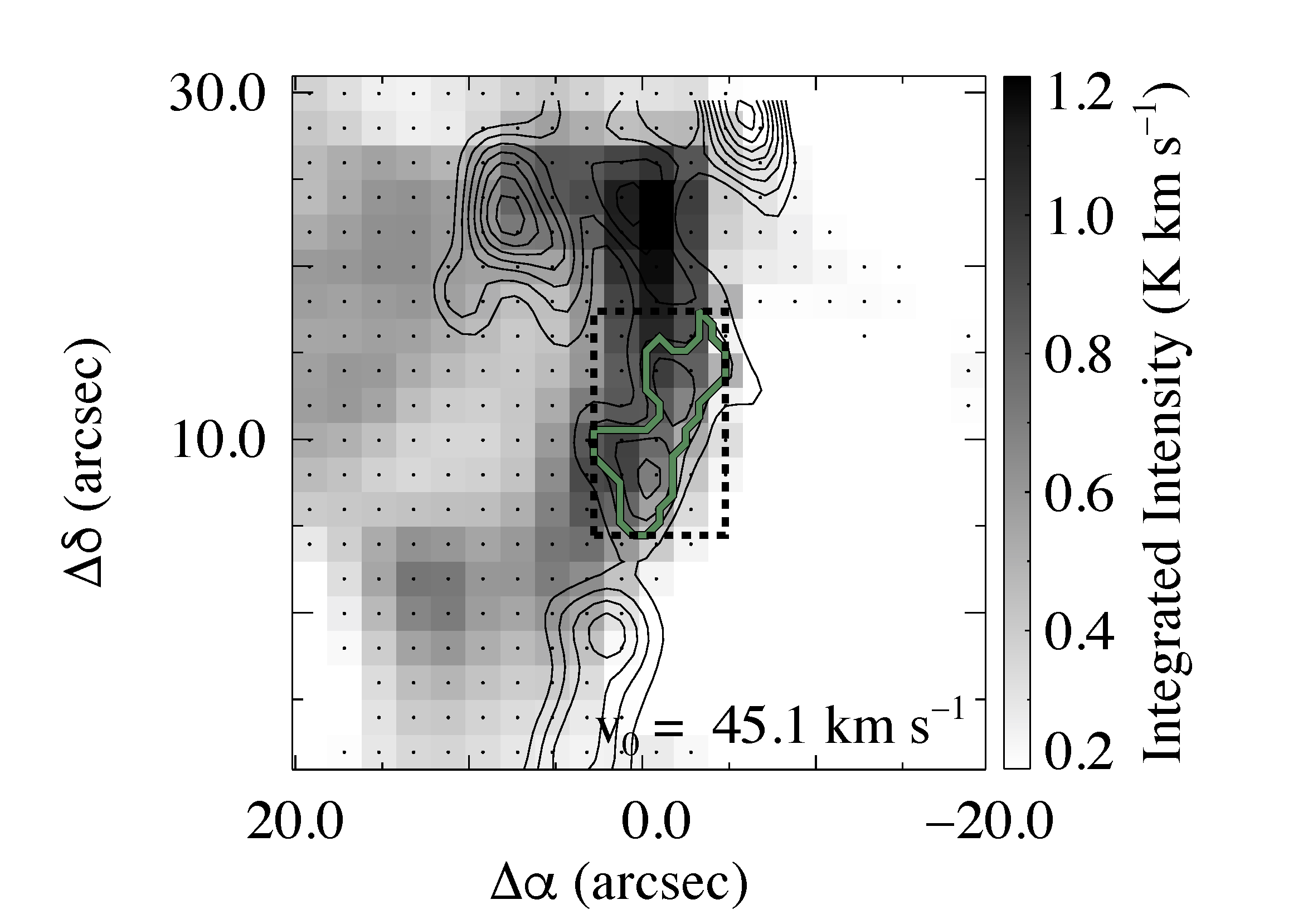}
\includegraphics[trim = 0mm 0mm 0mm 0mm, clip, width = 0.33\textwidth]{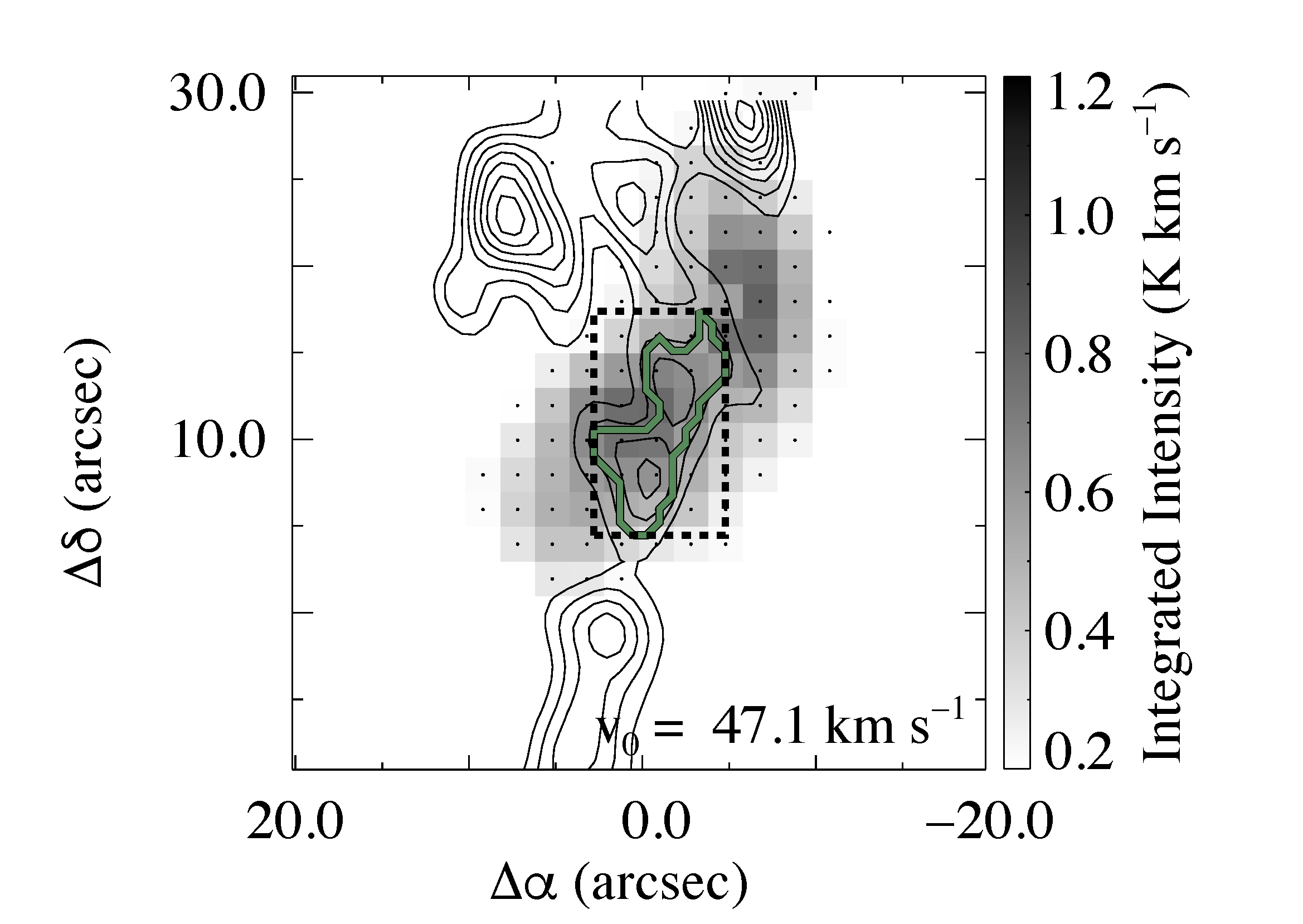}
\includegraphics[trim = 35mm 10mm 0mm 0mm, clip, width = 0.45\textwidth]{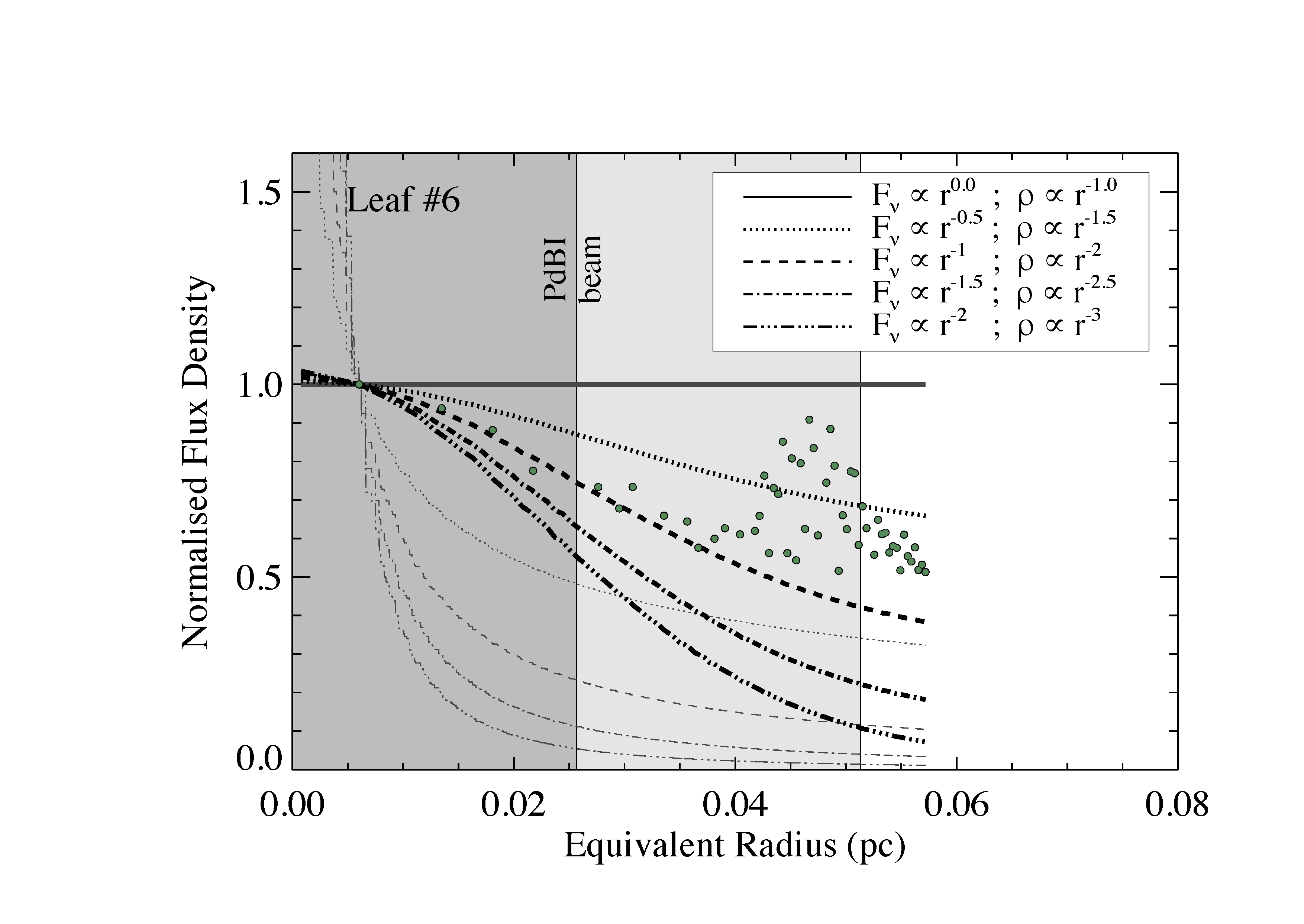}
\end{center}
\caption{The average spectrum, spatial distribution of integrated \ntwoh \ (1-0) emission, and the radial flux density profile of leaf~\#6. Black Gaussian profiles signify that the continuum flux cannot be unambiguously accredited to a single structure. }
\label{Figure:leaf6}
\end{figure*}

\noindent\textbf{Leaf~\#6:} is situated at $\{\Delta\alpha,\,\Delta\delta\}=\{-0.2\,{\rm arcsec},\,7.5\,{\rm arcsec}\}$. The leaf boundary has an hourglass-shaped profile, implying the presence of unresolved fragments (see also Fig.~\ref{Figure:leaf6}). The southern half of the hourglass profile is spatially coincident with extended 4.5, 8, and 24\,\micron \ emission (note this source appears as a ``hole'' in the mid-infrared-derived mass surface density map of \citealp{kainulainen_2013}; Fig.~\ref{Figure:msd_map}). This leaf was also identified in the \emph{Herschel} 70\,\micron \ images and described as a ``protostellar massive dense core'' by \citet[core \#18; their table~1]{nguyen_2011}. The mass estimated from the \emph{Herschel} images is $\sim20\pm9$\,\solar, which is consistent with our 3.2\,mm continuum-derived masses, $M_{\rm c}~\sim~17$\,\solar \ and $M^{\rm b}_{\rm c}~\sim~4$\,\solar \ (see \S~\ref{Section:analysis}). A cursory inspection of the \emph{Herschel} images show that there may be a secondary 70\,\micron \ source within the leaf boundary (i.e. the northern portion of the hourglass). Only when we set ${\rm min\_npix}=15$ (i.e. below the resolution limit of our observations) does the dendrogram algorithm identify the two individual structures. Higher angular resolution observations are required to confirm whether or not this is the case. 

Fig.~\ref{Figure:leaf6} highlights the distribution of \ntwoh \ (1-0) emission associated with this feature. There are two distinct velocity components spatially coincident with leaf~\#6. The measured centroid velocities and FWHM line-widths are $v_{\rm 0,1}=45.07\,{\rm km\,s^{-1}}\pm0.01\,{\rm km\,s^{-1}}$ and $v_{\rm 0,2}=47.11\,{\rm km\,s^{-1}}\pm0.01\,{\rm km\,s^{-1}}$ and $\Delta v_{1}~=~0.80\,{\rm km\,s^{-1}}\pm0.02\,{\rm km\,s^{-1}}$ and $\Delta v_{2}~=~0.95\,{\rm km\,s^{-1}}\pm0.03\,{\rm km\,s^{-1}}$, respectively. Since the continuum flux cannot be unambiguously accredited to a single structure, leaf~\#6 is rejected from the analysis in \S~\ref{Section:analysis} (other than the mass estimation). \\

\noindent\textbf{Leaf~\#7:} is situated at $\{\Delta\alpha,\,\Delta\delta\}=\{7.4\,{\rm arcsec},\,22.7\,{\rm arcsec}\}$, close to the peak in extinction (H6 is located at $\{\Delta\alpha,\,\Delta\delta\}=\{3\farc0,\,21\farc1\}$; \citealp{butler_2012}). Leaf~\#7 has the smallest aspect ratio of the identified leaves, $\AR=1.46$ and an equivalent radius of 3.81\,arcsec \ (corresponding to a physical radius of $\sim0.05$\,pc at a distance of 2900\,pc). Although the continuum emission extends both to the south-west and south-east, leaf~\#7 appears to be monolithic (at the spatial resolution of our PdBI observations). It is dark at 8 and 24\,\micron, and was identified and classified as a ``IR-quiet massive dense core'' by \citet[core \#6; their table~1]{nguyen_2011}. The mass estimated from \emph{Herschel} observations is $\sim20\pm12$\,\solar, which is consistent with our continuum-derived masses, $M_{\rm c}~\sim~22$\,\solar \ and $M^{\rm b}_{\rm c}~\sim~9$\,\solar \ (within the factor of 2 uncertainty; see \S~\ref{Section:analysis}). 

\begin{figure*}
\begin{center}
\includegraphics[trim = 0mm 0mm 0mm 0mm, clip, width = 0.33\textwidth]{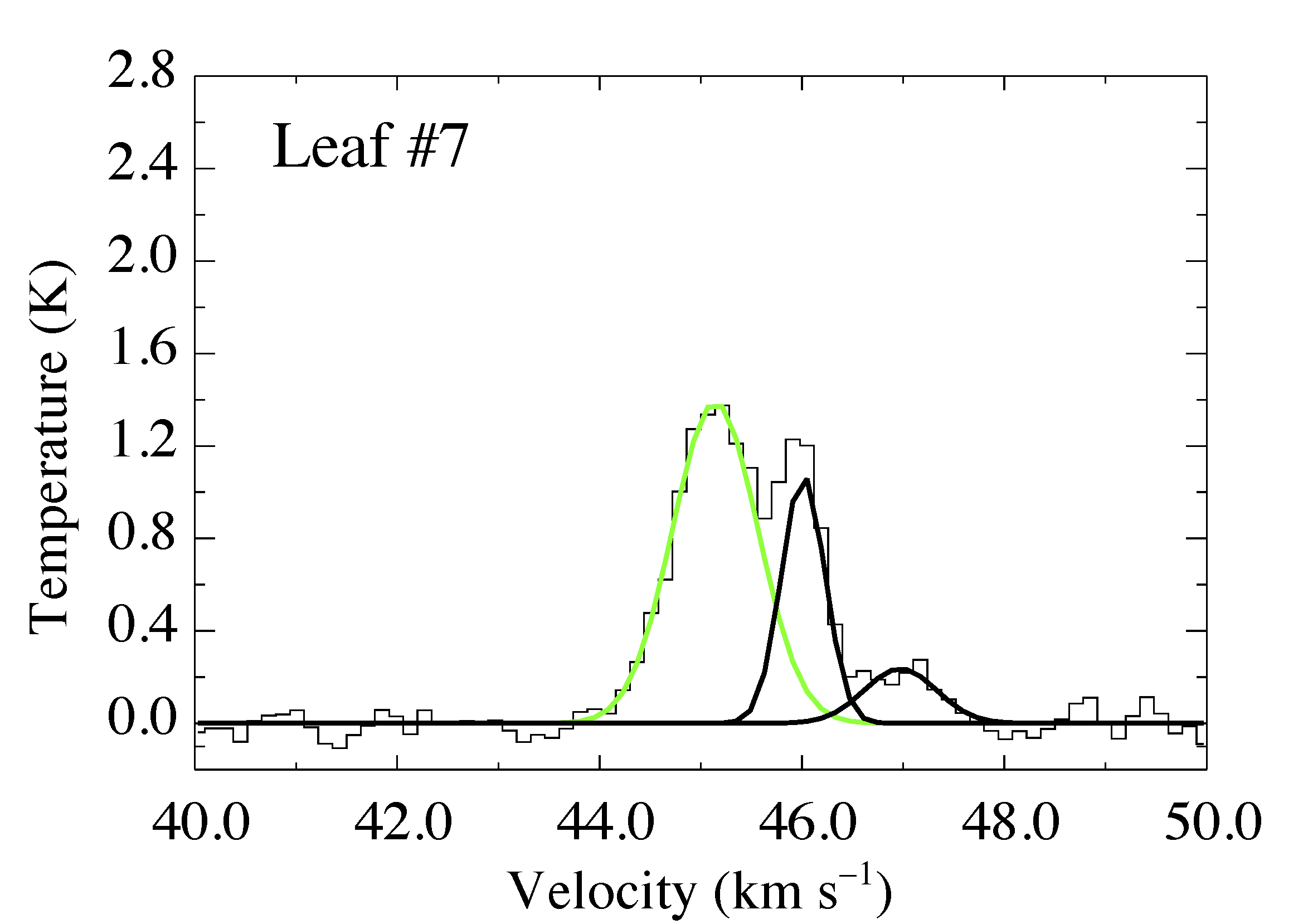}
\includegraphics[trim = 0mm 0mm 0mm 0mm, clip, width = 0.33\textwidth]{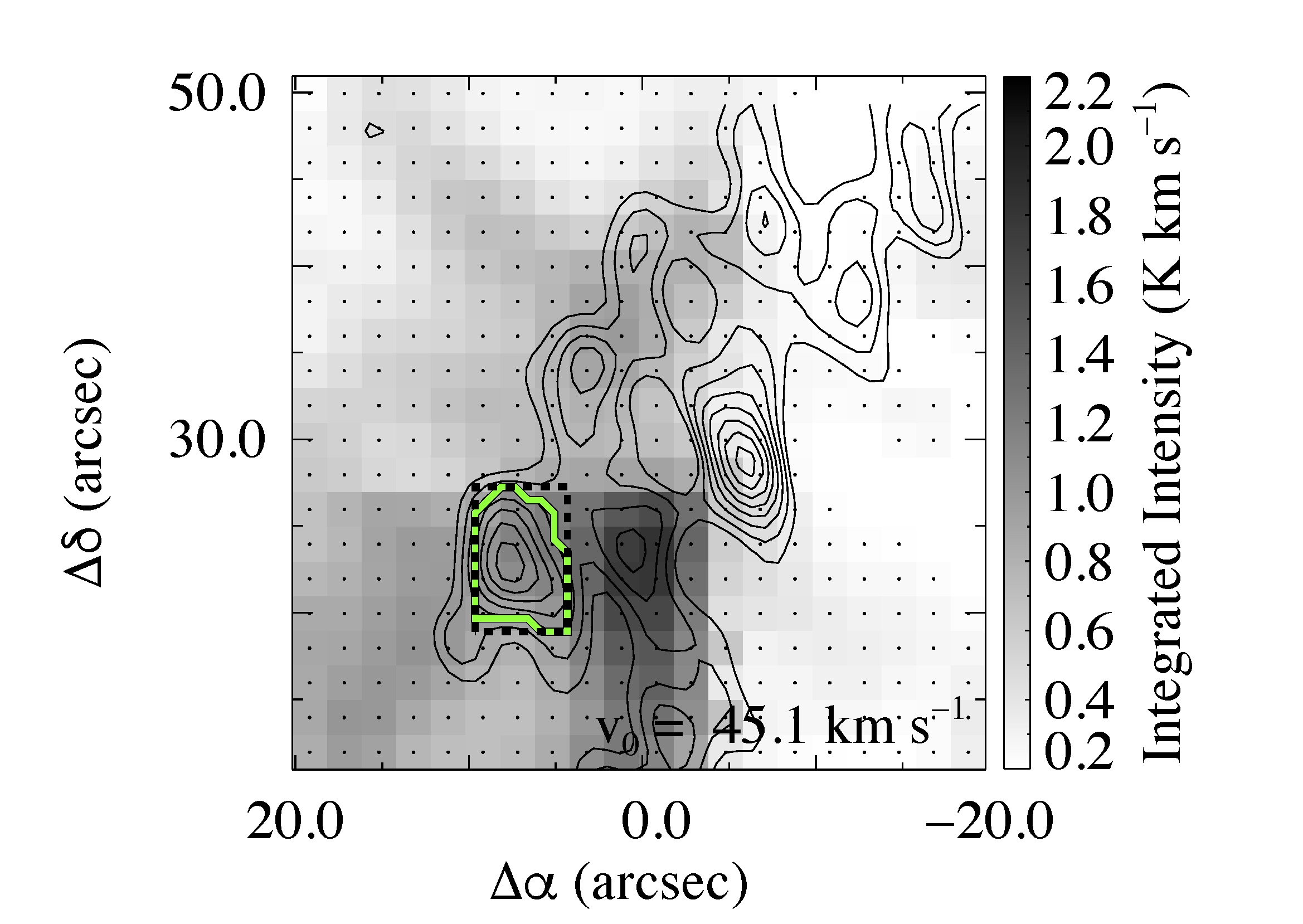}
\includegraphics[trim = 0mm 0mm 0mm 0mm, clip, width = 0.33\textwidth]{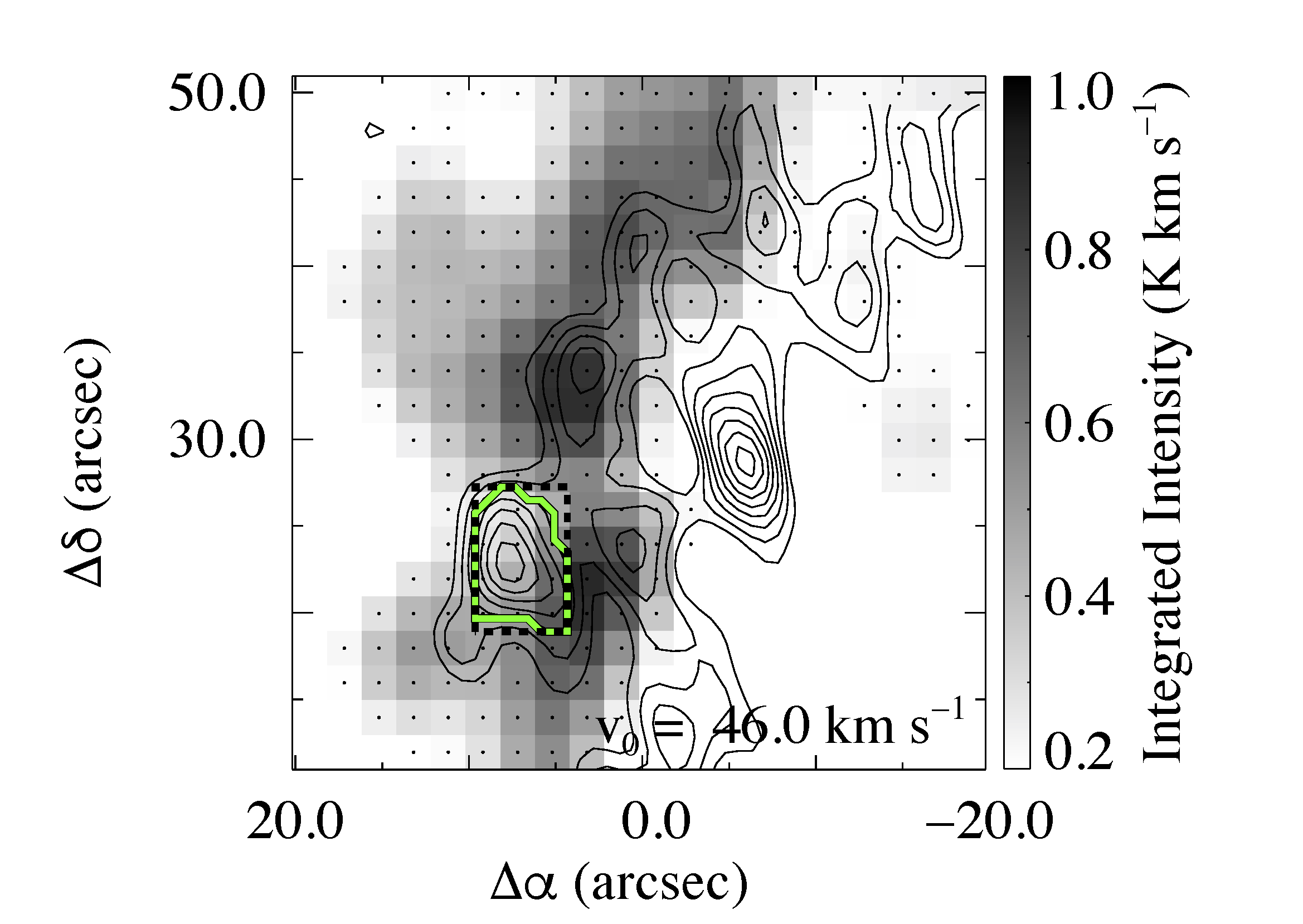}
\includegraphics[trim = 0mm 0mm 0mm 0mm, clip, width = 0.33\textwidth]{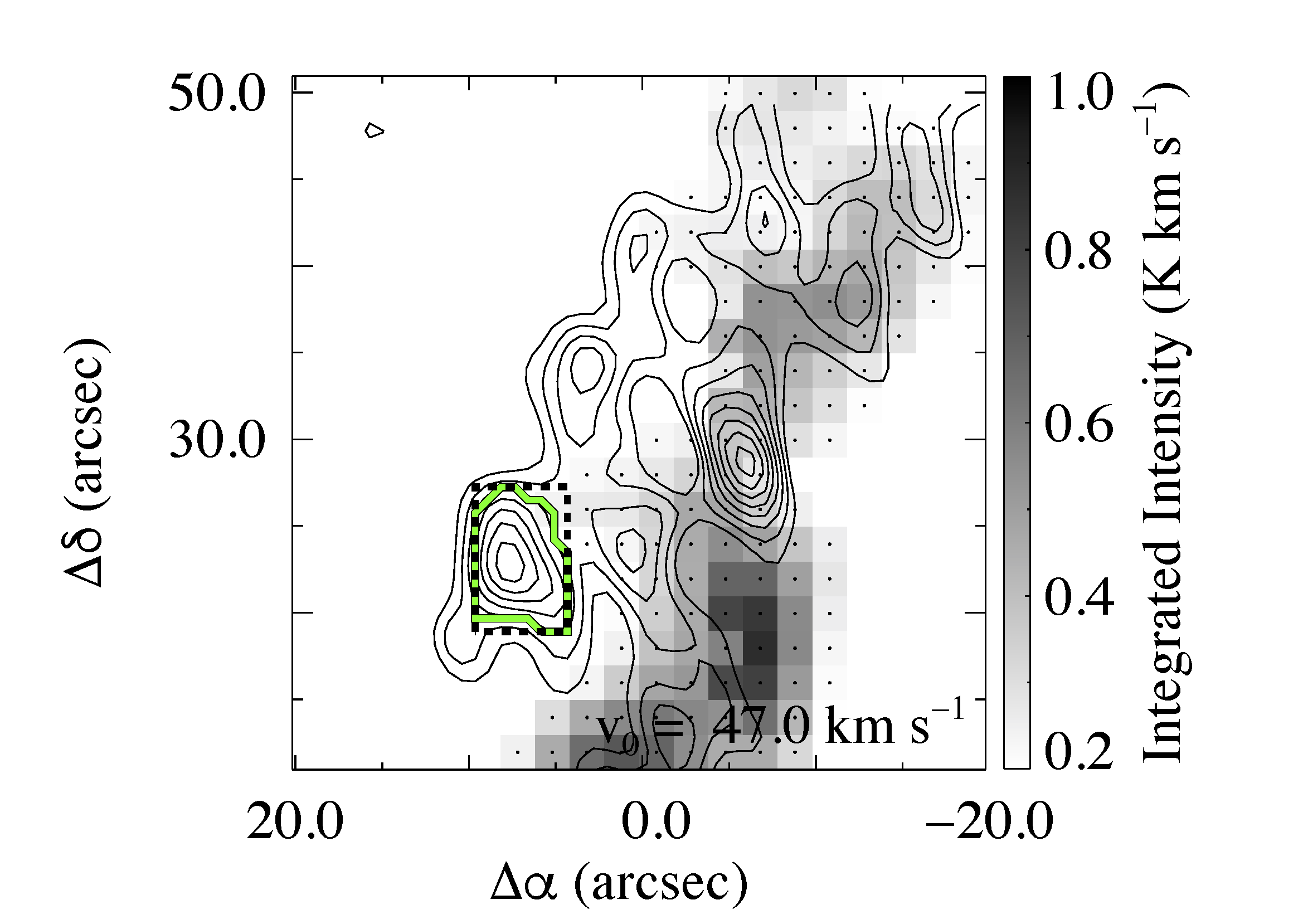}\\

\includegraphics[trim = 35mm 10mm 0mm 0mm, clip, width = 0.45\textwidth]{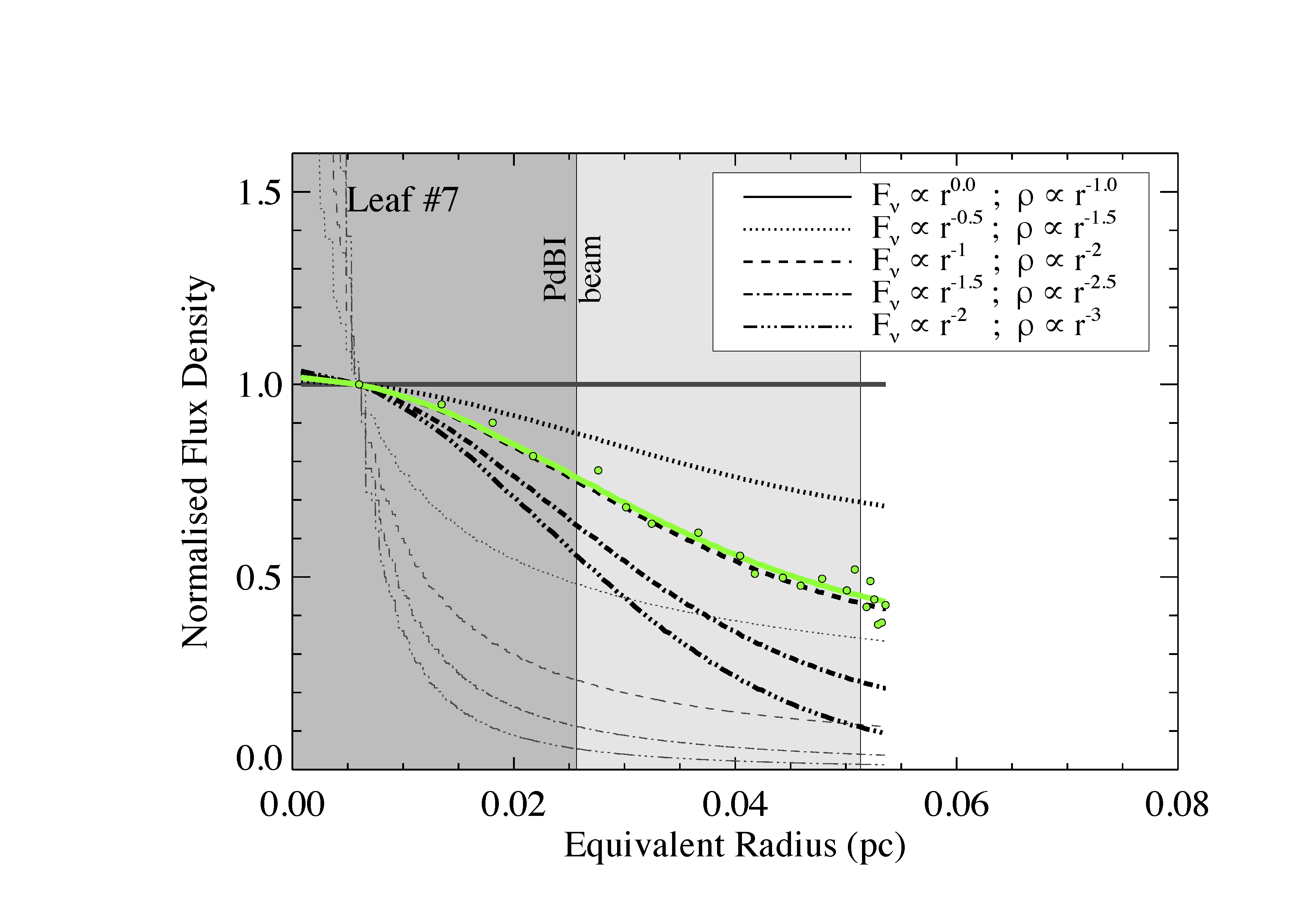}
\end{center}
\caption{The average spectrum, spatial distribution of integrated \ntwoh \ (1-0) emission, and the radial flux density profile of leaf~\#7. Note the difference in scaling, which is selected to enhance the features of individual components. }
\label{Figure:leaf7}
\end{figure*}

Fig.~\ref{Figure:leaf7} shows the distribution of \ntwoh \ (1-0) emission associated with leaf~\#7. The best-fitting solution to the spatially-averaged spectrum requires a three-component model. The third component, at $\sim47.0$\,\kms, is significant to the 5\,$\sigma_{\rm rms}$ level (a two-component fit increases the residuals by a factor of $\lesssim2$). The centroid velocities of the measured components are $v_{\rm 0,1}=45.14\,{\rm km\,s^{-1}}\pm0.02\,{\rm km\,s^{-1}}$, $v_{\rm 0,2}=46.01\,{\rm km\,s^{-1}}\pm0.01\,{\rm km\,s^{-1}}$, and $v_{\rm 0,3}=46.97\,{\rm km\,s^{-1}}\pm0.05\,{\rm km\,s^{-1}}$, respectively. The corresponding FWHM line-widths are $\Delta v_{1}~=~0.99\,{\rm km\,s^{-1}}\pm0.04\,{\rm km\,s^{-1}}$, $\Delta v_{2}~=~0.50\,{\rm km\,s^{-1}}\pm0.03\,{\rm km\,s^{-1}}$, and $\Delta v_{3}~=~0.84\,{\rm km\,s^{-1}}\pm0.15\,{\rm km\,s^{-1}}$, respectively. Inspecting the spatial distribution of the emission associated with each component indicates that the low-velocity component dominates over the other two, which are more prominent towards the north and west of the cloud, respectively. \\

\begin{figure*}
\begin{center}
\includegraphics[trim = 0mm 0mm 0mm 0mm, clip, width = 0.33\textwidth]{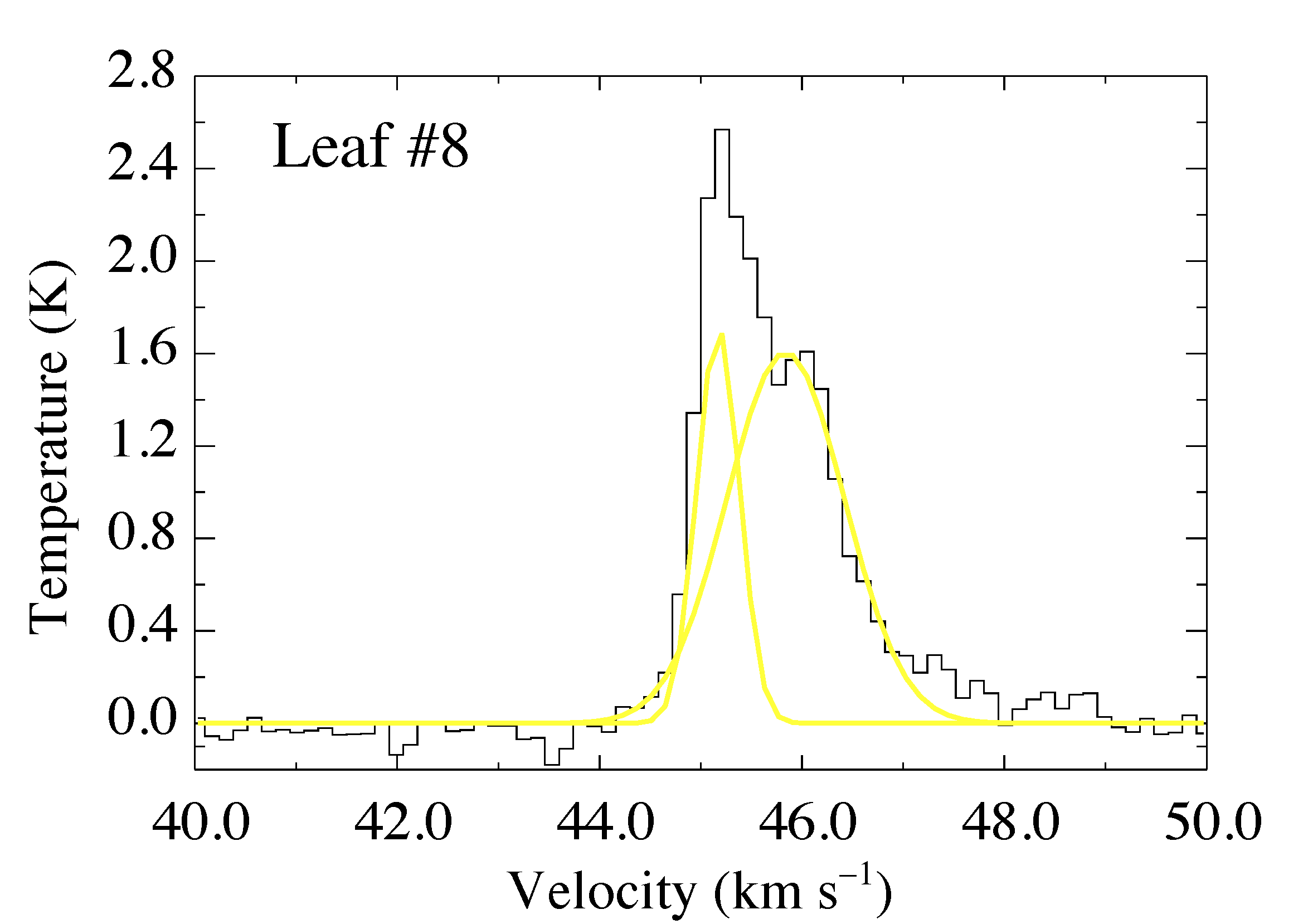}
\includegraphics[trim = 0mm 0mm 0mm 0mm, clip, width = 0.33\textwidth]{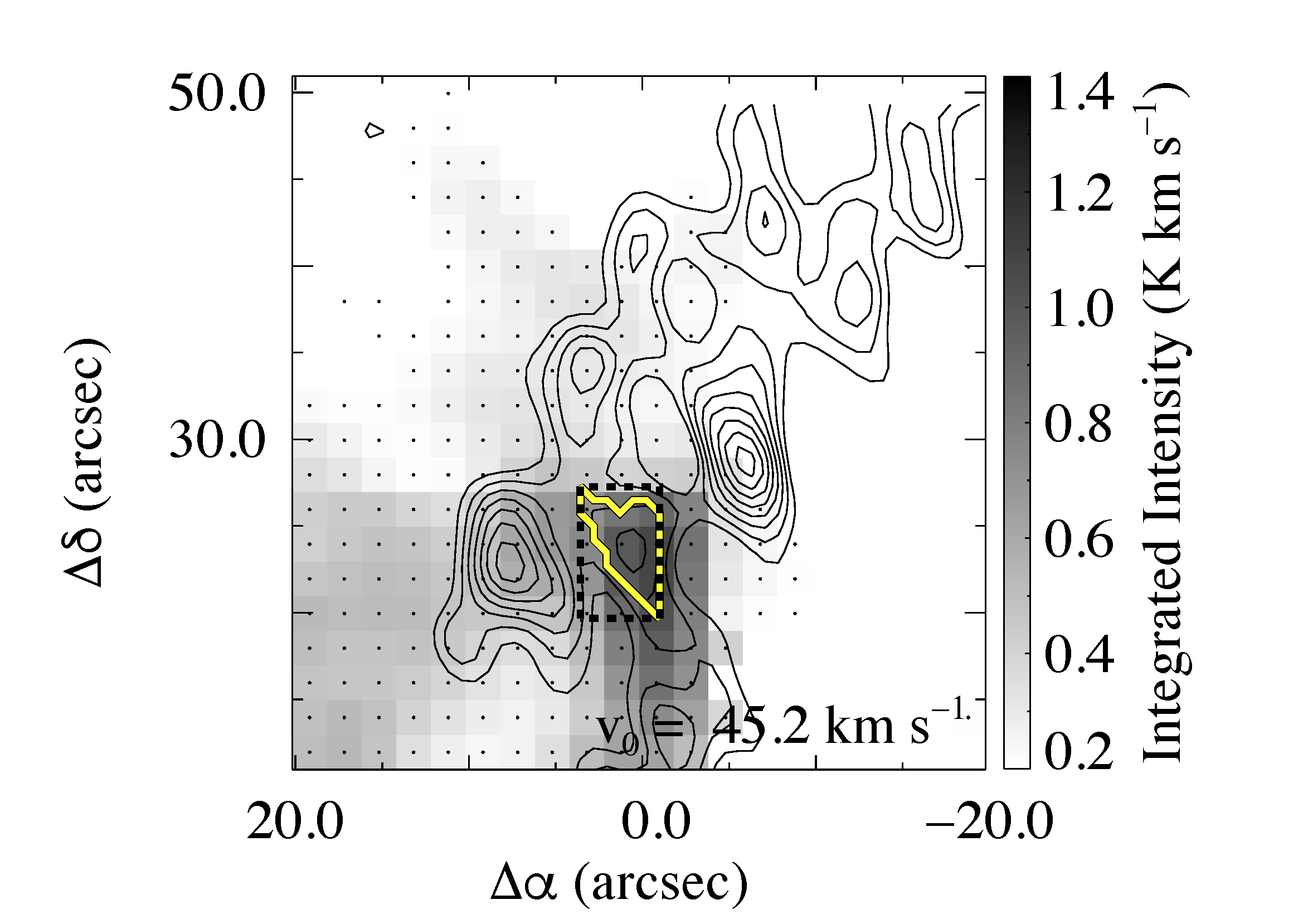}
\includegraphics[trim = 0mm 0mm 0mm 0mm, clip, width = 0.33\textwidth]{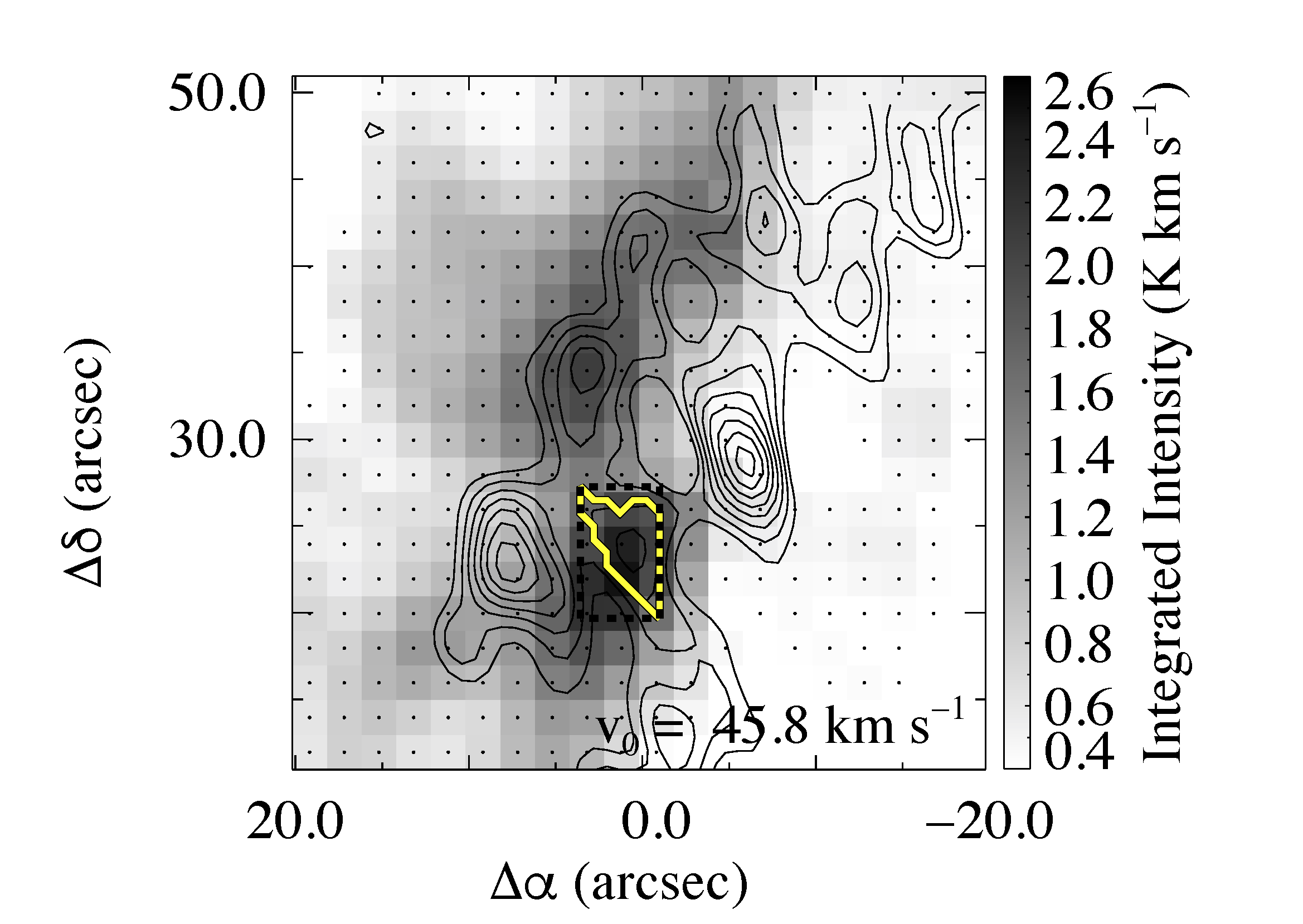}
\includegraphics[trim = 35mm 10mm 0mm 0mm, clip, width = 0.45\textwidth]{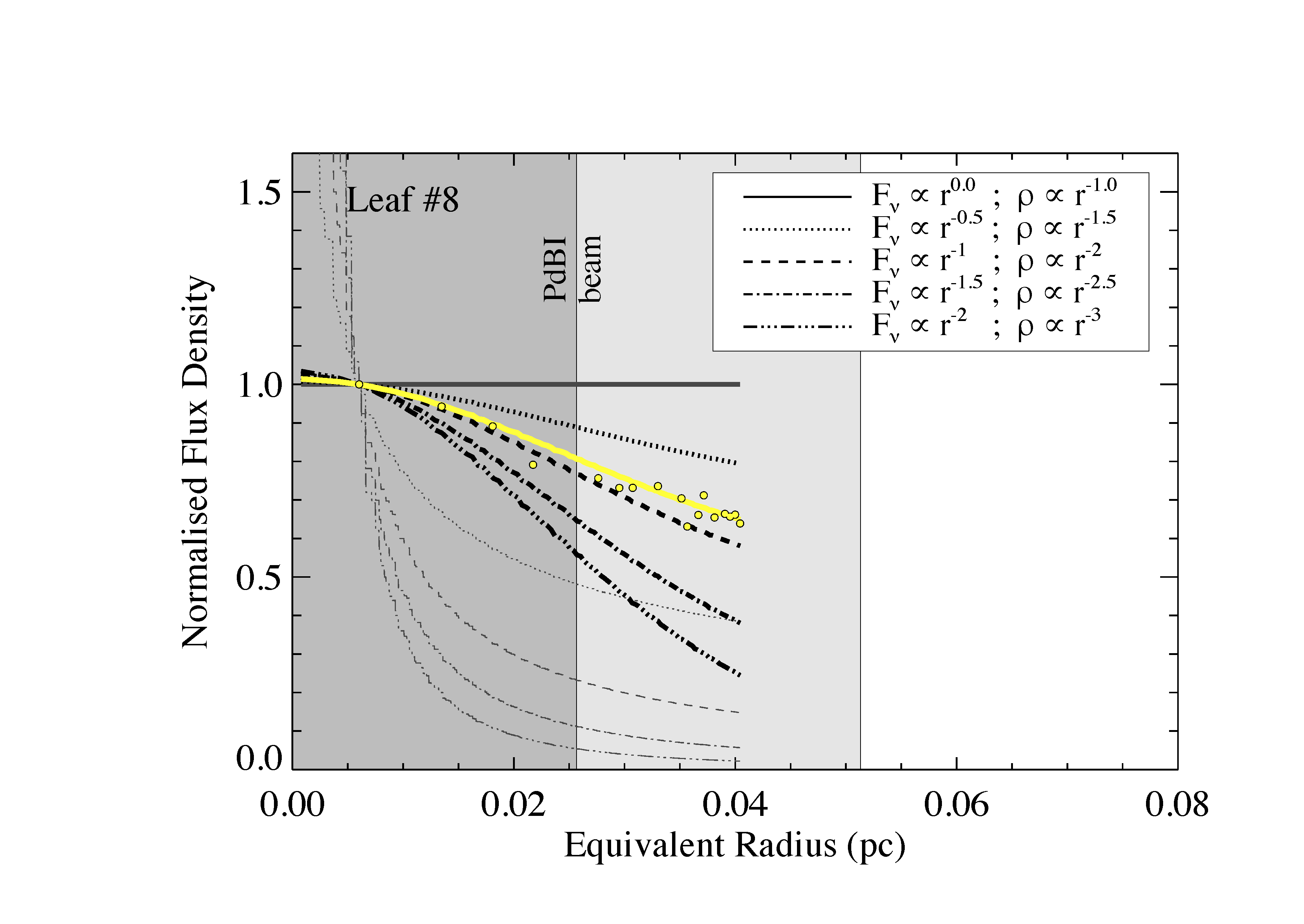}
\end{center}
\caption{The average spectrum, spatial distribution of integrated \ntwoh \ (1-0) emission, and the radial flux density profile of leaf~\#8. Both Gaussian components appear yellow since although leaf~\#8 appears to be monolithic in continuum, it cannot be unambiguously linked to either velocity component. Note the difference in scaling, which is selected to enhance the features of both components.  }
\label{Figure:leaf8}
\end{figure*}

\noindent\textbf{Leaf~\#8:} situated at $\{\Delta\alpha,\,\Delta\delta\}=\{0.5\,{\rm arcsec},\,23.5\,{\rm arcsec}\}$, leaf~\#8 has the smallest projected separation from the H6 extinction peak (located at $\{\Delta\alpha,\,\Delta\delta\}=\{3.0\,{\rm arcsec},\,21.1\,{\rm arcsec}\}$; \citealp{butler_2012}). Leaf~\#8 has an aspect ratio of $\AR=1.88$ and the continuum emission appears to be singly peaked. Fig.~\ref{Figure:leaf8} takes a closer look at the distribution of \ntwoh \ emission towards leaf~\#8. The spectrum indicates the presence of two velocity components. The measured centroid velocities and FWHM line-widths are $v_{\rm 0,1}=45.17\,{\rm km\,s^{-1}}\pm0.01\,{\rm km\,s^{-1}}$ and $v_{\rm 0,2}=45.83\,{\rm km\,s^{-1}}\pm0.02\,{\rm km\,s^{-1}}$ and $\Delta v_{1}~=~0.49\,{\rm km\,s^{-1}}\pm0.02\,{\rm km\,s^{-1}}$ and $\Delta v_{2}~=~1.37\,{\rm km\,s^{-1}}\pm0.03\,{\rm km\,s^{-1}}$, respectively. We are unable to unambiguously relate either velocity component to the continuum emission. Note that this is different to the cases of leaf~\#5 and \#6, where the continuum emission may be attributed to two independent structures. In this example we use both measurements of the FWHM line-width for studying the dynamical properties of leaf~\#8 in \S~\ref{Section:analysis}. \\

\noindent\textbf{Leaf~\#9:} is situated at $\{\Delta\alpha,\,\Delta\delta\}=\{-6.3\,{\rm arcsec},\,28.8\,{\rm arcsec}\}$. The continuum emission appears monolithic and, under the assumptions made in \S~\ref{Section:analysis}, it is the most massive of the identified leaves with $M_{\rm c}~\sim~24$\,\solar \ and $M^{\rm b}_{\rm c}~\sim~12$\,\solar. Leaf~\#9 is dark at 70\,\micron \ and not identified in \citet{nguyen_2011}. The figures describing the \ntwoh \ emission associated with leaf~\#9 can be found in the main text (top panels Fig.~\ref{Figure:spec_mom}). Two spectral components are evident. The measured centroid velocities and FWHM line-widths are $v_{\rm 0,1}=45.37\,{\rm km\,s^{-1}}\pm0.02\,{\rm km\,s^{-1}}$ and $v_{\rm 0,2}=46.63\,{\rm km\,s^{-1}}\pm0.01\,{\rm km\,s^{-1}}$ and $\Delta v_{1}~=~0.79\,{\rm km\,s^{-1}}\pm0.04\,{\rm km\,s^{-1}}$ and $\Delta v_{2}~=~0.62\,{\rm km\,s^{-1}}\pm0.03\,{\rm km\,s^{-1}}$, respectively. As discussed in \S~\ref{Section:results_line}, the high-velocity component dominates over the low-velocity counterpart. \\

\noindent\textbf{Leaf~\#10:} is situated at $\{\Delta\alpha,\,\Delta\delta\}=\{2.8\,{\rm arcsec},\,34.1\,{\rm arcsec}\}$. To the north-west of leaf~\#10 is extended 4.5, 8, and 24\,\micron \ emission, implying the presence of an internal heating source. This appears as a ``hole'' in the mid-infrared-derived mass surface density map of \citet{kainulainen_2013} (Fig.~\ref{Figure:msd_map}). This was identified in the \emph{Herschel} 70\,\micron \ images and described as a ``protostellar massive dense core'' by \citet[core \#28; their table~1]{nguyen_2011}. The mass estimated from \emph{Herschel} observations is $\sim55\pm11$\,\solar. This is substantially different to our continuum-derived masses, $M_{\rm c}~\sim~8$\,\solar \ and $M^{\rm b}_{\rm c}~\sim~2$\,\solar. However, we note that leaf~\#10 is the smallest of the identified leaves, with an angular radius of $\sim2\,{\rm arcsec}$ (corresponding to a physical radius of $\sim0.03$\,pc at a distance of 2900\,pc) and that it is slightly offset in position from the location of 4.5, 8, and 24\,\micron \ emission. The difference in mass may therefore reflect a difference in source definition. 

\begin{figure*}
\begin{center}
\includegraphics[trim = 0mm 0mm 0mm 0mm, clip, width = 0.33\textwidth]{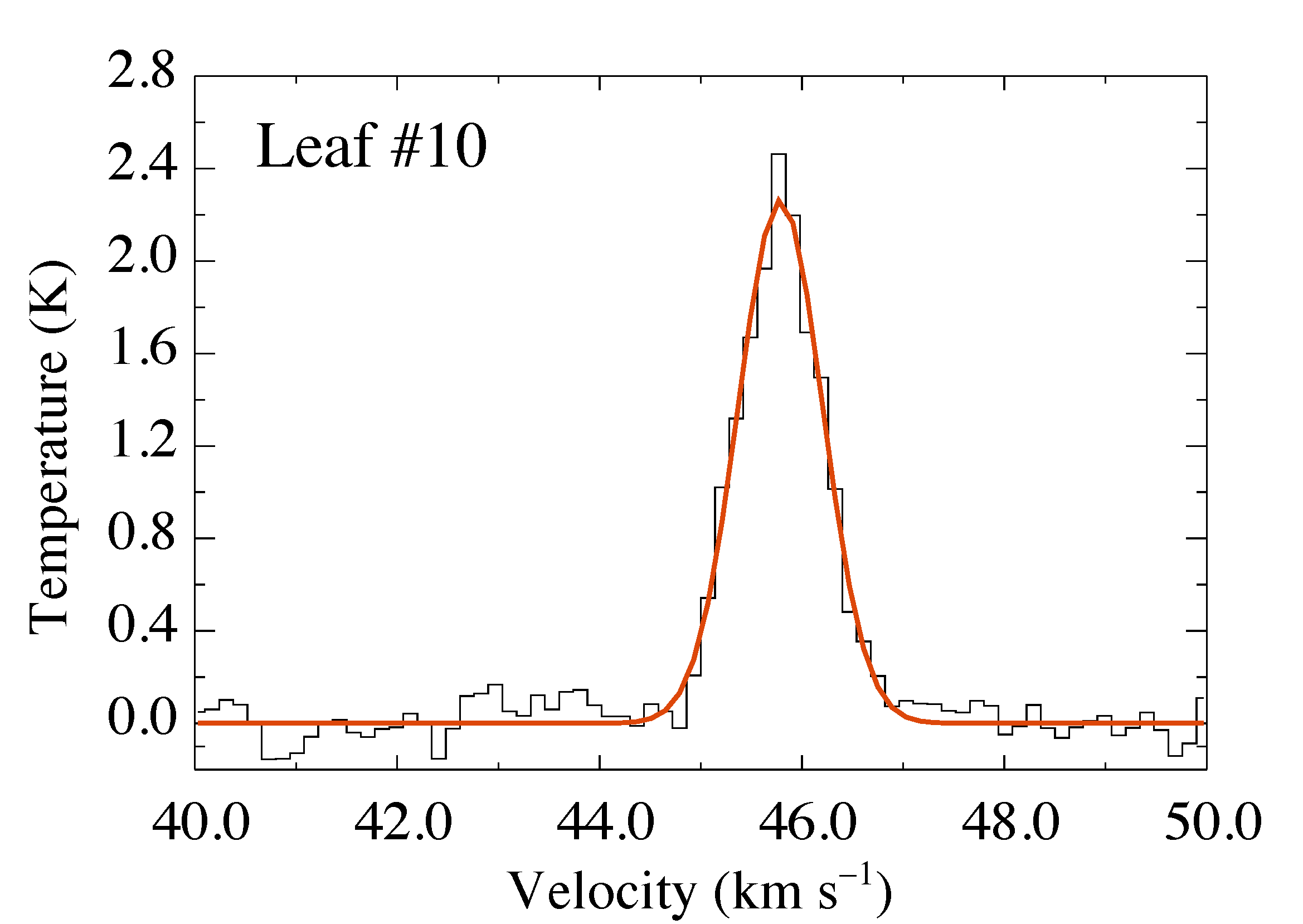}
\includegraphics[trim = 0mm 0mm 0mm 0mm, clip, width = 0.33\textwidth]{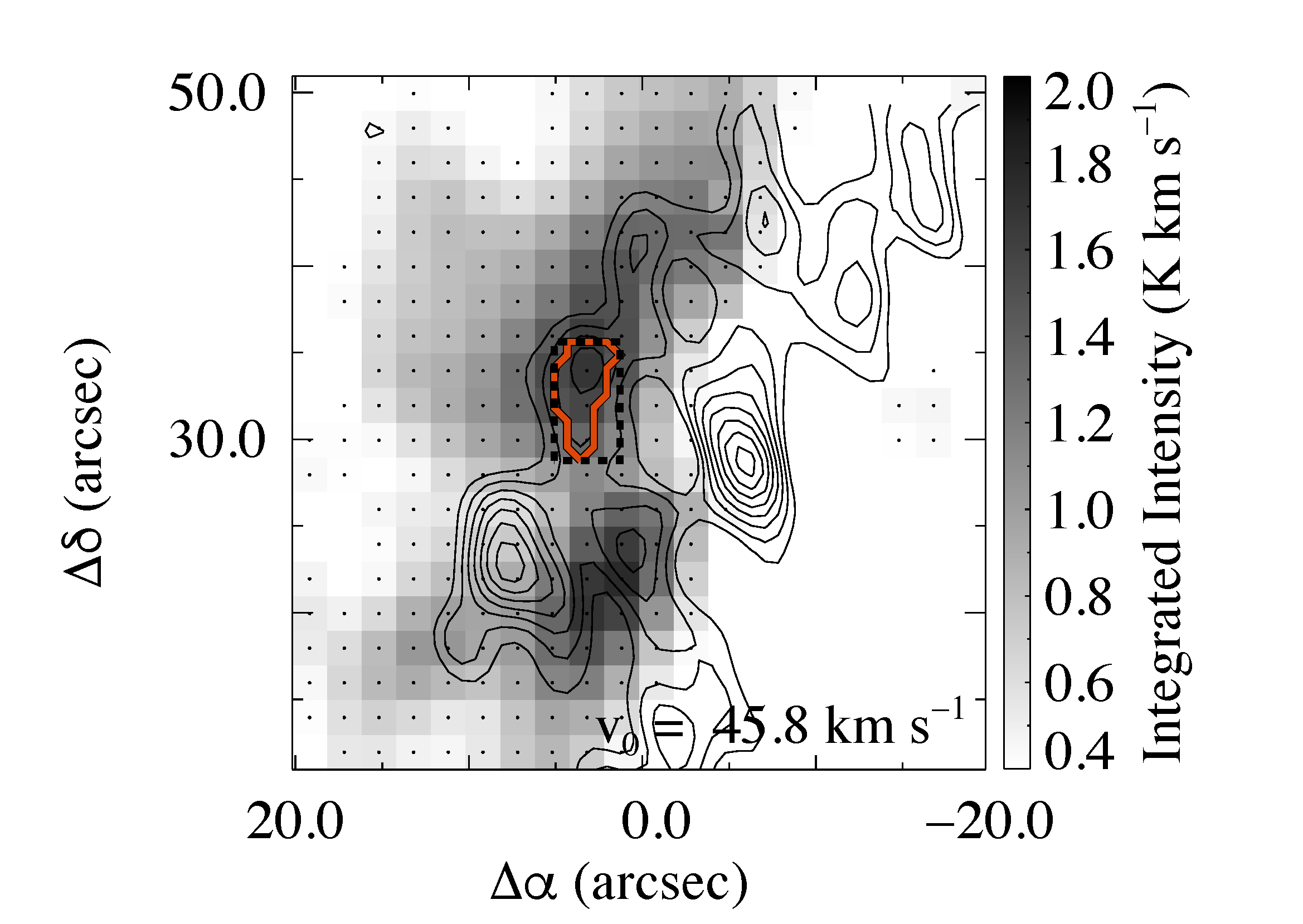}
\includegraphics[trim = 35mm 10mm 0mm 0mm, clip, width = 0.45\textwidth]{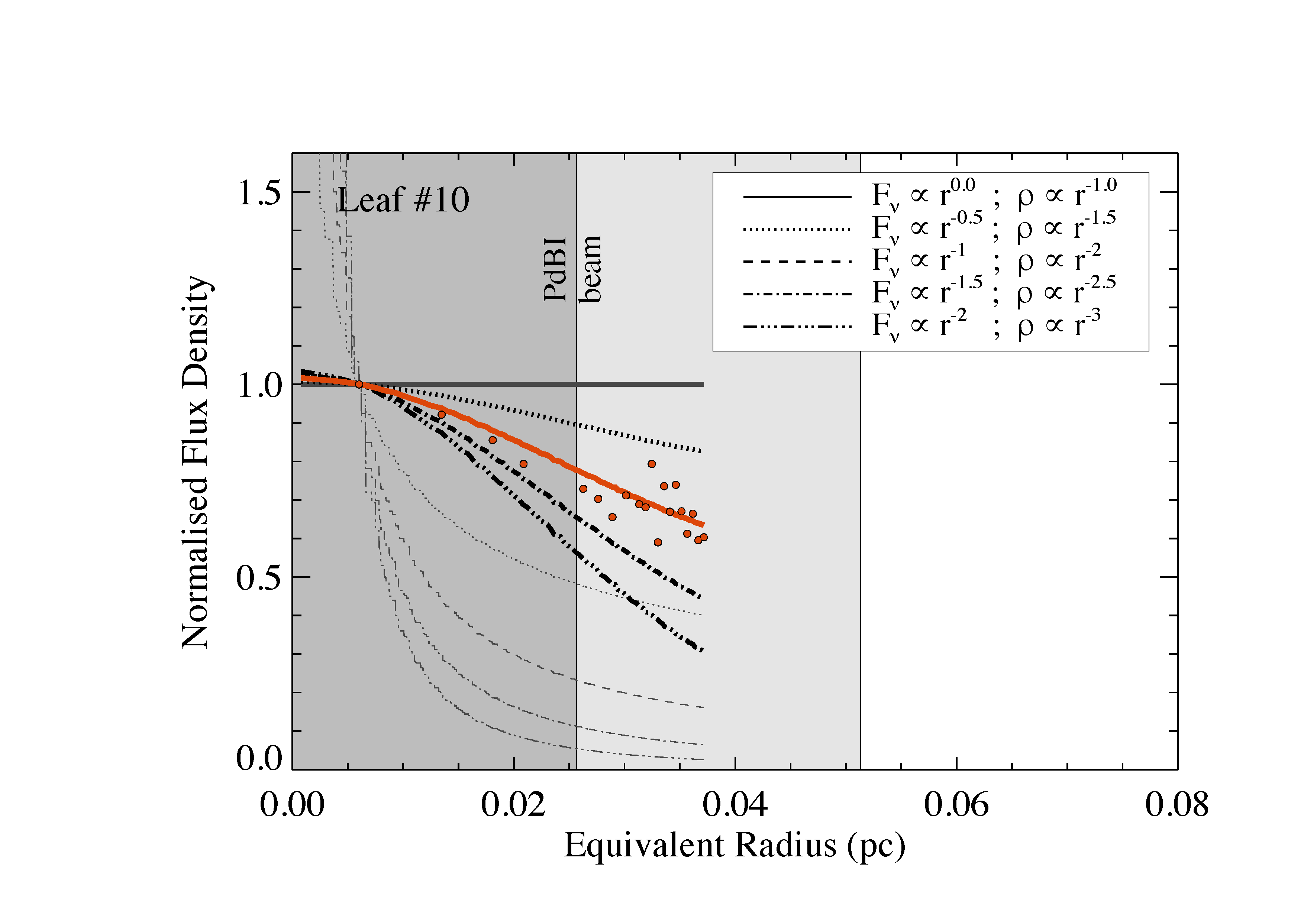}
\end{center}
\caption{The average spectrum, spatial distribution of integrated \ntwoh \ (1-0) emission, and the radial flux density profile of leaf~\#10.}
\label{Figure:leaf10}
\end{figure*}

Fig.~\ref{Figure:leaf10} shows the distribution of \ntwoh \ (1-0) emission associated with leaf~\#10. The emission is singly peaked. The measured centroid velocity and FWHM line-width of this component are $v_{\rm 0}=45.79\,{\rm km\,s^{-1}}\pm0.01\,{\rm km\,s^{-1}}$ and $\Delta v~=~0.98\,{\rm km\,s^{-1}}\pm0.02\,{\rm km\,s^{-1}}$, respectively.  \\

\begin{figure*}
\begin{center}
\includegraphics[trim = 0mm 0mm 0mm 0mm, clip, width = 0.33\textwidth]{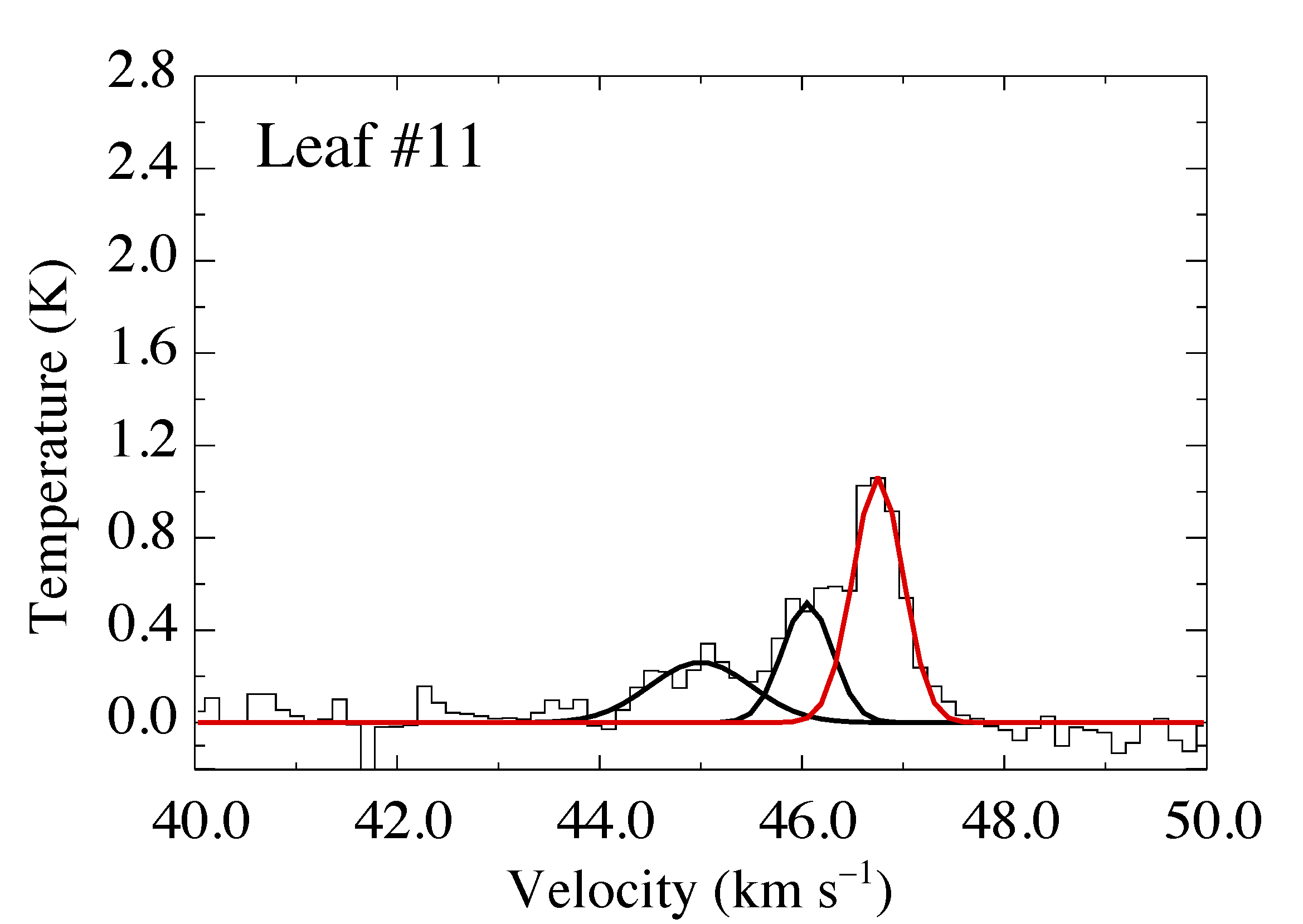}
\includegraphics[trim = 0mm 0mm 0mm 0mm, clip, width = 0.33\textwidth]{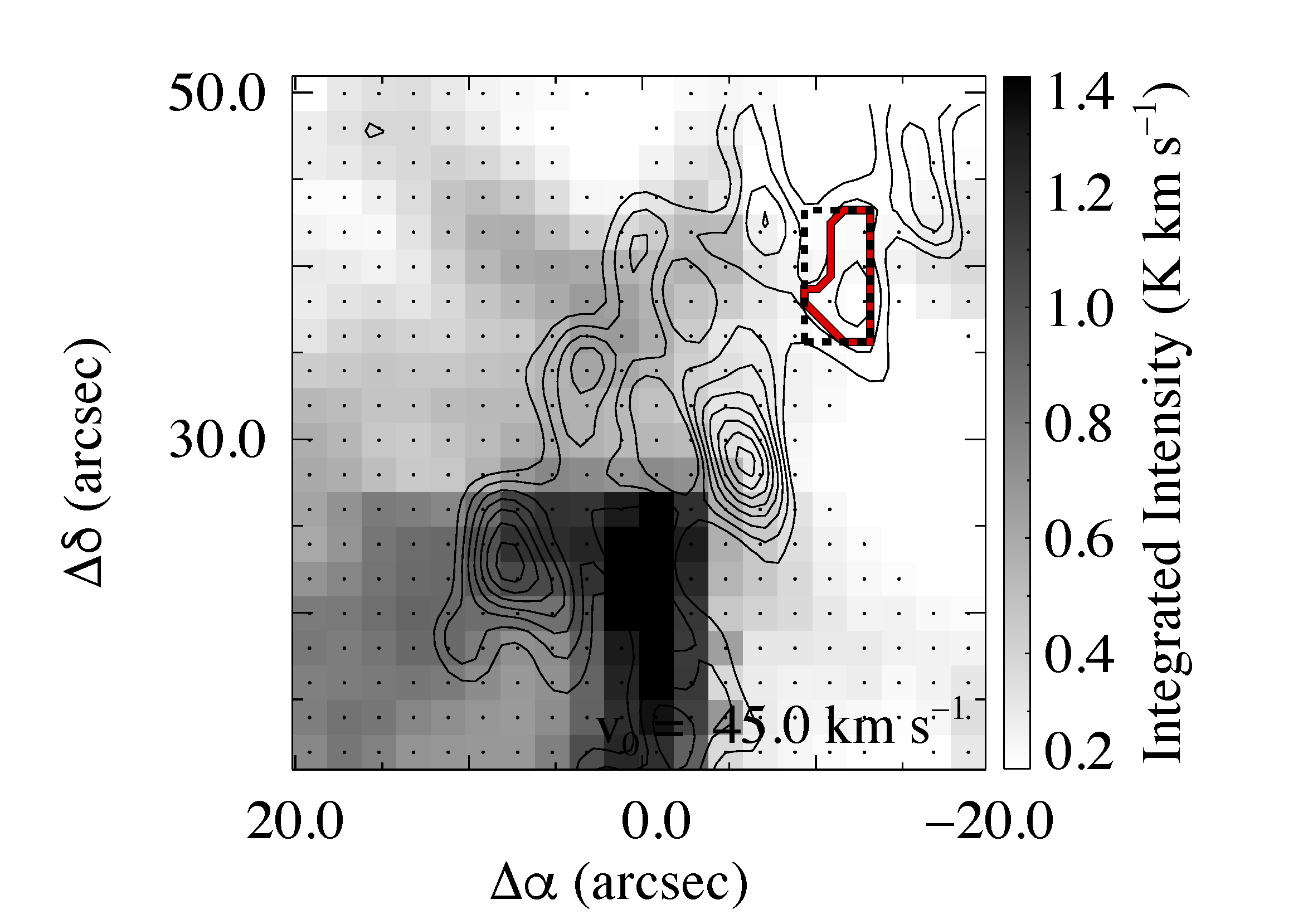}
\includegraphics[trim = 0mm 0mm 0mm 0mm, clip, width = 0.33\textwidth]{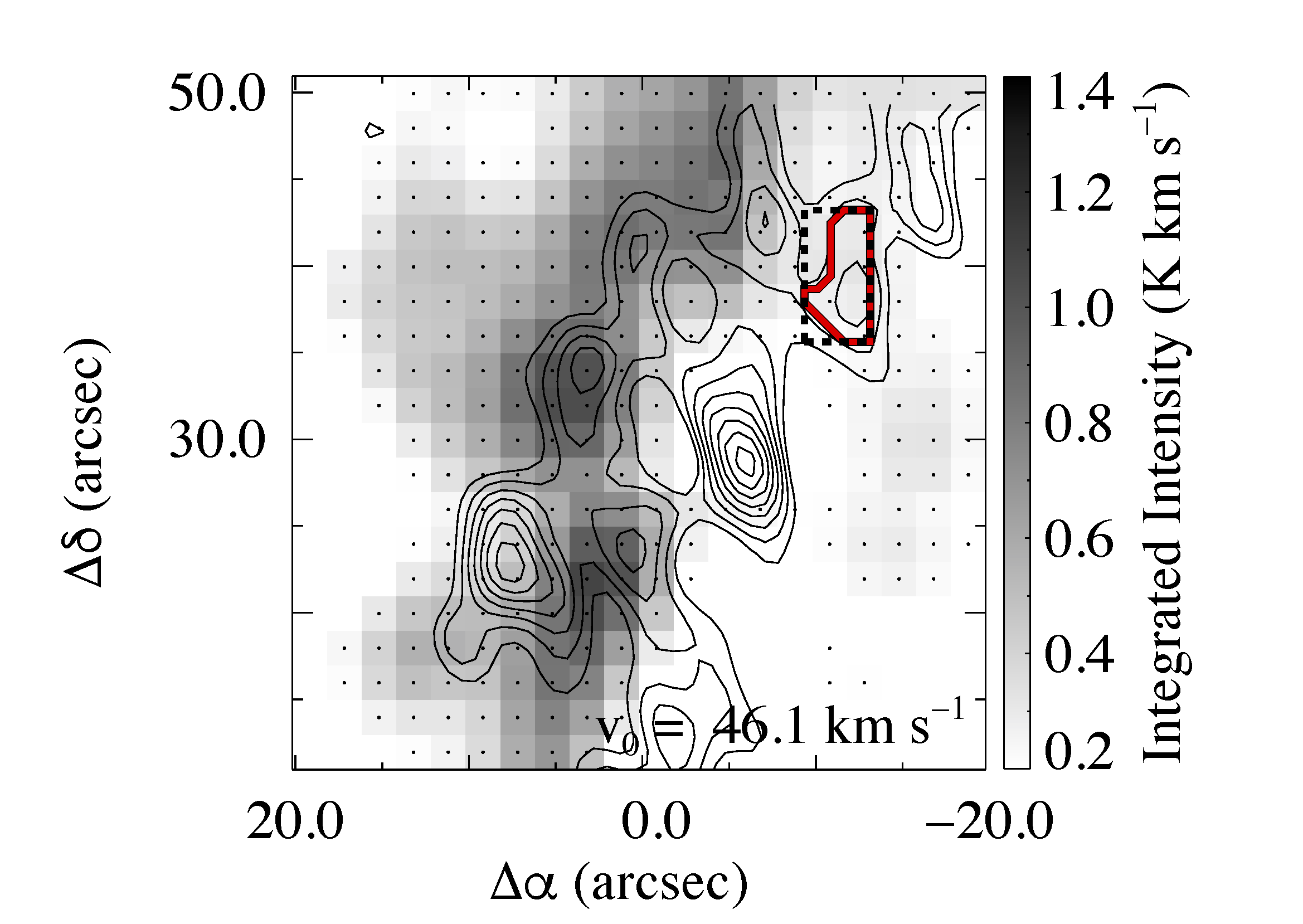}
\includegraphics[trim = 0mm 0mm 0mm 0mm, clip, width = 0.33\textwidth]{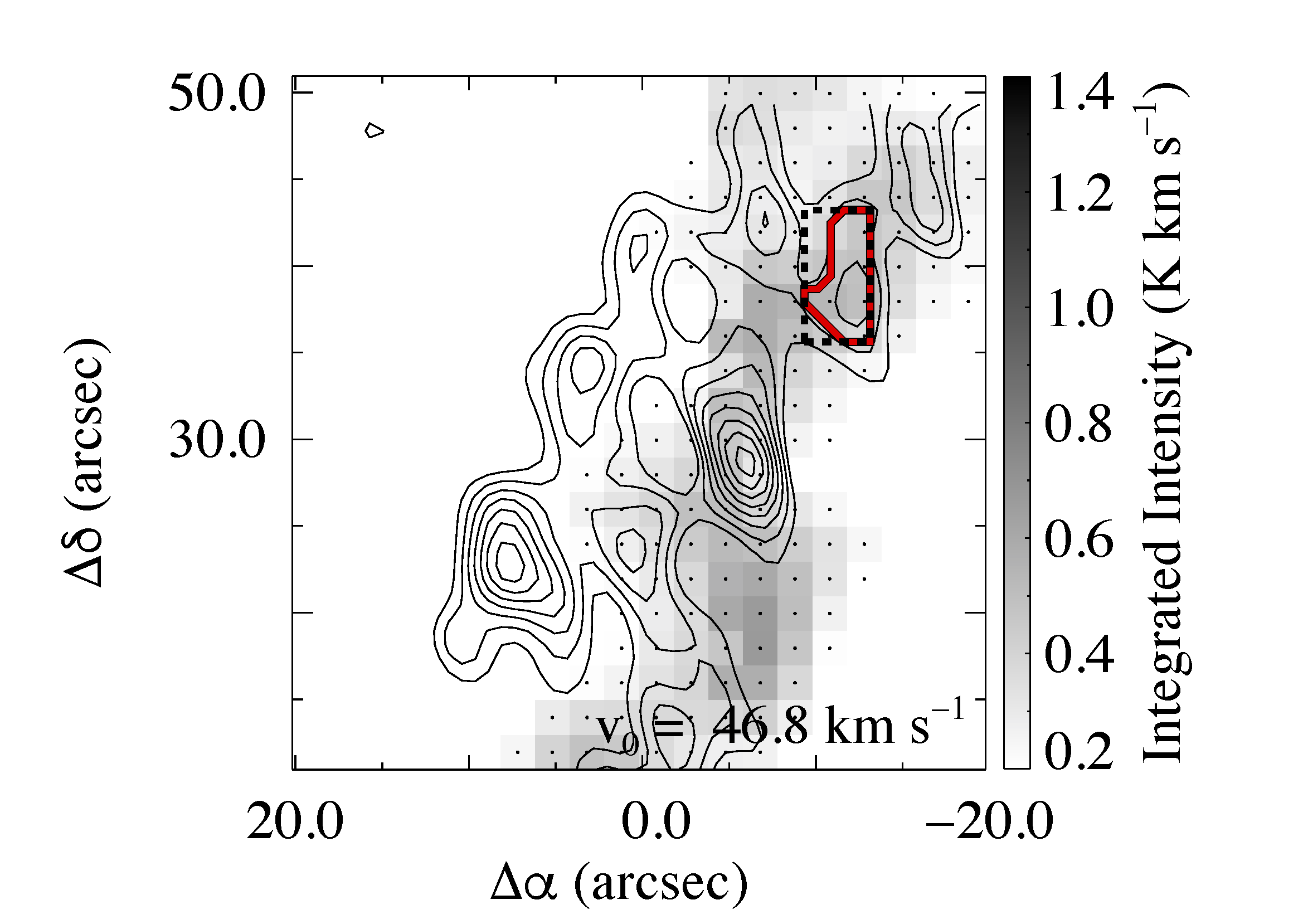}\\

\includegraphics[trim = 35mm 10mm 0mm 0mm, clip, width = 0.45\textwidth]{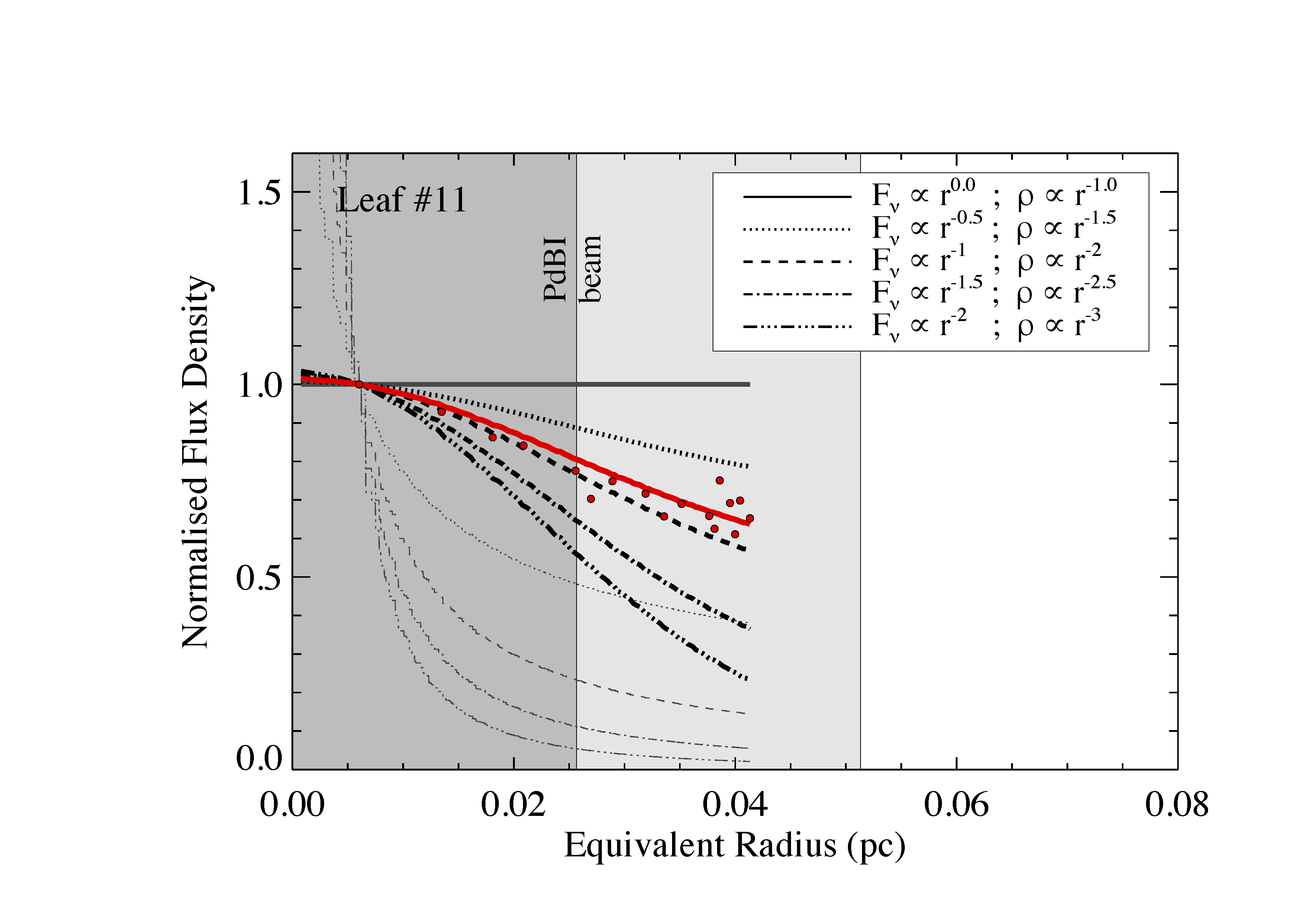}
\end{center}
\caption{The average spectrum, spatial distribution of integrated \ntwoh \ (1-0) emission, and the radial flux density profile of leaf~\#11. }
\label{Figure:leaf11}
\end{figure*}

\noindent\textbf{Leaf~\#11:} is situated at $\{\Delta\alpha,\,\Delta\delta\}=\{-12.4\,{\rm arcsec},\,38.7\,{\rm arcsec}\}$. Although the continuum emission appears monolithic, three spectral components are identified in the spatially-averaged spectrum extracted from within the leaf boundary. The centroid velocities of the measured components are $v_{\rm 0,1}=45.00\,{\rm km\,s^{-1}}\pm0.14\,{\rm km\,s^{-1}}$, $v_{\rm 0,2}=46.05\,{\rm km\,s^{-1}}\pm0.06\,{\rm km\,s^{-1}}$, and $v_{\rm 0,3}=46.75\,{\rm km\,s^{-1}}\pm0.03\,{\rm km\,s^{-1}}$, respectively. The corresponding FWHM line-widths are $\Delta v_{1}~=~1.18\,{\rm km\,s^{-1}}\pm0.33\,{\rm km\,s^{-1}}$, $\Delta v_{2}~=~0.58\,{\rm km\,s^{-1}}\pm0.16\,{\rm km\,s^{-1}}$, and $\Delta v_{3}~=~0.58\,{\rm km\,s^{-1}}\pm0.06\,{\rm km\,s^{-1}}$, respectively. Although low in intensity, the low-velocity component is significant to $>3\,\sigma_{\rm rms}$. Inspecting the spatial distribution of integrated emission associated with each component (Fig.~\ref{Figure:leaf11}), it is evident that the high-velocity component dominates over the others. \\

\begin{figure*}
\begin{center}
\includegraphics[trim = 0mm 0mm 0mm 0mm, clip, width = 0.33\textwidth]{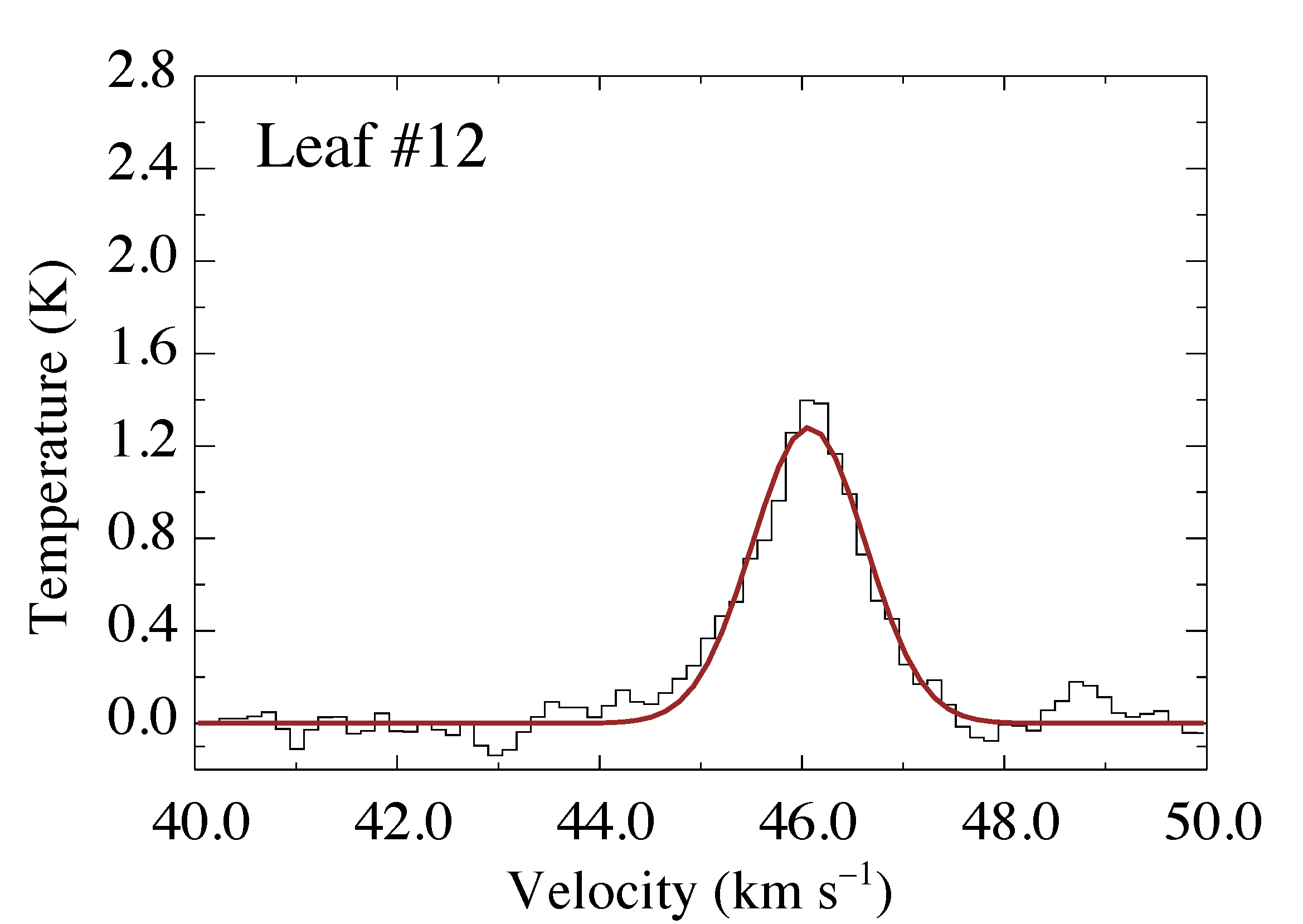}
\includegraphics[trim = 0mm 0mm 0mm 0mm, clip, width = 0.33\textwidth]{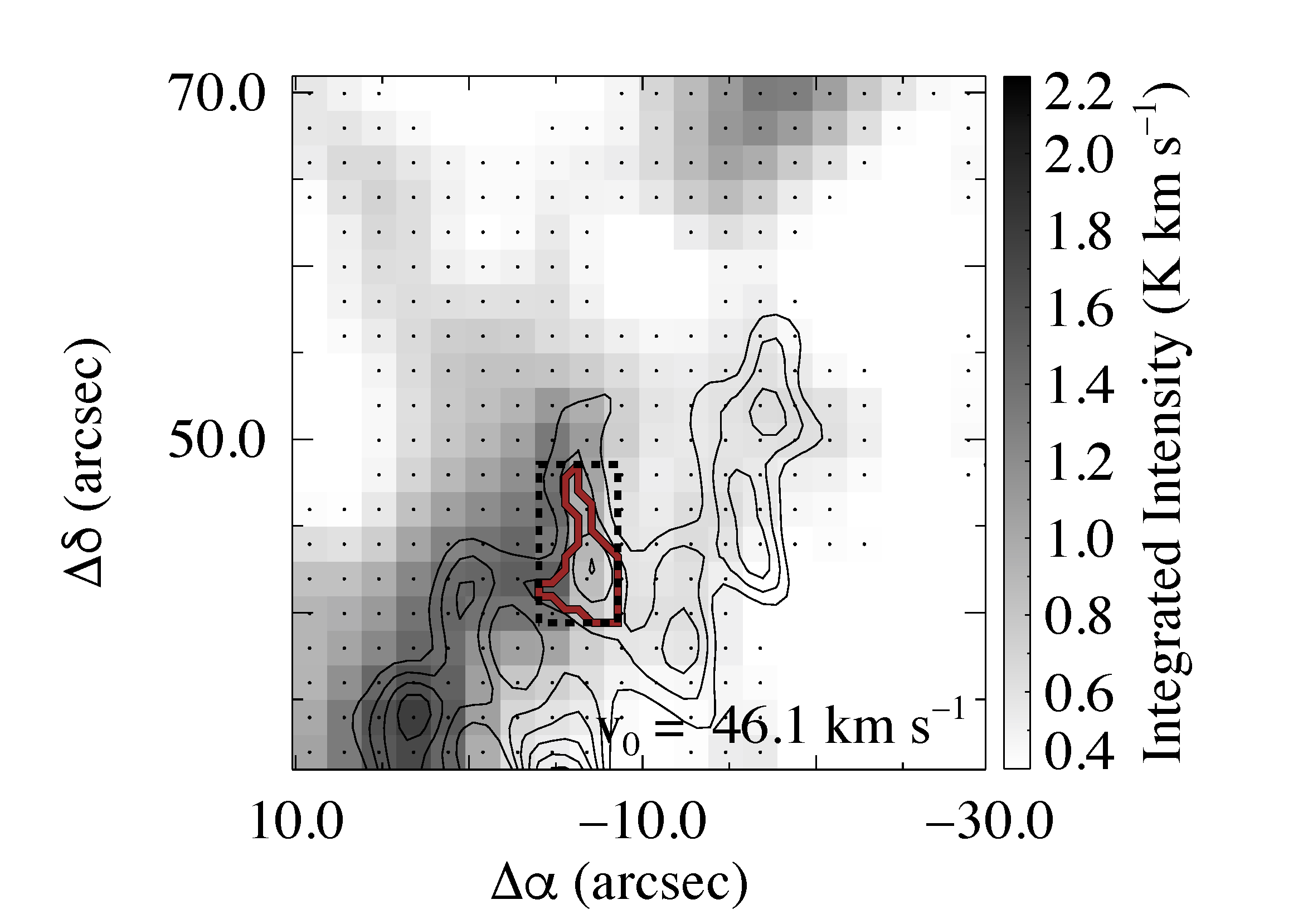}
\includegraphics[trim = 35mm 10mm 0mm 0mm, clip, width = 0.45\textwidth]{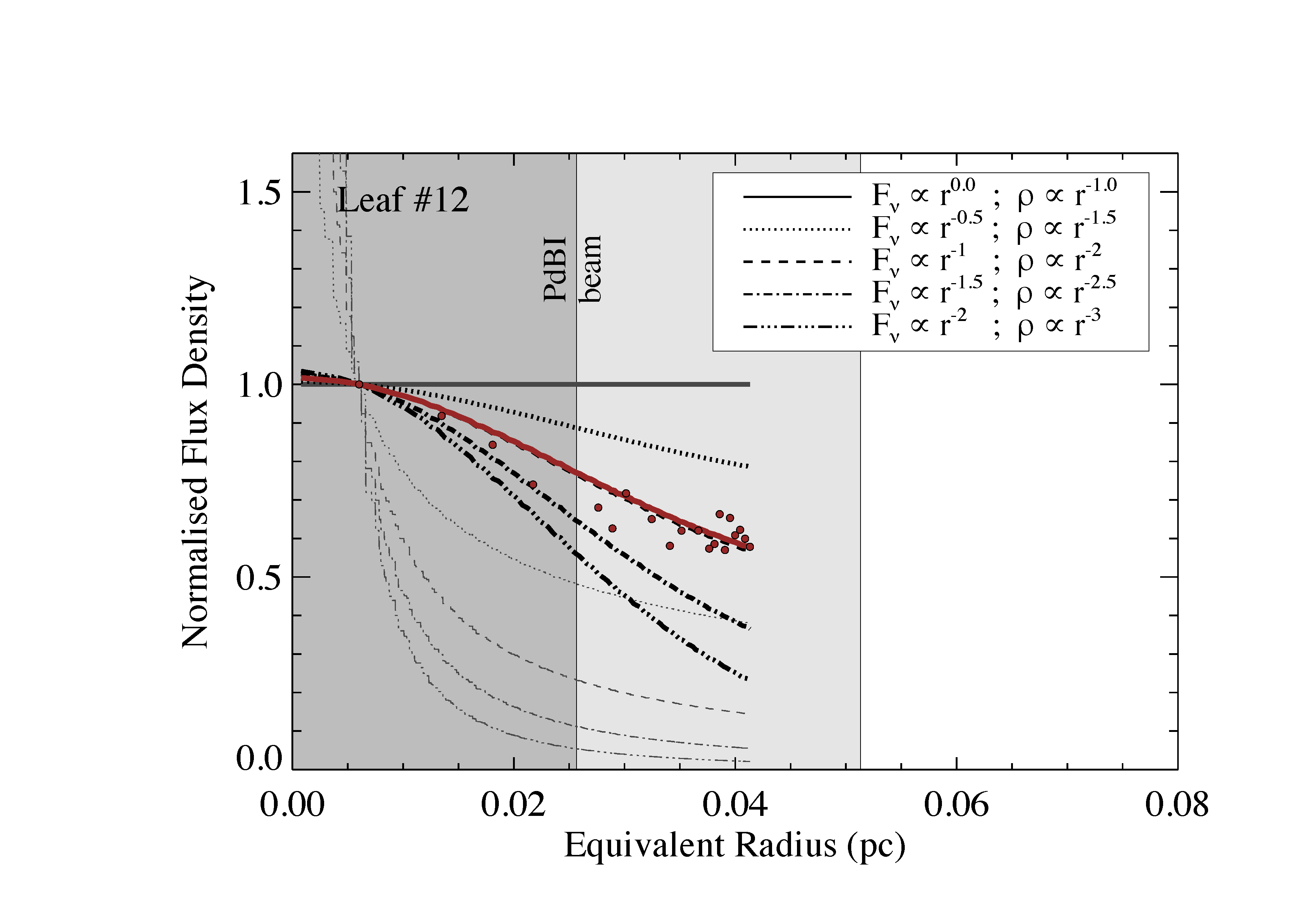}
\end{center}
\caption{The average spectrum, spatial distribution of integrated \ntwoh \ (1-0) emission, and the radial flux density profile of leaf~\#12. }
\label{Figure:leaf12}
\end{figure*}

\noindent\textbf{Leaf~\#12:} is situated at $\{\Delta\alpha,\,\Delta\delta\}=\{-7.1\,{\rm arcsec},\,42.5\,{\rm arcsec}\}$. The leaf has an irregular-shaped boundary, with an extension towards the north. This leads to an artificially-high aspect ratio of $\AR=2.15$. Fig.~\ref{Figure:leaf12} shows the distribution of \ntwoh \ emission associated with leaf~\#12. Only one spectral component is identified. The measured centroid velocity and FWHM line-width of this component are $v_{\rm 0}=46.07\,{\rm km\,s^{-1}}\pm0.01\,{\rm km\,s^{-1}}$ and $\Delta v~=~1.32\,{\rm km\,s^{-1}}\pm0.02\,{\rm km\,s^{-1}}$, respectively.  \\

\begin{figure*}
\begin{center}
\includegraphics[trim = 0mm 0mm 0mm 0mm, clip, width = 0.33\textwidth]{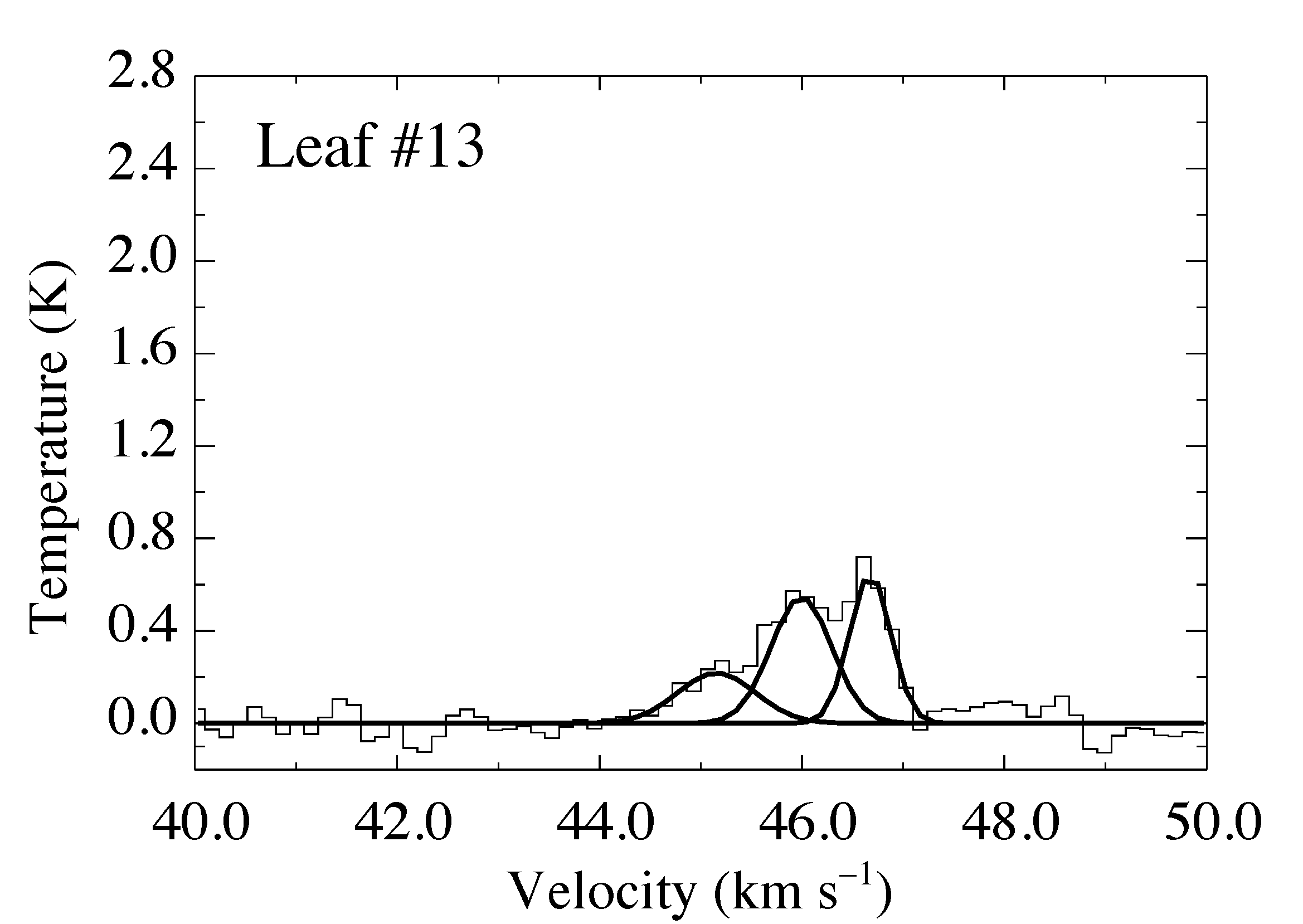}
\includegraphics[trim = 0mm 0mm 0mm 0mm, clip, width = 0.33\textwidth]{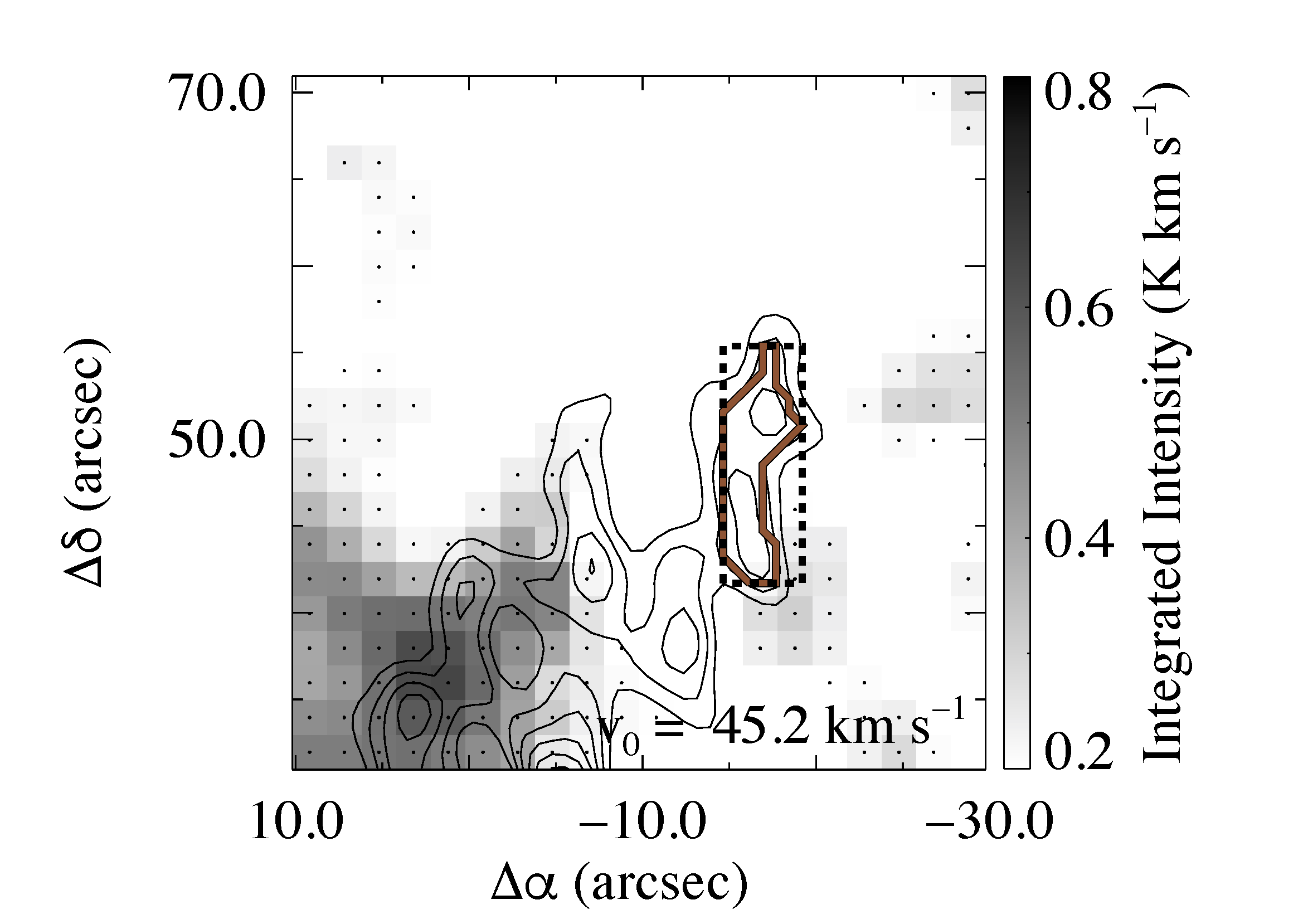}
\includegraphics[trim = 0mm 0mm 0mm 0mm, clip, width = 0.33\textwidth]{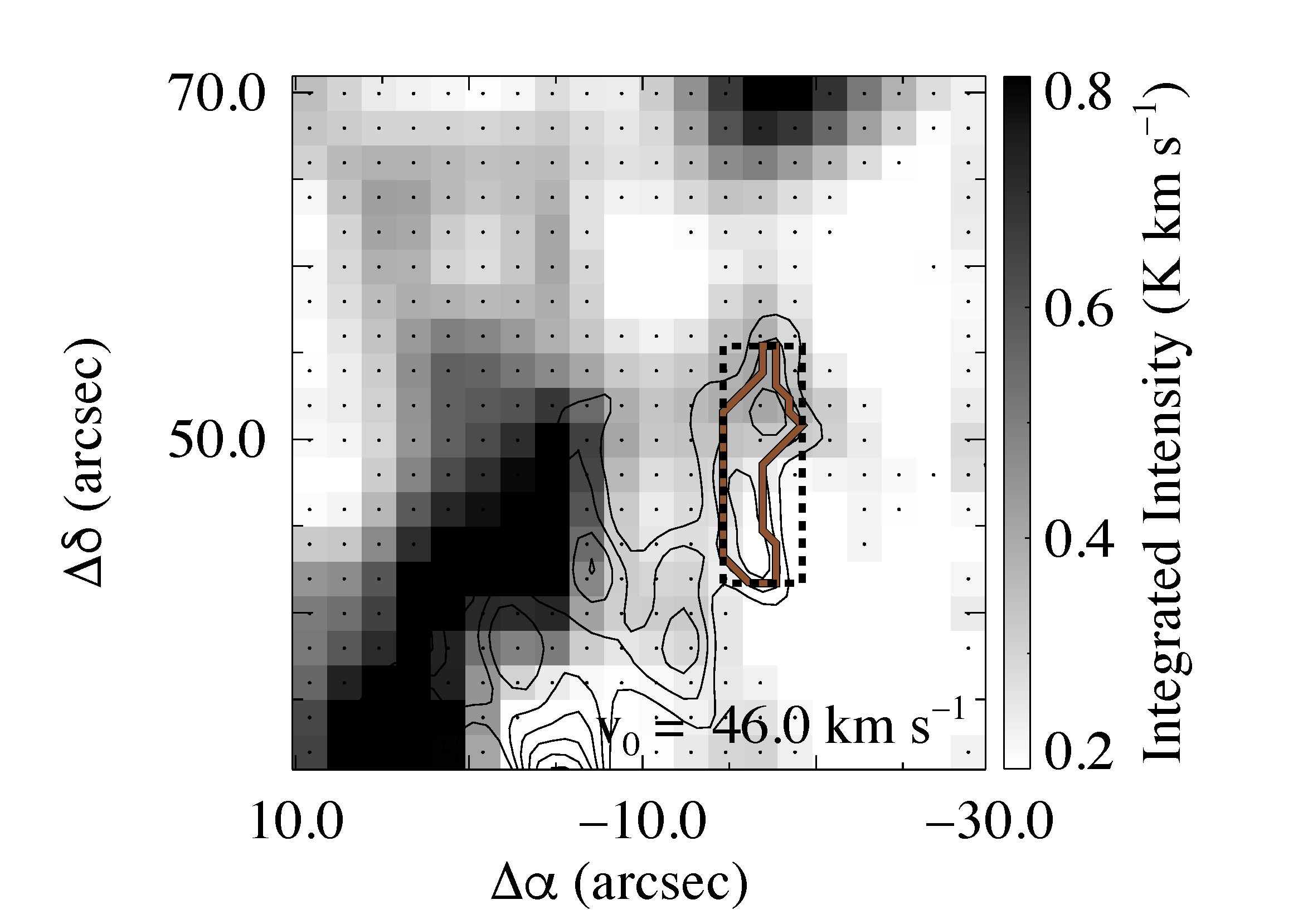}
\includegraphics[trim = 0mm 0mm 0mm 0mm, clip, width = 0.33\textwidth]{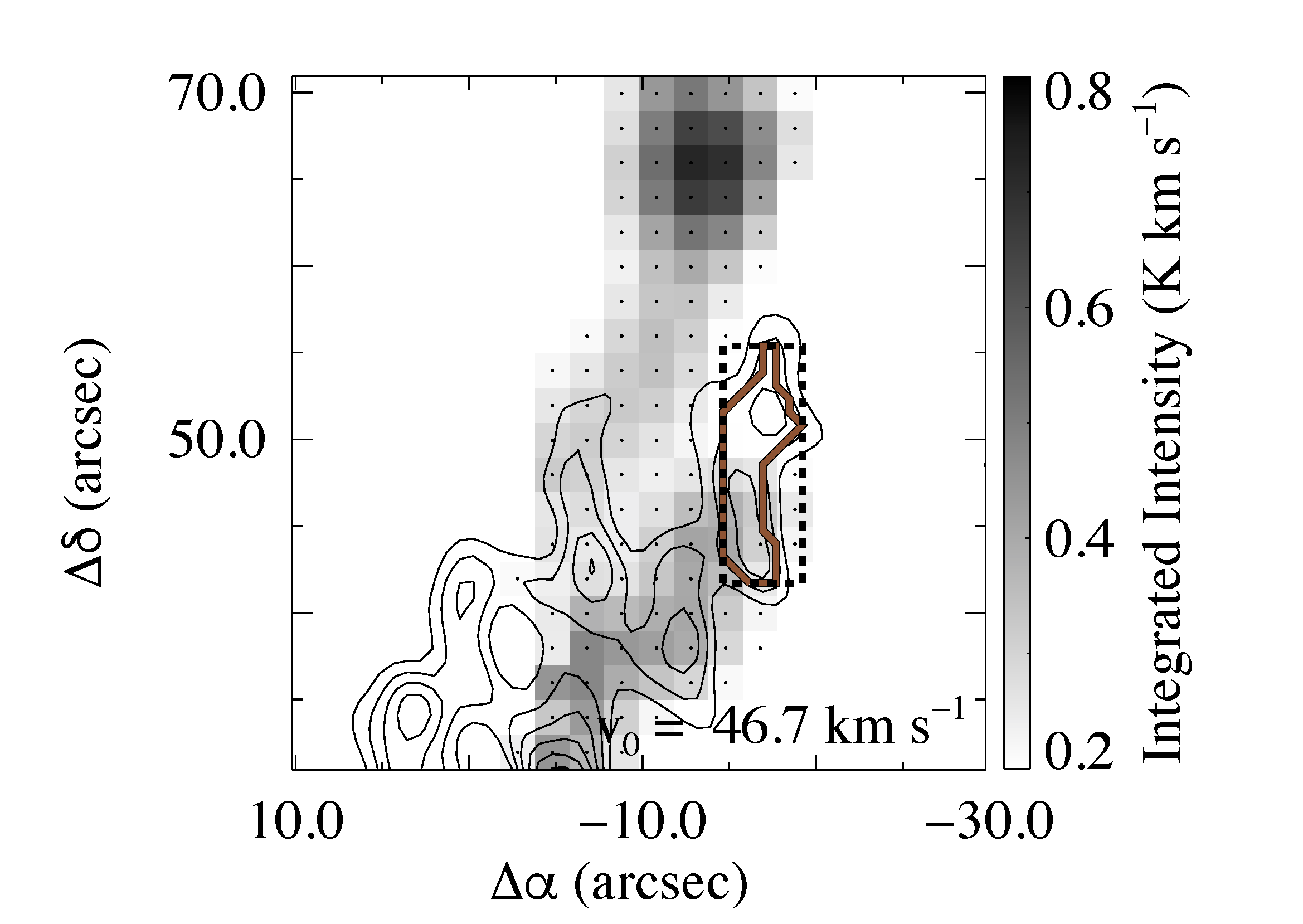}\\

\includegraphics[trim = 35mm 10mm 0mm 0mm, clip, width = 0.45\textwidth]{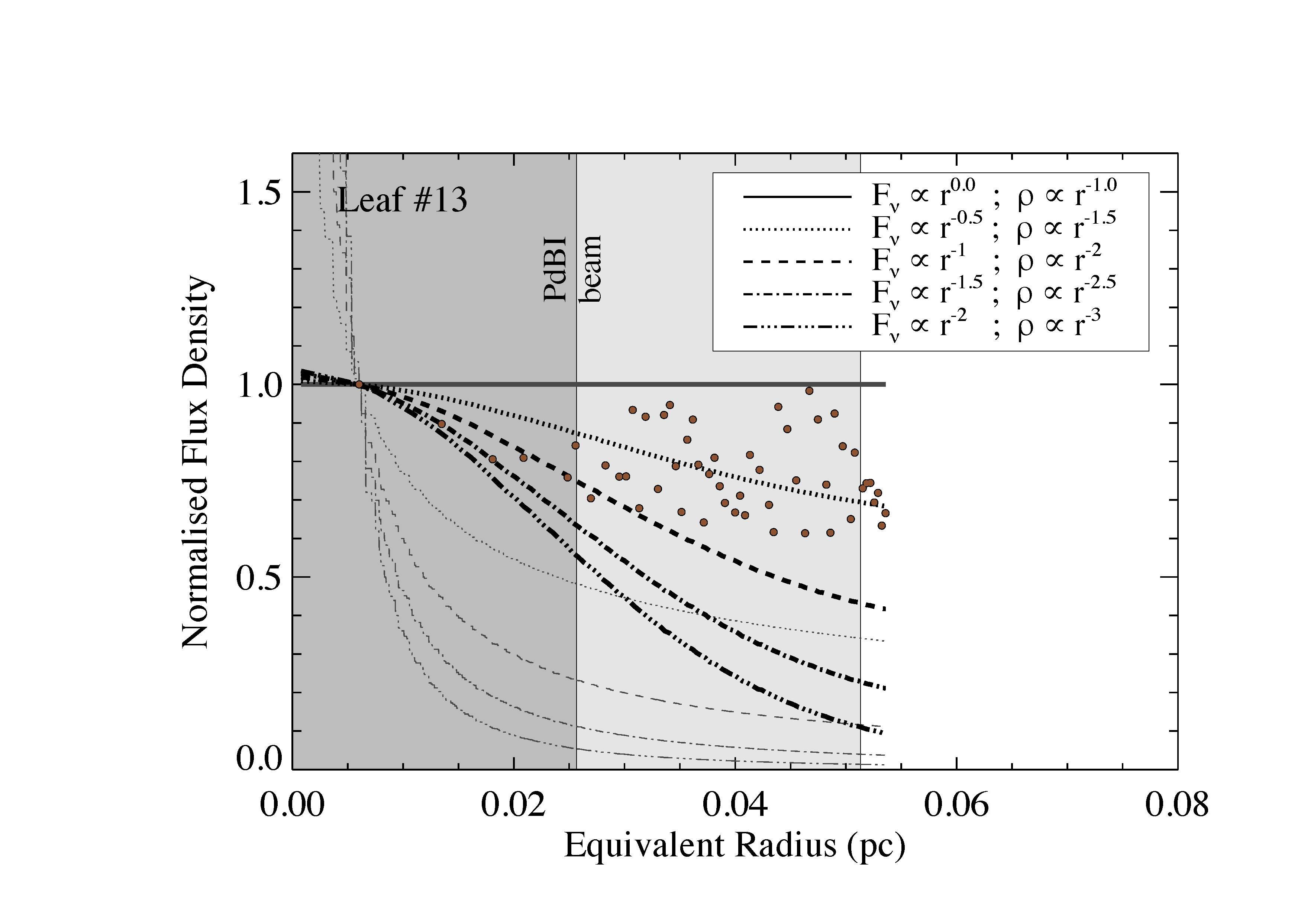}
\end{center}
\caption{The average spectrum, spatial distribution of integrated \ntwoh \ (1-0) emission, and the radial flux density profile of leaf~\#13. Black Gaussian profiles reflect the fact that the continuum flux accredited to leaf~\#13 cannot be unambiguously attributed to a single structure.  }
\label{Figure:leaf13}
\end{figure*}

\noindent\textbf{Leaf~\#13:} is situated at $\{\Delta\alpha,\,\Delta\delta\}=\{-16.9\,{\rm arcsec},\,43.2\,{\rm arcsec}\}$, and is the northernmost of the identified leaves. It has a high aspect ratio of $\AR=3.55$, an irregular-shaped boundary, and the continuum emission has two peaks (see Fig.~\ref{Figure:cont_dendro}), possibly indicating the presence of unresolved fragments (see also Fig.~\ref{Figure:leaf13}). The top-left panel of fig.~\ref{Figure:leaf13} shows the spatially-averaged \ntwoh \ spectrum extracted from within the leaf boundary. Three spectral components are evident. The centroid velocities of the measured components are $v_{\rm 0,1}=45.16\,{\rm km\,s^{-1}}\pm0.26\,{\rm km\,s^{-1}}$, $v_{\rm 0,2}=45.99\,{\rm km\,s^{-1}}\pm0.06\,{\rm km\,s^{-1}}$, and $v_{\rm 0,3}=46.68\,{\rm km\,s^{-1}}\pm0.03\,{\rm km\,s^{-1}}$, respectively. The corresponding FWHM line-widths are $\Delta v_{1}~=~0.91\,{\rm km\,s^{-1}}\pm0.42\,{\rm km\,s^{-1}}$, $\Delta v_{2}~=~0.71\,{\rm km\,s^{-1}}\pm0.22\,{\rm km\,s^{-1}}$, and $\Delta v_{3}~=~0.48\,{\rm km\,s^{-1}}\pm0.06\,{\rm km\,s^{-1}}$, respectively. Visual inspection of the spatial distribution of each component reveals that, similar to leaf~\#5, it appears as though the northern portion of the leaf is associated with one velocity component ($\sim46$\,\kms) whereas the southern portion is associated with another ($\sim47$\,\kms). As a consequence, leaf~\#13 is rejected analysis in \S~\ref{Section:analysis} (other than the mass estimation).  \\

\bsp	
\label{lastpage}
\end{document}